\documentclass[11pt, a4paper]{article}
\pdfoutput=1

\usepackage{jcappub}
\usepackage{subfigure}
\usepackage{bm}
\usepackage{amssymb}
\usepackage{epsfig}
\usepackage{float}
\usepackage{caption}
\usepackage{color}
\usepackage{upgreek}
\usepackage{amsmath}
\usepackage{graphicx}
\usepackage{float}
\usepackage[utf8]{inputenc} 

\newcommand{\be}{\begin{equation}}
\newcommand{\ee}{\end{equation}}
\newcommand{\bea}{\begin{eqnarray}}
\newcommand{\eea}{\end{eqnarray}}

\newcommand{\Rm}[1]{\mathrm{#1}}
\newcommand{\Bf}[1]{\bm{#1}}

\begin{document}
\setcounter{tocdepth}{2}

\title{Multi-Scale Perturbation Theory I: \\Methodology and Leading-Order Bispectrum Corrections in the Matter-Dominated Era}

\author[a]{Christopher S. Gallagher,}
\emailAdd{c.s.gallagher@qmul.ac.uk}

\author[a]{Timothy Clifton,}
\emailAdd{t.clifton@qmul.ac.uk}

\author[a,b,c]{Chris Clarkson}
\emailAdd{chris.clarkson@qmul.ac.uk}

\affiliation[a]{School of Physics and Astronomy, Queen Mary University of London, UK.}
\affiliation[b]{Department of Physics \& Astronomy, University of the Western Cape, Cape Town 7535, South Africa}
\affiliation[c]{Department of Mathematics \& Applied Mathematics, University of Cape Town, Cape Town 7701, South Africa}

\keywords{cosmological perturbation theory, non-gaussianity, power spectrum, gravity}
\arxivnumber{1910.04894}

\abstract{
Two-parameter perturbation theory is a scheme tailor-made to consistently include nonlinear density contrasts on small scales ($<100\; \Rm{Mpc}$), whilst retaining a traditional approach to cosmological perturbations in the long-wavelength universe. In this paper we study the solutions that arise from this theory in a spatially-flat dust-filled cosmology, and what these imply for the bispectrum of matter. This is achieved by using Newtonian perturbation theory to model the gravitational fields of nonlinear structures in the quasi-linear regime, and then using the resulting solutions as source terms for the cosmological equations. We find that our approach results in the leading-order part of the cosmological gravitational potentials being identical to those that result from standard cosmological perturbation theory at second-order, while the dark matter bispectrum itself yields some differences on Hubble scales.  This demonstrates that our approach is sufficient to capture most leading-order relativistic effects, but within a framework that is far easier to generalize. We expect this latter property to be particularly useful for calculating leading-order relativistic corrections to the matter power spectrum, as well as for deriving predictions for relativistic effects in alternative theories of gravity.}

\maketitle
\flushbottom

\bibliographystyle{unsrt}

\section{Introduction}
\label{Introduction} 

Physical cosmology is founded on the Cosmological Principle, that the large-scale universe exhibits \textit{homogeneity} and \textit{isotropy}. This naturally leads one to consider a perturbed Friedmann-Lema\^{i}tre-Robertson-Walker (FLRW) solution as the geometry to model space-time. Provided all fluctuations around the FLRW background are small, we expect to be able to model their behaviour under the influence of gravity by linearising the relevant Einstein field equations. The ``background'' FLRW solution can then be thought of as the space on which the small perturbations live.  

This approach forms the basis of regular relativistic Cosmological Perturbation Theory (or ``CPT" for short) \cite{Bardeen:1980gic}. If greater precision is required, one can consider second-order perturbation theory, where products of first-order quantities are allowed to act as sources \cite{Malik:2008im}. One can then naturally extend this by considering third, fourth, and even $n$-th order perturbation theory, with each order expected to generate smaller corrections than the last \cite{Carlson:2009clc}. At any point in this process, we expect to be able to calculate the correlation functions and power spectra associated with the quantities involved, and hence to determine observables to any given level of accuracy.

Modelling the universe in this way has been enormously successful, especially with regards to observations of the Cosmic Microwave Background (CMB) radiation. It has culminated in the much-celebrated $\Rm{\Lambda}$CDM model of the universe; a model so successful that it can accurately predict the statistical properties of numerous observables with only six free parameters. However, the CMB is not our only probe of inhomogeneity in the universe - one could also hope to model and measure inhomogeneities in the distribution of matter in the universe. This is a key science goal of next-generation galaxy surveys such as the SKA, Euclid and LSST \cite{SKA, Euclid, LSST}.

Predicting the statistical properties of the distribution of matter in the universe turns out to be more complicated than modelling the anisotropy of the CMB, as the gravitational collapse of matter tends to produce very large density contrasts at late times. As the density contrast is treated perturbatively in the approach described above, this means that the source terms for second-order perturbation theory may include terms that are in fact the same size as, or even larger than, those that occur at first-order. Such a result implies that it may not be consistent to exclude these terms at the first level of approximation, and would naively appear to imply that we may need to consider terms at all orders if we want accurate results \cite{Clifton:2010fr, Clarkson:2011zq, Carlson:2009clc}.

One way to try and argue oneself out of this conundrum might be to propose a smoothing scale or cut-off, below which perturbations can be ignored. Such an approach, although of great practical utility, is unsatisfactory on a fundamental level as general relativity is a \textit{nonlinear} theory, which means that perturbations on different length scales can couple together \cite{Carlson:2009clc}. This can be conceptually illustrated in Fourier space - second-order product terms can be transformed into convolution integrals over the whole range of $k$-space, which in turn implies that the evolution equation for a single Fourier mode is dependent on \textit{all} other modes. It therefore seems quite conceivable that some short-scale physics could be affecting evolution even at large scales.

Another response is to argue that although situations with large density contrast are formally outside of the domain of applicability of CPT, one could approach such situations by considering arbitrarily high orders in perturbation theory, and then using the results to extrapolate to the desired configurations. This is motivated by noticing that convergent power series approximations can be used to extrapolate functions away from the point where the expansion is performed (for example in ``analytic continuation"). It does, however, rely on being able to perform perturbation theory to high orders, which at our current level of understanding is not particularly practical (equations would very quickly become unmanageably large).

An approach that takes this problem seriously, and is designed to deal with the effects of large-$k$ modes, is {\it effective field theory} \cite{Baumann:2010tm}. Here the various couplings that appear in the theory are renormalized, to fit the results of measurements or simulations \cite{Carrasco:2012cv}. The hope is that all effects can be incorporated into the renormalized coupling parameters, and that the data will be sufficient to provide the required information about them. These techniques have a long history of success in particle physics, where the short-wavelength (i.e. high energy) physics is assumed to be unknown, and the long-wavelength (low energy) physics is what is required by experimentalists.

However, while it may be the case that the UV completion of a given theory of particle physics is unknown, the short-wavelength physics required to describe nonlinear structures in cosmology (such as galaxies and clusters of galaxies) is very well understood. In particular, the post-Newtonian expansion provides a well-defined way to treat slowly-changing, weak, short-wavelength gravitational fields even when the density contrast is extremely large. These expansions have a long and storied history of successfully describing the gravitational field of nonlinear astrophysical systems, as long as the characteristic velocities in the system remain small (compared to the speed of light) \cite{Will:1981, Poisson:2014}. In other words, unlike in particle physics, we already have the theoretical basis in place to describe the short-wavelength physics, and so it is not necessary to resort to agnostic techniques involving renormalisation. 

The post-Newtonian (PN) expansion is derived by writing the Einstein equations in the form of a null wave equation, and then Taylor expanding the retarded null Green's function solution over short distances. Direct linearisation of the energy density is not required, though velocity fields and time derivatives are taken to be ``small''. This approach forces the various different quantities that appear in the field equations into a hierarchy of different sizes relative to the small peculiar velocity, in direct contrast to what happens in cosmological perturbation theory (where all perturbative quantities are forced to have the same size, as a result of direct linearisation). The question then arises; is it possible to model the effects of short-wavelength gravitational physics in cosmology using the post-Newtonian expansion, whilst retaining a CPT description of long-wavelength physics?

This question was studied in the context of an approach named {\it two-parameter perturbation theory} (``2PPT'', for short) in Refs. \cite{Goldberg:2016lcq, Goldberg:2017gsm, Gallagher:2018bdl}. These papers have shown that this set-up has a well-defined gauge problem, in which the field equations and conservation can be written in terms of gauge-invariant variables. They have shown how the cosmological expansion is related to the gravitational field of the nonlinear structures, and they have verified that the constraint equations are maintained under time evolution. The equations that result reproduce the expected Newtonian equations at leading-order on small scales, and provide a well-defined and self-consistent way to study the effects of these structures on the large-scale cosmological perturbations.

Unfortunately, the features that render the cosmological equations in the two-parameter theory interesting also present severe challenges when it comes to finding solutions. In particular, one is forced to consider spatially inhomogeneous linear differential operators at leading-order in cosmological perturbations (something that does not occur at all in CPT). This makes eigenfunctions difficult to find, as they are dependent on the nonlinear solutions to the Newtonian equations, which are themselves dependent on spatially inhomogeneous and stochastic initial data. Added to this, we have the extra complication that taking derivatives in this formalism is non-trivial, as space and time derivatives do not act in the same way in the two different sectors of the theory.

In this paper, we consider the problem of finding solutions to the equations of two-parameter perturbation theory in an Einstein-de Sitter universe, and using them to calculate the bispectrum of matter. This is achieved by the key assumption that we can use Eulerian perturbation theory in the quasi-linear regime, in order to find approximate solutions to the leading-order post-Newtonian equations. Each two-parameter perturbation in the system is then expanded using the same approach, leading to a hierarchy of linear equations that can be solved order-by-order to get successively more accurate approximations to the original two-parameter equations. The solutions obtained can then be used to calculate the statistical properties of the matter distribution, and hence observables.

We use Greek letters to represent space-time indices, and Latin letters for spatial indices. Dashes refer to differentiation with respect to conformal time, spatial derivatives are denoted by $\partial_i$, and $\nabla^2$ refers to the Laplacian operator associated with spatial partial derivatives in comoving coordinates. We choose to work in geometrized units throughout, in which $G = c = 1$. We also work in longitudinal gauge in both sectors of the two-parameter perturbation theory.

\section{Two-parameter perturbation theory in Einstein-de Sitter Universes}
\label{sec:2p}

In this section we will introduce the two-parameter perturbation theory (2PPT) constructed in Refs. \cite{Goldberg:2016lcq, Goldberg:2017gsm, Gallagher:2018bdl}, as well as the method we will use to solve them. The 2PPT approach simultaneously expands the metric and matter fields in both Cosmological Perturbation Theory (CPT) and post-Newtonian (PN) approximations, and uses the labels $\epsilon\sim 10^{-4}$ and $\eta\sim 10^{-2}$ to label the smallness parameter in each of these two expansions, respectively. All variables $Q$ can then be written as a sum of terms $Q^{(n,m)} \sim \epsilon^n \eta^m$, and equated to each other by expanding the field or conservation equations order-by-order in these parameters. The resulting equations are given in  Appendix \ref{FieldEquationsGaugeInvariantVariables}.

\subsection{Metric and stress-energy tensors}

The 2PPT approach in an Einstein-de Sitter universe was first presented in Ref. \cite{Goldberg:2016lcq}, and we refer the reader to that paper for further details on what we present  here. For the purposes of this paper, we will consider the following metric:
\begin{equation} \label{FLRW1}
ds^2 = a^2(\tau) \Big[ -(1 + 2 U+ 2 \phi) d\tau^2 + \big( (1-2 U - 2 \psi)\delta_{ij}  \big) dx^i dx^j \Big] \, ,
\end{equation}
where the perturbations can be related to the gauge-invariant variables from Ref. \cite{Goldberg:2017gsm} as $U \equiv - {\textstyle \frac{1}{2}} \left( \Phi^{(0,2)}\right)$, $\phi \equiv - {\textstyle \frac{1}{2}} \left( \Phi^{(1,0)}  \right)$ and $\psi \equiv {\textstyle \frac{1}{2}} \left(\Psi^{(1,0)} \right)$.  Roughly speaking, one can refer to $U$ as the Newtonian gravitational potential, and to $\phi$ and $\psi$ as the cosmological perturbations.

Other terms that occur in the full 2PPT treatment, such as vectors and tensors, higher-order PN corrections, and mixed-order terms (e.g. $\Phi^{(1,1)}$) have been neglected in Eq. (\ref{FLRW1}), but are included in the full equations given in Appendix \ref{FieldEquationsGaugeInvariantVariables}. The reader may also note that Eq. (\ref{FLRW1}) is written in longitudinal gauge. This gauge takes on a special status in the 2PPT approach, as it turns out to be the only commonly used gauge in which both the PN and CPT expansions can be simultaneously performed. This issue is explained further in Ref. \cite{GoldbergThesis}, and is examined in detail in Ref. \cite{Clifton:2020oqx}. It is used to construct gauge-invariant variables, and to explore the properties of this formalism under gauge transformations, in Ref. \cite{Goldberg:2016lcq}.

Correspondingly, the stress-energy tensor for dust can be written in the form 
\begin{align}
T^{\mu\nu} = (\rho_{\Rm{N}} + \rho)\,u^{\mu}u^{\nu} \;,
\end{align}
where $u^{\mu} = \frac{1}{a}(1-U - \phi, \; v_{\Rm{N}}^i + v^i)$. The matter perturbations are related to those used in Ref. \cite{Goldberg:2016lcq} by $\rho_{\Rm{N}} \equiv  \Bf{\rho}^{(0,2)}$, $\rho \equiv  \Bf{\rho}^{(1,0)}$, $v_{\Rm{N}i} \equiv \mathbf{v}^{(0,1)}_i$ and $v_{i} \equiv \mathbf{v}^{(1,0)}_i$, and where we have again neglected mixed-order and higher-order PN quantities. Such quantities will be considered further in a follow-up paper \cite{Gallagher:2019lcd}, where we will also use methods similar to those deployed in Refs. \cite{Villa:2015ppa, Nakamura:2007} to include vector and tensor modes.

\subsection{Action of derivatives}
\label{secder}

The defining assumption of the PN expansion is that the velocity is small, $v_{\Rm{N}} \sim \eta$. This, in turn, mandates that time derivatives of PN quantities are small with respect to spatial derivatives, i.e.
\begin{align}
\frac{dx}{dt} \sim \frac{1}{\nabla} \cdot \frac{d}{dt} \sim \eta \;. 
\end{align}
Accordingly, we treat time derivatives of a quantity that are perturbed in the PN sector as having an extra factor of $\eta$, as compared to spatial derivatives of the same object. Quantities that are exclusively perturbed in the CPT sector do not have this property, as derivatives do not change the size of quantities in this approach (this is required of CPT, if it is to be applicable to both sub and super-horizon scales). 

Mathematically, for PN perturbed quantities we have
\be \label{ndiff}
 N_,i \sim \frac{N}{L_{\Rm{N}}}
\qquad {\rm and} \qquad
N^\prime \sim \eta \, \frac{ N}{L_{\Rm{N}}} \; ,
\ee
where $L_{\Rm{N}}$ is the length scale associated with the Newtonian system, while for CPT we have
\be
{C}_{,i} \sim \frac{C}{L_{\Rm{C}}}
\qquad {\rm and} \qquad
C^\prime \sim \frac{C}{L_{\Rm{C}}} \; ,
\ee
where $L_{\Rm{C}}$ is the length scale associated with the cosmology (i.e. the horizon). The reader may note the use of symbol ``$N$" to denote a generic PN quantity, and symbol ``$C$'' to denote a generic cosmological quantity.

In the 2PPT approach we wish to consider both types of perturbation together, in the same equations. This requires choosing a set of units. If we choose to write all of derivatives in units of $L_{\Rm{N}}$, then we need to additionally choose a relationship between $L_{\Rm{N}}$ and $L_{\Rm{C}}$. Following Ref. \cite{Goldberg:2016lcq}, we choose
\be \label{choice}
L_{\Rm{N}} = \eta L_{\Rm{C}} \;,
\ee
which is consistent with $L_{\Rm{C}} \sim 30 \, \Rm{Gpc}$ and $L_{\Rm{N}} \sim 100 \, \Rm{Mpc}$ (these are the length scale of the particle horizon and typical superclusters, respectively). This choice also results in square of the Hubble factor being the same order of magnitude as the leading order PN density of dust $\mathcal{H}^2\sim \rho$, which in turn yields the nice property that the cosmological expansion is sourced by the average Newtonian mass density in the universe.

This can all be encapsulated in the following rules of thumb for dealing with derivatives:
\begin{itemize} \label{diffsizing}
\item[(i)] Taking spatial derivatives of CPT quantities adds a factor of $ \displaystyle \frac{\eta}{L_{\Rm{N}}}$.
\item[(ii)] Taking spatial derivatives of PN quantities adds a factor of $\displaystyle \frac{1}{L_{\Rm{N}}}$. 
\item[(iii)]  Taking time derivatives adds factors of $\displaystyle \frac{\eta}{L_{\Rm{N}}}$ to both CPT and PN perturbed quantities.
\end{itemize}
For further details of post-Newtonian gravity the reader is referred to the textbooks by Will \cite{Will:1981} and Poisson \& Will \cite{Poisson:2014}, and for cosmological perturbation theory to the review by Malik \& Wands \cite{Malik:2008im}.

\subsection{Field Equations}

Using the metric given in Eq. (\ref{FLRW1}), and taking the rules from Section \ref{diffsizing} into account, one finds that the leading-order field equations are given at order $\sim\eta^2/L_{\Rm{N}}^2$ by
\begin{align} \label{leadingorder}
\mathcal{H}' &= -\frac{4\pi a^2}{3}\rho_{\Rm{N}} - \frac{1}{3}\nabla^2 U \;, \\[5pt]
\mathcal{H}^2 &= \frac{8\pi a^2}{3}\rho_{\Rm{N}} + \frac{2}{3}\nabla^2 U  \;, 
\end{align}
where $\mathcal{H}=a^{\prime}/a$ is the conformal Hubble rate, and primes denote differentiation with respect to conformal time, $\tau$. By averaging these equations it can be seen that we obtain
\begin{align} 
\mathcal{H}' &= -\frac{4\pi a^2}{3}\bar{\rho}  \;, \label{accel} \\
\mathcal{H}^2 &= \frac{8\pi a^2}{3}\bar{\rho} \label{friedmann}  \;,
\end{align}
which leaves the fluctuations around the average given by
\begin{align} 
\nabla^2  U &= 4\pi a^2 \updelta \rho_{\Rm{N}}  \label{NewtonPoisson}\;,
\end{align}
where $\bar{\rho} $ denotes the mean of $\rho_{\Rm{N}}$, and $\updelta \rho_{\Rm{N}} $ denotes the fluctuation around the mean. This average value of $\rho_{\Rm{N}}$ must be the same at all points in the Universe, otherwise these equations are inconsistent with the initial assumption of a background FRW metric with $a=a(\tau)$. 


Equations (\ref{accel}) and (\ref{friedmann}) are identical to the Friedmann equations for an Einstein-de Sitter (EdS) universe, and admit the well-known solution
\begin{align}
    a &= \tau^2 \;, \\
    \mathrm{which \;implies} \quad \mathcal{H} &= \frac{2}{\tau} \;.
\end{align}
 Likewise, Eq. (\ref{NewtonPoisson}) can be seen to be identical to the Poisson equation of Newtonian gravity on an expanding background, and correspondingly the solutions for $U$ must be given by the linear sum of Newtonian gravitational potentials of all matter fields.

As was demonstrated in Ref. \cite{Gallagher:2018bdl}, conservation of the Einstein constraint equations under time evolution demands that $\updelta_{\Rm{N}} \equiv \updelta \rho_{\Rm{N}} / \bar{\rho}$ and $v_{\Rm{N}i}$ satisfy the continuity equation and Euler equations:
\begin{align}
\updelta_{\Rm{N}}' + \partial^i(v_{\Rm{N}i}\big(1+\updelta_{\Rm{N}})\big) &= 0\;, \label{continuity}\\[5pt]
v_{\Rm{N}i}' + \mathcal{H} v_{\Rm{N}i} + \partial_iU + v_{\Rm{N}j}\partial^j v_{\Rm{N}i} \label{euler} &=0 \;.
\end{align}
Under the assumption of vanishing vorticity, these expressions form a closed \textit{nonlinear} system for the three Newtonian perturbations $\{U, \updelta_{\Rm{N}},v_{\Rm{N}i} \}$. Their solutions should be understood as the leading-order contribution to the PN sector of the theory, with subsequent higher-order corrections representing relativistic effects. Techniques for finding solutions to this system (for a given initial matter distribution) are usually obtained using either Newtonian N-body simulations, or Newtonian perturbation theory (or ``NPT" for short). We will use the latter in this study, though the reader may wish to keep in mind that an all-orders resummed NPT solution still only constitutes the leading-order contribution to the gravitational field in the 2PPT set up.

The next order of field equations is at $\sim {\eta^4}/{L_{\Rm{N}}^2}$. Neglecting vectors and tensors, the evolution equation for the scalar degree of freedom and
the trace-free $ij$ field equation give

\begin{align} 
(\psi+ U)'' + 3\mathcal{H}(\psi+ U)' =&\;  \frac{4\pi a^2\bar{\rho} }{3}(1+ \updelta _{\Rm{N}})v_{\Rm{N}}^2 + \mathcal{H}(\psi'-\phi')  + \frac{1}{3}\nabla^2 (\psi - \phi) 
\nonumber \\ &   
 + \frac{7}{6}(\nabla U)^2 + \frac{2}{3}(\phi + \psi + 2U)\nabla^2 U \;,  \label{evol} \\
\partial^i \partial_j (\psi - \phi) + 2 \partial^i U \partial_j U + 2 (\psi + \phi + 2 U)& \partial^i \partial_j U - \frac{1}{3} \delta^i_{\;j} \bigg[ \nabla^2  (\psi - \phi) + 2  ( \nabla U )^2  + 2 (\psi + \phi + 2 U) \nabla^2 U \bigg] \nonumber \\
&= 8\pi a^2\bar{\rho} \; (1 +  \updelta_{\Rm{N}})\big( v_{\Rm{N}}^i v_{\Rm{N}j} - \frac{1}{3}  \delta^i_{\;j}  v_{\Rm{N}}^2 \big) \label{tracefreeij} \;,
\end{align}
while the generalised Poisson and momentum constraint equations give
\begin{align}
\frac{1}{3} \nabla^2 \psi - \mathcal{H}(\psi' + U') - \mathcal{H}^2(\phi+U) =&\; \frac{4\pi a^2 \bar{\rho}}{3}\updelta + \frac{4\pi a^2 \bar{\rho}}{3}(1 + \updelta_{\Rm{N}})v_{\Rm{N}}^2  \nonumber \\ 
&- \frac{1}{2} (\nabla U)^2 - \frac{4}{3}(\psi + U) \nabla^2 U \; ,  \label{genPoisson} \\
\partial_i \big( \psi'  + \mathcal{H}\phi \big)  =& - \frac{3\mathcal{H}^2}{2} (1 + \updelta_{\Rm{N}}) v_i \; \label{momentum} \;.
\end{align}
These equations can be seen to contain quadratic and even cubic products of lower-order perturbations, as well as products of (unsolved-for) cosmological perturbations and (solved-for) Newtonian perturbations, in ways that simply cannot occur in standard CPT.

For the rest of the paper, we will refer to Eqs. (\ref{evol})--(\ref{momentum}) as the \textit{2PPT field equations}. Although these equations are not the only field equations that can be derived using the 2PPT formalism, they do contain the critical physics that the formalism seeks to investigate; the effects of small-scale nonlinearities on large-scale cosmological perturbations. In fact, one could think of these equations as a set that describe {first-order cosmological perturbations} on top of a universe that already contains nonlinear structure on small scales. In this sense, they model cosmological back-reaction of small-scale structure on the large-scale Universe, within a well-defined framework.
   
\subsection{The utility of Newtonian perturbation theory} \label{Necessity}

Equations (\ref{evol}) and (\ref{tracefreeij}) are difficult to solve. There are a number of reasons for this, including the fact that $\updelta_{\Rm{N}}$, $U$ and $v_{\Rm{N}i}$ are themselves the solutions to nonlinear differential equations (the Eulerian equations of fluid dynamics), and as such are complicated functions of initial conditions. This renders the linear differential operators on the left-hand sides of these equations dependent on spatial position, which makes it is unclear what set of eigenfunctions should be used as a basis for constructing solutions. 

We may compare this to the situation in CPT, where the first-order equations can be expressed heuristically as
\begin{align}
\hat{\mathcal{L}}_{\rm CPT}(\tau) \, \bm{u}_1 = 0 \; , \label{1heuristic}
\end{align}
where $\hat{\mathcal{L}}_{\rm CPT}(\tau)$ is a matrix-valued linear differential operator containing both spatial and temporal derivatives, {but which functionally depends only on conformal time}. The $\bm{u}_1$ in this equation is intended to denote all first-order quantities ($\phi$, $\psi$, $\updelta$, $\dots$) arranged into a column vector.
This homogeneous matrix-valued differential equation can easily be diagonalised in either real space or Fourier space, as $\hat{\mathcal{L}}_{\rm CPT}(\tau)$ does not depend on space. 

Similarly, higher-order CPT equations can be written as
\begin{align}
\hat{\mathcal{L}}_{\rm CPT}(\tau) \, \bm{u}_2 &\sim \bm{u}_1^2 \;  \label{2heuristic} \\[3pt]
\hat{\mathcal{L}}_{\rm CPT}(\tau) \, \bm{u}_3 &\sim \bm{u}_1\bm{u}_2 + \bm{u}_1^3 \; , \label{3heuristic}
\end{align}
where numerical subscripts denote the order of a quantity in the CPT expansion. The key point to note here is that at each order the linear differential operator $\hat{\mathcal{L}}_{\rm CPT}(\tau)$ remains the same, so successive approximations can be found by identifying particular solutions for given source terms and then simply adding them to the original first-order solution. 

It is immediately apparent that two-parameter perturbation theory does not follow this structure: The leading-order evolution equations are \textit{nonlinear}, and the sub-leading field equations (\ref{evol})--(\ref{momentum})) cannot be written in the form of Eq. (\ref{2heuristic}).  Instead, what we have is an equation of the form
\begin{align} \label{2Pheuristic}
\hat{\mathcal{L}}_{\rm 2PPT}(\tau, U, \updelta_{\Rm{N}}, v_{\Rm{N}i}) \, \bm{u}_{\eta^4} 
 = \bm{u}_{\eta^2}^2 \;,
\end{align}
where $\bm{u}_{\eta^2}$ is a column vector of the leading-order nonlinear solutions ($\sim {\eta^2}/{L_{\Rm{N}}^2}$) and $\bm{u}_{\eta^4}$ is a column vector of the sub-leading-order solutions ($\sim {\eta^4}/{L_{\Rm{N}}^2}$). It can be seen that the linear operator in this equation, $\hat{\mathcal{L}}_{\rm 2PPT}(\tau, U, \updelta_{\Rm{N}}, v_{\Rm{N}i})$, is a function of the nonlinear solutions to the leading-order field equations, which themselves are complicated functions of stochastic initial conditions.


Solving Eq. (\ref{2Pheuristic}) requires care; the usual strategy for diagonalising the linear operator in equations of this type involves taking spatial derivatives of the trace-free $ij$-field equation (\ref{tracefreeij}), and using the result to eliminate derivatives of the combination $\psi - \phi$ from the evolution equation (\ref{evol}). In the case of 2PPT, however, taking spatial derivatives will affect post-Newtonian and cosmological terms in different ways, as explained in Section \ref{secder}. We must be careful to ensure that this operation is performed consistently, and that product terms that can exist at higher orders do not influence the results. 

Let us demonstrate this with an example; differentiating the first term in the trace-free $ij$-field equation (\ref{tracefreeij}) results in
\begin{align}
\partial_i \partial^j (\partial^i \partial_j (\psi -\phi))= \nabla^4(\psi-\phi) \sim \frac{\eta^6}{L_{\Rm{N}}^4} \; .
\end{align}
The original equation was order $\sim {\eta^4}/{L_{\Rm{N}}^2}$, while this term is now at order $\sim {\eta^6}/{L_{\Rm{N}}^4}$; We say that the spatial derivatives have ``promoted" this term to higher order. This is potentially problematic, as terms in the trace-free $ij$-field equation at order $\sim {\eta^6}/{L_{\Rm{N}}^2}$ will also appear at order $\sim {\eta^6}/{L_{\Rm{N}}^4}$ after differentiation (e.g. $U \delta \rho_{\Rm{N}} v_{\Rm{N}i} v_{\Rm{N}j}$). Such terms therefore need to be considered at the same time, in any consistent treatment. Similar issues arise when using ``inverse Laplacians", as the action of inverse derivatives can also affect a quantity's size.

The net effect of this is that applying $\partial_i\partial^j$ to Eq. (\ref{tracefreeij}) results in an equation with the schematic form
\begin{align}
\nabla^4 (\phi - \psi) + \mathcal{I}(\tau, x) + \mathcal{T}^{ij}(\tau, x)v_i v_j  = \mathcal{S}(\tau, x) \label{divtracefreeij}\;,
\end{align}
where $\mathcal{I}(\tau, U, \updelta_{\Rm{N}}, v_{\Rm{N}i}, \phi, \psi, v_{i}, \updelta)$ and $ \mathcal{T}^{ij}(\tau, U, \updelta_{\Rm{N}}, v_{\Rm{N}i})$ are functions of both Newtonian and cosmological perturbations. 
%
%
Equation (\ref{divtracefreeij}) is nonlinear in $v_i$, and has a particularly complex operator structure (differential operators depend inhomogeneously on the leading-order solutions of the nonlinear Eulerian equations). This means that simply applying an inverse Laplacian, as one might do in CPT, will not be sufficient here.

\begin{figure*}
\centering
\includegraphics[width=0.78\linewidth]{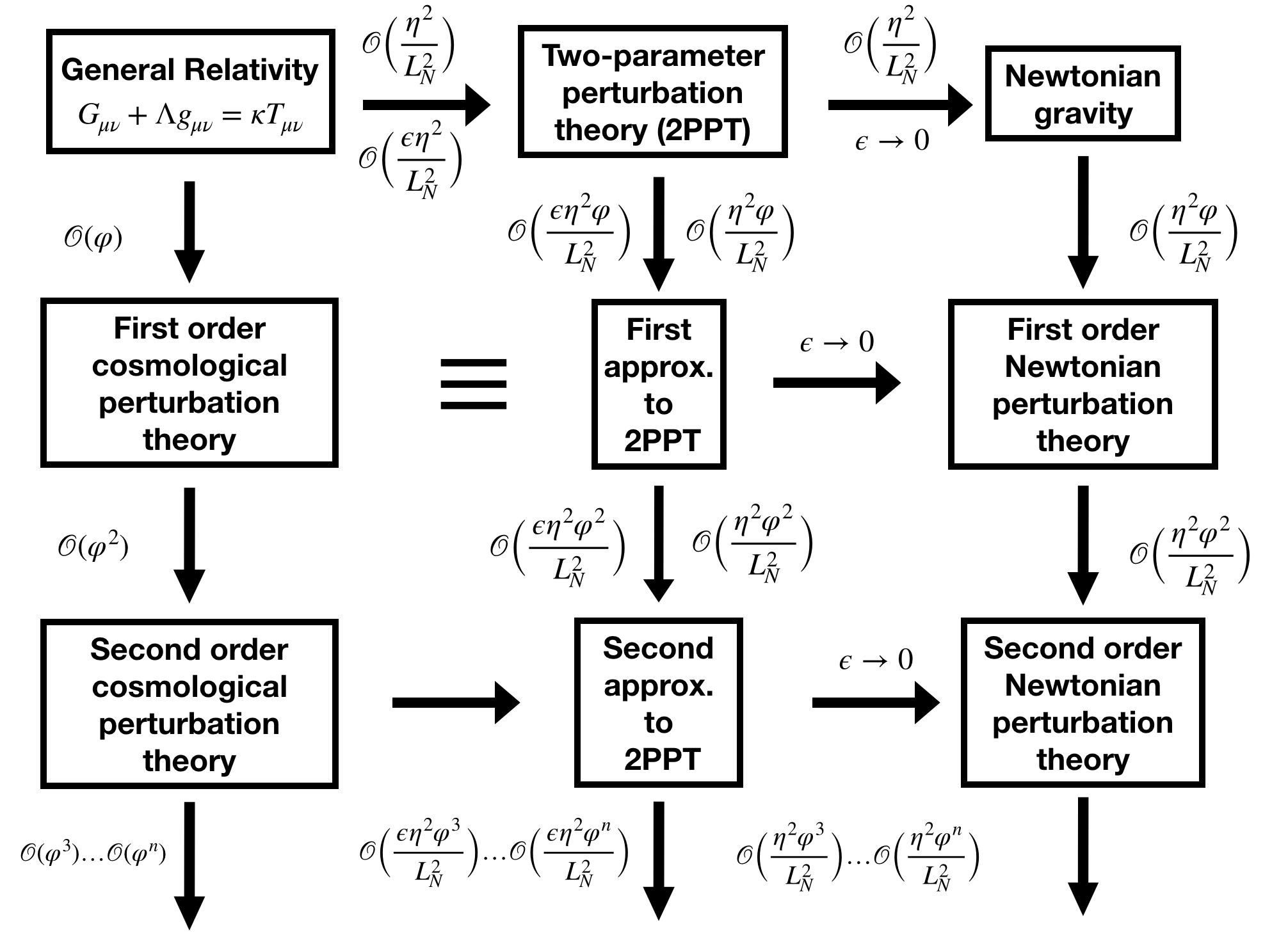}
\caption{A flowchart illustrating the differences and relationships between between CPT and 2PPT.}
\label{flowchart}
\end{figure*}

In order to proceed analytically, it is therefore useful to make further approximations. To this end, we will use NPT to solve the continuity and Euler equations (\ref{continuity})-(\ref{euler}). This works by performing a series expansion on all quantities in the Newtonian density contrast  $\updelta_{\Rm{N}}$, which in the present case can be equivalently given in terms of the seed gravitational potential, $\varphi$. This means that we write, for example, the Newtonian gravitational potential as
\begin{align} \label{Npertseries}
U &= U^{(1)} + \frac{1}{2} U^{(2)} + \dots = \sum_{n=1}^{\infty} \frac{U^{(n)}}{n!}  \;, 
\end{align}
where the superscript here denotes the order in $\varphi$. Similar expansions apply to all other variables in Eqs. (\ref{continuity})-(\ref{euler}), which can then be solved for order-by-order to get approximate solutions in the nonlinear regime of structure formation. The series solutions for Newtonian quantites can then be used to solve Eqs. (\ref{evol}), (\ref{genPoisson}) and (\ref{divtracefreeij}), where cosmological quantities are similarly expanded in $\varphi$, and the equations are again solved order-by-order in $\varphi$.

This further expansion should formally be regarded as a third (and separate) expansion to the two that have already been performed in 2PPT, this time associated with the linear fluctuations $\varphi$. In what follows, we will refer to terms of order $\sim \varphi^n$ as the ``$n$\textit{-th approximations}" to whatever equations they are intended to solve (e.g. the ``$2^{\Rm{nd}}$ approximation to the 2PPT evolution equation" or the ``$1^{\Rm{st}}$ approximation to the Newtonian Euler equation"). It is important to note that taking derivatives will not alter the powers of quantities in $\varphi$, as the series expansion associated with $\varphi$ does not require making any assumptions about spatial or temporal scales.

The fundamental differences and relationships between the approach we propose, and the standard CPT and NPT approaches, are illustrated in Figure (\ref{flowchart}). At the top-left of this diagram we have the full, unperturbed theory of general relativity. CPT can be found within the solutions of the full general theory by hypothesising the existence of a background FLRW solution, and then considering perturbations around this background that are expanded in powers of the initial fluctuations, $\varphi$. First, second and higher-order CPT is obtained by simply working to higher and higher orders in $\varphi$, which corresponds to moving down the figure. On the other hand, the Newtonian limit of general relativity is obtained by performing an expansion in $\eta$. This is illustrated in the figure by first performing the 2PPT expansion (which corresponds to the first step right), and then taking the limit $\epsilon\rightarrow 0$ (which corresponds to the second step right). Both 2PPT and Newtonian theories can then also be expanded in $\varphi$, which again corresponds to moving down the figure.


We will demonstrate subsequently that first-order CPT and the first approximation to 2PPT are identical, and that the $\epsilon \rightarrow 0$ limit of 2PPT reproduces the first-order results of NPT. At second approximation, however, there is no longer an exact correspondence between CPT and 2PPT; The constraint equations differ, and the resulting phenomenology is therefore no longer the same. This will be explained in more detail in the sections that follow, followed by some discussion of exactly what effects are being included or neglected in each approach, as well as the corresponding benefits and drawbacks.


\subsection{Outline of the paper}

The method that we are proposing to find analytic solutions to the 2PPT field equations depends critically on NPT. We will therefore give a concise pedagogical description of this subject in Section \ref{npt}. Following that, we will describe the techniques that have been developed for solving the second-order CPT field equations in Poisson gauge in Section \ref{cpt}, as these techniques turn out to be precisely the ones that will also be required to find solutions to the 2PPT field equations at second approximation. The second-order CPT results will also provide us with a baseline against which we can compare our own results.


After calculating solutions in NPT, second-order CPT and the second approximation to 2PPT, we will calculate the tree-level dark matter bispectrum in Poisson gauge as a demonstration of the similarities and differences between all three approaches. This statistic has the benefit of being relatively easy to calculate, unlike direct observables and some other integrated quantities \cite{Umeh:2016nuh, Jolicoeur:2017nyt, Jolicoeur:2017eyi, Jolicoeur:2018blf}, and therefore provides a useful window through which we can view the predictions of the various different perturbation theories that are considered in this paper. 

Some of our results are conceptually similar to those found in Ref. \cite{Castiblanco:2018qsd}, which also utilizes a weak-field equations on a Friedmann background. However, the details are significantly different, due to the different structure of the field equations.

\section{Newtonian Perturbation Theory} \label{npt}
  
Newtonian perturbation theory is a method for finding analytic solutions of the Newtonian continuity and Euler equations (\ref{continuity})--(\ref{euler}),
used principally in the weakly nonlinear regime. In particular, we will give an introduction to \textit{Goroff's method} \cite{Goroff:1986ep}, which allows one to find expressions for the fastest growing mode of the solution for Einstein-de Sitter universes at \textit{all orders}. These  solutions are given in terms of \textit{kernels}, which encapsulate the mode-coupling behaviour induced by the nonlinearity of the equations \cite{Bernardeau:2001qr}. 

\subsection{Finding solutions}

After expanding all quantities as in Eq. (\ref{Npertseries}), Goroff's method proceeds by inserting the resulting series into the nonlinear Newtonian evolution equations (\ref{continuity})--(\ref{euler}) and the Newton-Poisson equation (\ref{NewtonPoisson}). Discarding all terms that are quadratic and higher-order in $\varphi$, one can then solve the linearised equations for the leading-order terms,
which in Fourier space give the following:
\begin{align} \label{linearnewtonpert}
\updelta_{\Rm{N}}^{(1)\prime\prime}(\Bf{k},\tau) + \mathcal{H} \updelta_{\Rm{N}}^{(1)\prime}(\Bf{k},\tau) - \frac{3}{2}\mathcal{H}^2 \updelta_{\Rm{N}}^{(1)}(\Bf{k},\tau) =0 \;.
\end{align}
Performing a separation of variables, such that $\updelta_{\Rm{N}}^{(1)}(\Bf{k},\tau) = \updelta_{0}^{(1)}(\Bf{k}) \mathcal{D}(\tau)$, one finds solutions described by a spatial initial condition, $\updelta_{0}^{(1)}(\Bf{k})$, and a growth factor, $\mathcal{D}(\tau)$. We are then free to take $\updelta_{0}^{(1)}(\Bf{k})$ to be a Gaussian random field, and can find that the growing solution in EdS has $\mathcal{D}(\tau) = a$. 

To find the solution for the velocity field, it is convenient to write the velocity divergence as $\theta_{\Rm{N}} \equiv \partial^i v_{\Rm{N}i}\;,$ which allows the linearized continuity equation (\ref{continuity}) to be written as
\begin{align}
\updelta_{\Rm{N}}^{(1)\prime} + \theta_{\Rm{N}}^{(1)} = 0 \;,
\end{align}
which in turn implies
\begin{align}
\theta_{\Rm{N}}^{(1)} (\Bf{k},\tau)= -\mathcal{H}(\tau) a(\tau) \updelta_{0}^{(1)}(\Bf{k}) \;.
\end{align}
Likewise, the leading-order Newton-Poisson equation (\ref{NewtonPoisson}) can be linearized and written as
\begin{align}
\nabla^2 U^{(1)} = \frac{3\mathcal{H}^2}{2}\updelta_{\Rm{N}}^{(1)} \;,
\end{align}
which has the solution
\begin{align}
U^{(1)}(\Bf{k}) = -\frac{6}{k^2}\updelta_{0}^{(1)}(\Bf{k}) 
\;,
\end{align}
where we have used $\mathcal{H}^2a = 4$ (valid in EdS). The gravitational potential in this expression can be seen to time-independent, and we therefore have $U^{(1)}(\Bf{k})=\varphi$, where $\varphi$ is the initial condition for the gravitational potential.
 
Having obtained the linear solutions, one can then proceed to find second and higher-order solutions to Eqs. (\ref{NewtonPoisson}), (\ref{continuity}) and (\ref{euler}). Such equations will contain inhomogeneous source terms, which are constructed from quadratic (or higher) products of the solutions found at the previous order. Such equations have general solutions that are given by solutions to the linear equations added to a particular solution for the equations with the extra source terms. This process can be repeated iteratively to find successively higher-order corrections, each (one may hope) providing a better approximation to the full solutions than the last. 

Given that the linear growth factor in EdS is given by $\mathcal{D} = a$, it is not hard to convince oneself that the time-dependence of the particular solution for the second-order quantities should be $\sim a^2$. Continuing the same logic to third-order we find time dependence $\sim a^3$, and to $n$-th order $\sim a^n$. All that is left is then to determine the spatial dependencies at each order, which contains information about the mode coupling. This motivates us to adopt the following expansions for our perturbations:
\begin{align}
\updelta_{\Rm{N}}(\Bf{k},\tau) &= \sum_{n=1}^{\infty} \frac{\updelta_{\Rm{N}}^{(n)}(\Bf{k})}{n!}\; a^n  \label{deltaNsplit} \\
\theta_{\Rm{N}}(\Bf{k},\tau) &= -\mathcal{H}\sum_{n=1}^{\infty} \frac{\theta_{\Rm{N}}^{(n)}(\Bf{k})}{n!}\; a^n \label{thetaNsplit} \;,
\end{align}

\noindent
where it can be found that $\updelta_{\Rm{N}}^{(n)}(\Bf{k})$ and $\theta_{\Rm{N}}^{(n)}(\Bf{k})$ are given solutions of the form
\begin{align} \label{kernels}
\updelta_{\Rm{N}}^{(n)}(\Bf{k}) = \int \bigg( \prod_{i=1}^n \frac{\Rm{d}^3 q_i}{(2\pi)^{3i}} \updelta_{0}^{(1)}(\Bf{q}_i)   \bigg)(2\pi)^3 \delta^{(3)}\left(\Bf{k} - \sum_{i=1}^n \Bf{q}_i\right) F^{(s)}_{n}(\Bf{q}_1,\Bf{q}_2,...,\Bf{q}_n) \;, \\
\theta_{\Rm{N}}^{(n)}(\Bf{k}) = \int \bigg( \prod_{i=1}^n \frac{\Rm{d}^3 q_i}{(2\pi)^{3i}} \updelta_{0}^{(1)}(\Bf{q}_i)   \bigg)(2\pi)^3 \delta^{(3)}\left(\Bf{k} - \sum_{i=1}^n \Bf{q}_i \right) G^{(s)}_{n}(\Bf{q}_1,\Bf{q}_2,...,\Bf{q}_n) \;,
\end{align}
and where $F^{(s)}_n$ and $G^{(s)}_n$ are the symmetrised versions of
\begin{align}
F_n(\Bf{k}_{1...n} ) =& \sum_{m=1}^{m=n-1} {n \choose m} \frac{G_{n-m}(\Bf{k}_{m...n}) }{(2n+3)(n-1)}\bigg\{ (2n+1) \alpha(\Bf{k}_{1:m},\Bf{k}_{m:n}) F_m(\Bf{k}_{1...m}) 
\nonumber \\ &\;\;\;\;\;\;\;\;\;\;\;\;\;\;
+2 \beta(\Bf{k}_{1:m},\Bf{k}_{m:n})G _m(\Bf{k}_{1...m})\bigg\}\; , \\
G_n(\Bf{k}_{1...n} )  =& \sum_{m=1}^{m=n-1} {n \choose m} \frac{G_{n-m}(\Bf{k}_{m...n}) }{(2n+3)(n-1)}\bigg\{ 3 \alpha(\Bf{k}_{1:m},\Bf{k}_{m:n}) F_m(\Bf{k}_{1...m}) 
\nonumber \\&\;\;\;\;\;\;\;\;\;\;\;\;\;\;
+2n \beta(\Bf{k}_{1:m},\Bf{k}_{m:n})G _m(\Bf{k}_{1...m})\bigg\}\; .
\end{align}

\noindent
where we have used the shorthand notation $F_n(\Bf{k}_{1...n} ) = F_{n}(\Bf{k}_1,\Bf{k}_2,...,\Bf{k}_n)$ and $\Bf{k}_{i:j} = \Bf{k}_i + \Bf{k}_{i+1} + \dots +  \Bf{k}_{j-1} + \Bf{k}_{j}$, and where $\alpha$ and $\beta$ are given by
\begin{align}
\alpha(\Bf{p}_1, \Bf{p}_2) \equiv \frac{\Bf{k} \cdot \Bf{p}_2}{p_2^2} \;\;\; {\rm and} \;\;\;
\beta(\Bf{p}_1, \Bf{p}_2) \equiv  \frac{k^2\; \Bf{p}_1\cdot \Bf{p}_2}{2\;p_1^2\;p_2^2} \;,
\end{align}
with $\Bf{k} = \Bf{p}_1 + \Bf{p}_2$ enforced by the Dirac delta function. The quantities $F_n$ and $G_n$ are referred to as \textit{kernels}, and $\alpha$ and $\beta$ are the \textit{vertex couplings}. The details of their derivation are given in Appendix \ref{PTkernels}, for the interested reader. One may note an additional factor of ${n \choose m}$ compared to expressions for these quantities that exist elsewhere in the literature; These come from our choice to include factors on $\displaystyle \frac{1}{n!}$ in the series expansions, so as to match up with the approach used in relativistic CPT. This choice has no effect on the physics.

The linear ($n=1$) solutions have no nonlinear sources, so we must have $F_1 = G_1 = 1$, by definition. Given this information, one is able to recursively calculate all other $F_n$ and $G_n$ using the above relations, and symmetrise them to obtain $F^{(s)}_n$ and $G^{(s)}_n$. Starting with the $n=2$ case one finds $\displaystyle F_2(\Bf{k}_1,\Bf{k}_2) = \frac{2}{7}\big(5\alpha(\Bf{k}_1,\Bf{k}_2) + 2 \beta(\Bf{k}_1,\Bf{k}_2)\big)$, which upon symmetrising yields
\begin{align}
F_2^{(s)}(\Bf{k}_1,\Bf{k}_2) = \frac{10}{7} + \frac{4}{7}( \hat{\Bf{k}}_1 \cdot \hat{\Bf{k}}_2)^2+ \hat{\Bf{k}}_1 \cdot \hat{\Bf{k}}_2 \bigg(\frac{k_1}{k_2} + \frac{k_2}{k_1}\bigg) \;.
\end{align} 
Similarly, one finds $\displaystyle G_2(\Bf{k}_1,\Bf{k}_2) = \frac{2}{7}\big(3\alpha(\Bf{k}_1,\Bf{k}_2) + 4 \beta(\Bf{k}_1,\Bf{k}_2)\big)$, which, after symmetrising, gives
\begin{align}
G_2^{(s)}(\Bf{k}_1,\Bf{k}_2) = \frac{6}{7} + \frac{8}{7}( \hat{\Bf{k}}_1 \cdot \hat{\Bf{k}}_2)^2 + \hat{\Bf{k}}_1 \cdot \hat{\Bf{k}}_2 \bigg(\frac{k_1}{k_2} + \frac{k_2}{k_1}\bigg) \;.
\end{align}
%
%
Expressions for $F_3^{(s)}$ and $G_3^{(s)}$ can be found in Ref. \cite{Goroff:1986ep}, and unsymmetrised expressions for $F_4$ and $G_4$ have been calculated, but contain 8523 terms each. No solutions are currently known for $n>4$, due to the rapid increase in the number of terms.

\vspace{0.2cm}
\subsection{From solutions to statistics}

The expressions calculated in the previous section can be used to calculate a variety of useful statistics that describe the distribution of structures in the Universe.
Here we will focus on their application to the calculation of the $n$-point statistics of dark matter, for which galaxies are a biased tracer. One such statistic is the bispectrum $B(k_1,k_2,k_3)$ (the Fourier transform of the $3$-point correlation function), which can be written as follows: 
\begin{align}
&(2\pi)^3 B(k_1,k_2,k_3) \delta^{(3)}(\Bf{k}_1 + \Bf{k}_2 +\Bf{k}_3)  \nonumber\\=& \langle \updelta_{\Rm{N}}(\Bf{k}_1)  \updelta_{\Rm{N}} (\Bf{k}_2) \updelta_{\Rm{N}} (\Bf{k}_3)  \rangle  \; ,
 \end{align}
where each term in the ensemble average can in turn be expanded as $\displaystyle \updelta_{\Rm{N}}^{(1)}(\Bf{k}_1)+ \frac{1}{2}\updelta_{\Rm{N}}^{(2)}(\Bf{k}_1) +\dots$.
As Wick's theorem implies that all odd correlators between Gaussian random fields must vanish, the leading-order contribution must then have four factors of $\varphi$, and is given by
 \begin{align} \label{treelevelbispectrum}
&(2\pi)^3 B(k_1,k_2,k_3) \delta^{(3)}(\Bf{k}_1 + \Bf{k}_2 +\Bf{k}_3)  \nonumber \\
=&\frac{1}{2} \langle \updelta_{\Rm{N}}^{(1)}(\Bf{k}_1)\updelta_{\Rm{N}}^{(1)}(\Bf{k}_2)\updelta_{\Rm{N}}^{(2)}(\Bf{k}_3) \rangle + \Rm{2\; cycl.\; perms}  \;.
 \end{align}
The sub-leading
contributions will then be those with six factors of $\varphi$, and so on. 

\begin{figure*}
\centering
\includegraphics[scale=0.26]{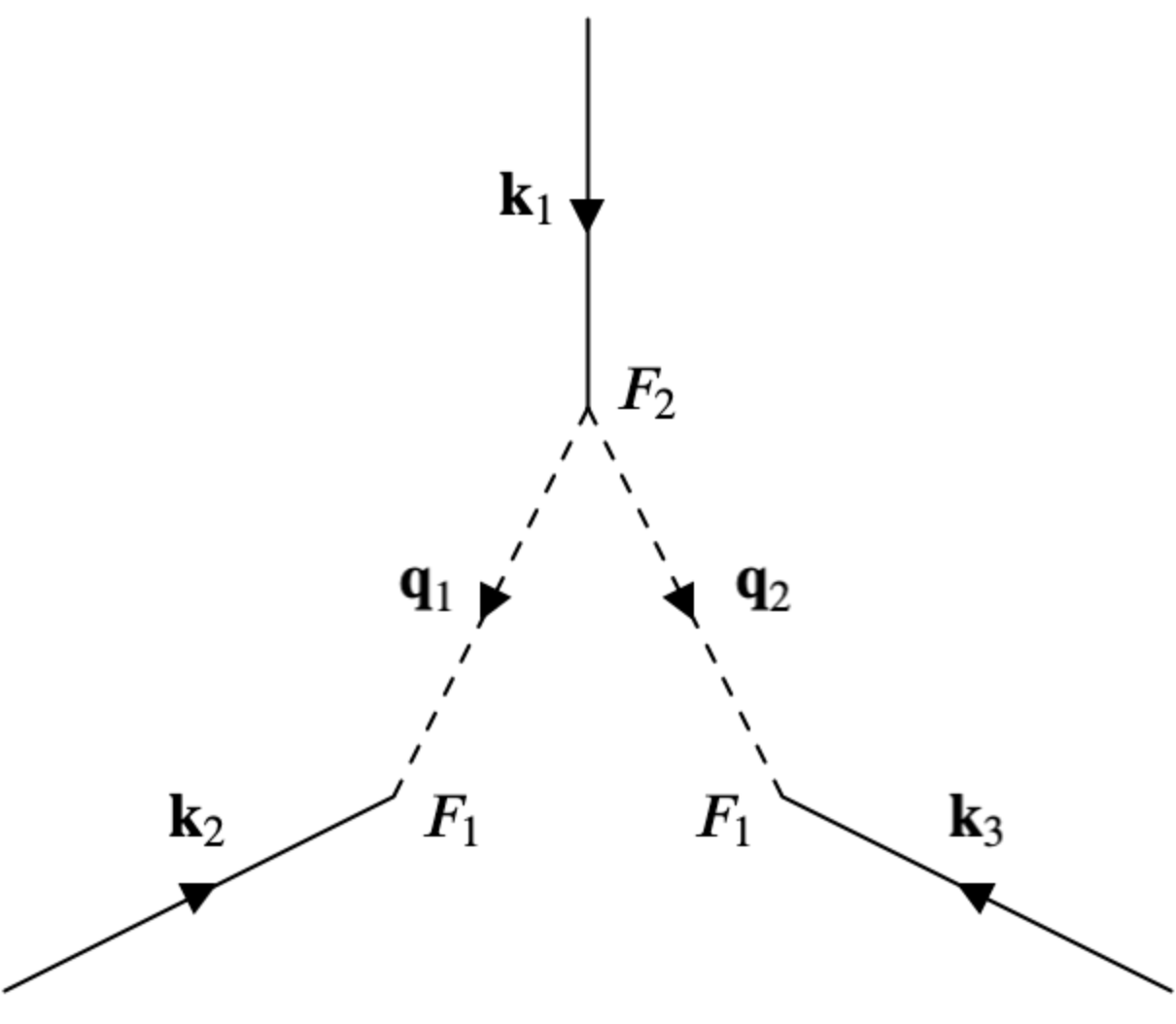} \;\;\;\;
\includegraphics[scale=0.26]{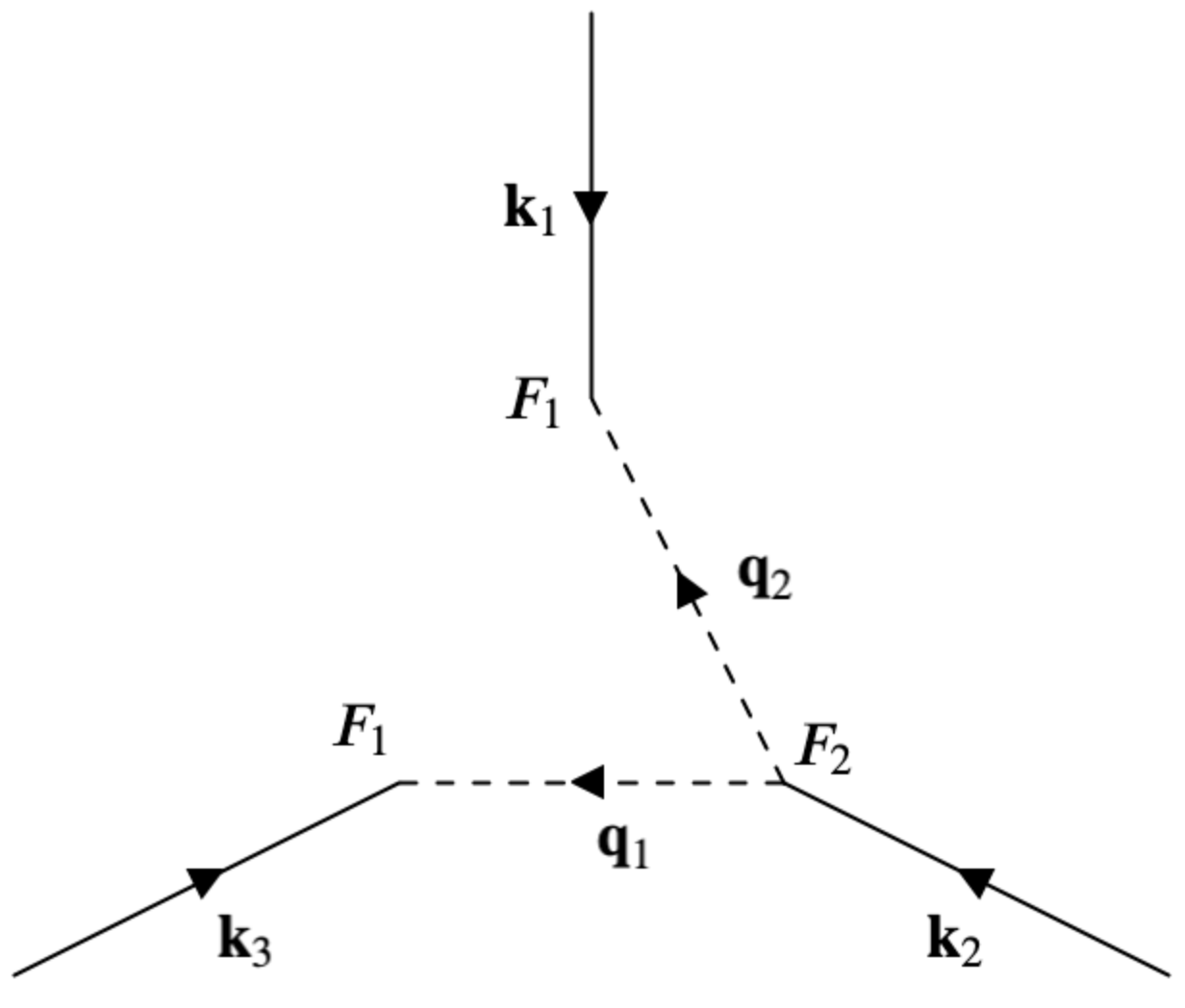} \;\;\;\;
\includegraphics[scale=0.26]{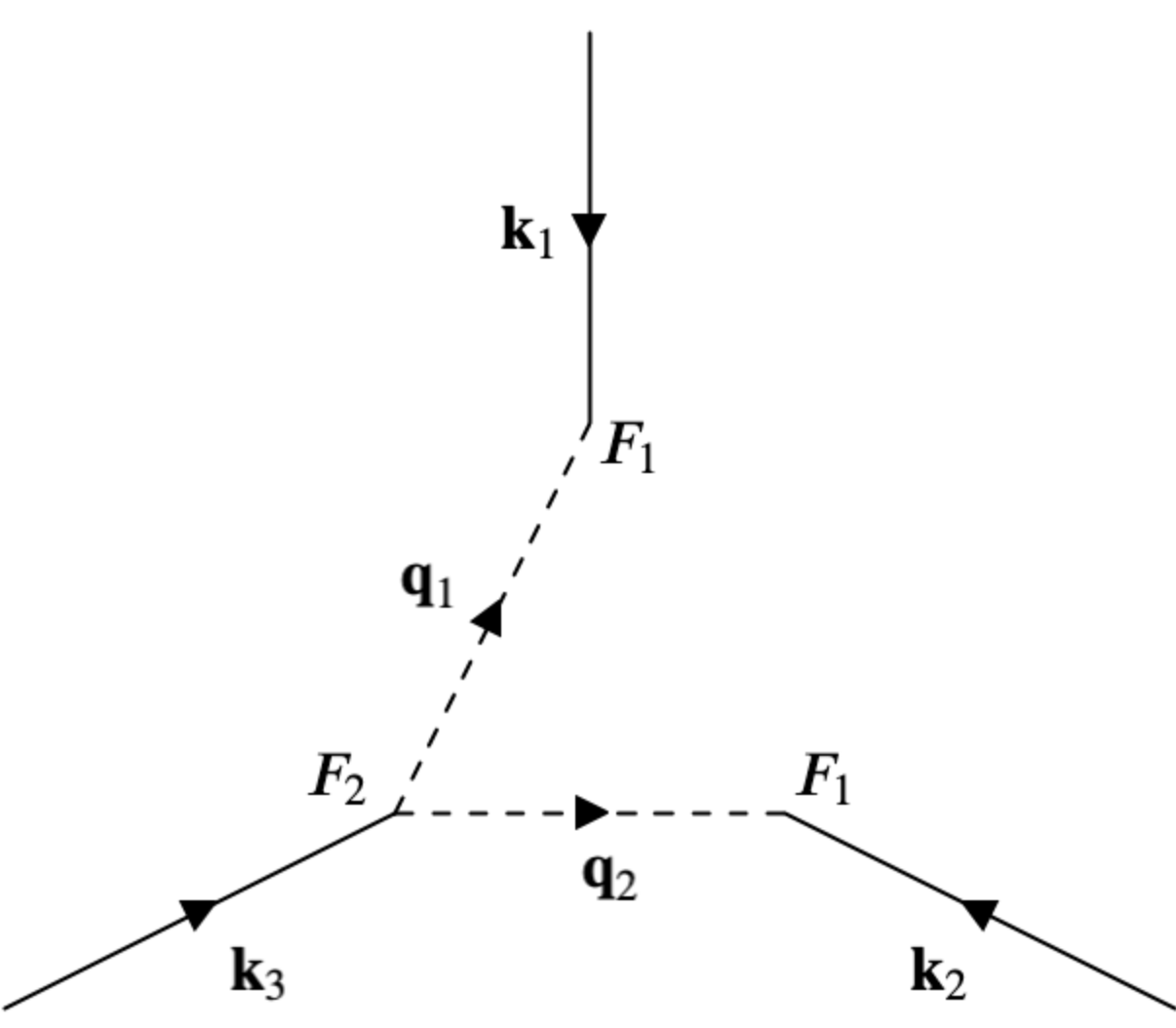}
\caption{The tree-level diagrams for the 3-point correlation function $\big \langle \updelta_{\Rm{N}}(\Bf{k}_1)\updelta_{\Rm{N}}(\Bf{k}_2)\updelta_{\Rm{N}}(\Bf{k}_3)\big \rangle$, which can be obtained from each other by cyclic permutations. The symmetry factors of these graphs is 2. }
\label{BispectrumFdiagram}
\end{figure*}

Now, using the solutions above, on finds correlators of the form $$\langle \updelta_{0}^{(1)}(\Bf{k}_1)\updelta_{0}^{(1)}(\Bf{k}_2)\updelta_{0}^{(1)}(\Bf{q}_1) \updelta_{0}^{(1)}(\Bf{q}_2)\rangle \;. $$ These can be evaluated using Wick's theorem, which gives 
\begin{align}
&\Big\langle \updelta_{0}^{(1)}(\Bf{k}_1)\updelta_{0}^{(1)}(\Bf{k}_2)\updelta_{0}^{(1)}(\Bf{q}_1) \updelta_{0}^{(1)}(\Bf{q}_2)\Big\rangle
\\
=&\Big\langle \updelta_{0}^{(1)}(\Bf{k}_1)\updelta_{0}^{(1)}(\Bf{k}_2)\Big\rangle \Big\langle\updelta_{0}^{(1)}(\Bf{q}_1) \updelta_{0}^{(1)}(\Bf{q}_2)\Big\rangle  + \Rm{2\; cycl.\; perms} \;. \nonumber
\end{align}
Writing $\langle \updelta_{0}^{(1)}(\Bf{k})  \updelta_{0}^{(1)} (\Bf{k}') \rangle = (2\pi)^3 P(k) \delta^{(3)}(\Bf{k} + \Bf{k}')$, and evaluating the relevant integrals, one can then finally write the bispectrum as
\begin{align} \label{b123}
B(k_1,k_2,k_3) = F^{(s)}_2(\Bf{k}_1,\Bf{k}_2) P(k_1)P(k_2) + \Rm{2\; cycl.\; perms} \;.
\end{align}
The linear matter power spectrum, $P(k)$, can be calculated using Boltzmann codes such as CLASS or SONG, in order to get quantitative results.

In deriving this expression for $B(k_1,k_2,k_3)$ we have discarded a term proportional to 
\begin{align}
\int \Rm{d}^3q \; F_2^{(s)}(\Bf{q},-\Bf{q} )\;P(q)\;P(k)\; \delta^{(3)}(\Bf{k}) \;.
\end{align}
This term can be seen to be zero everywhere, as $F_2^{(s)}(\Bf{q},-\Bf{q}) =0$ (except at $k=0$ where it diverges, but we take this limit to be unphysical).

Calculations of this type can be expressed in a very elegant fashion using Feynman diagram techniques \cite{Scoccimarro:1996jy}. We will briefly introduce these methods here, as they are an extremely useful timesaving device for dealing with the combinatorics involved in applying Wick's theorem. To this end, let us consider the $r^{th}$-order contribution to the $n$-point correlation function
\begin{align}
\big\langle Q_1(\Bf{k}_1) \dots Q_n(\Bf{k}_n) \big\rangle \;,
\end{align}
where $Q_i$ is either $\updelta(\Bf{k})$ or $\theta(\Bf{k})$. The Feynman rules that give an expression for this quantity in terms of $P(k)$, $F_m^{(s)}$ and $G_m^{(s)}$ are as follows:

\begin{itemize}
\item[(i)] Draw all connected diagrams that contain $n$ vertices connected by $r$ internal lines.
\item[(ii)] Label each external line with an external momentum vector, $\Bf{k}_i$, and each internal line with an internal momentum vector, $\Bf{q}_i\;$.
\item[(iii)] For each vertex with $m$ internal lines connected to it, assign a factor of $\delta^{(3)}(\Bf{k}_i + \sum_j^k \Bf{q}_j) K_m^{(s)}(\Bf{q}_{j...k})\;$.
\item[(iv)] Assign a factor of $P(q_i)$ for each internal line, where $\Bf{q}_i$ corresponds to the label for that line.
\item[(v)] Perform an integration over $q_i$ for every internal line labelled by $\Bf{q}_i\;$.
\item[(vi)] Multiply by $\displaystyle \frac{(2 \pi)^3}{\mathcal{S}} \prod_{\Rm{Vertices}} m!\;$. 
\end{itemize}
Here $K_m^{(s)}$ is the integral kernel (either $F_m^{(s)}$ or $G_m^{(s)}$, depending on whether one is considering $\updelta_{\Rm{N}}$ or $\theta_{\Rm{N}}$). The product in (vi) is taken over all vertices of order $m$ (i.e. for each vertex of order $m$, we add a factor of $m!$ into the product), and $\mathcal{S}$ is the symmetry factor (or multiplicity) of the diagram.

When applying the above rules, two diagrams are distinct if they cannot be deformed into one another without cutting any internal lines (sliding lines over other lines is allowed). By convention, the signs of the internal momenta are taken to have positive sign if the momentum is outgoing from the vertex, and multiplicity of a diagram is given by the number of ways in which you can permute the internal lines without altering the diagram.

These Feynman rules differ from those used in quantum field theory in that new vertices appear at each order in perturbations, rather than just increasing numbers of combinations of the same vertices. Although the diagrams' utility is somewhat restricted compared to in QFT, as the main computational challenge is in the calculation of the vertices themselves, this methodology does however give a clear representation of the key statistics, and is a very useful timesaving device when constructing typical integrals that occur in higher order corrections.

As an example, let us consider how to apply this technique to calculate the tree-level bispectrum (that is, calculate the bispectrum up to the level of accuracy immediately before loops start appearing in the internal lines of the Feynman diagrams). Figure (\ref{BispectrumFdiagram}) shows all three diagrams that appear at this level of accuracy. Applying the Feynman rules above then immediately gives the result found in Eq. (\ref{b123}), which we had previously calculated using Wick's theorem directly.

A variety of numerical techniques are available for numerically evaluating the integrals that result from higher loop calculations, such as the FAST-PT code in Ref. \cite{McEwen:2016fjn}. With all the preceding machinery in place, it is clear that the key physics of NPT is contained within the integration kernels $F_n^{(s)}$ and $G_n^{(s)}$. In the next section we will examine the relativistic generalisation of this perturbation theory, specifically in the Poisson gauge. We will show how changes in the terms that appear in the field equations result in changes to the kernels, and how these changes manifest in statistical quantities such as the matter bispectrum.

\section{Cosmological Perturbation Theory in Poisson Gauge} \label{cpt}
 
In this section we will give a brief introduction to the methods used to find second-order solutions for scalar quantities in relativistic CPT. This will proceed in Poisson gauge, as this is the one commonly used gauge that remains valid in 2PPT \cite{GoldbergThesis, Clifton:2020oqx}. The reader is referred to Refs. \cite{Villa:2015ppa, Bartolo:2005kv} for a more extensive description. 

The geometry in this case is specified by a line-element of the form given in Eq. (\ref{FLRW1}), with the Newtonian potentials set to zero (such that $U=0$). The cosmological perturbations are then expanded as
\begin{align}
\phi &= \phi^{(1)} + \frac{1}{2}\phi^{(2)} + \dots\\
\psi &= \psi^{(1)} + \frac{1}{2}\psi^{(2)} + \dots \, .
\end{align}
Likewise, the Newtonian contributions to the stress-energy are removed (such that $v_{\Rm{N}i}=0=\rho_{\Rm{N}}$), and the cosmological perturbations are written
\begin{align}
\updelta &= \updelta^{(1)} + \frac{1}{2}\updelta^{(2)} + \dots\\
v^i &= v^{(1)i} + \frac{1}{2}v^{(2)i} + \dots \, .
\end{align}
The superscripts in these expressions correspond to their expansion in $\varphi$, and the ellipses denote terms of higher than second-order.


 
For the above metric and stress-energy tensor components, the first-order Einstein equations for this system are well-known to be given by
%
%
\begin{align}
\psi^{(1)} &= \phi^{(1)} = \varphi\\
\updelta^{(1)} &= \frac{2}{3\mathcal{H}^2} \left(\nabla^2 \phi^{(1)} - 3\mathcal{H}^2 \phi^{(1)} \right) \;, \label{cpt1density}\\
\theta^{(1)} &= -\frac{2}{3\mathcal{H}}\nabla^2 \phi^{(1)} \; \label{cpt1velocity}\;,
\end{align}
where $\theta^{(1)}=v^{(1)i}_{\phantom{(1)i},i}$ is again the velocity divergence, and where we have used the result that $\phi^{(1)}$ has a growing mode that is constant in time in order to write it in terms of the initial fluctuation $\varphi$.
%
It is clear that the density contrast is not separable in this gauge, which means that it is often more practical to work directly with the gravitational potentials themselves when performing higher-order calculations. 

Let us now consider the methods that can be used to find second-order solutions in CPT. We will begin by outlining the procedure to calculate the gravitational ``slip'', given by $\psi^{(2)}-\phi^{(2)}$. We will then move on to discuss initial conditions, the evolution of scalars, and the calculation of statistics.
 
\subsection{The second-order slip}
 
We can now write down the following second-order trace-free $ij$-field equation, where $(v^{(1)})^2$ is the square of the peculiar velocity defined by $v^{(1)}_iv^{(1)i} =(v^{(1)})^2$:

\begin{align}
\frac{1}{2}\partial^i \partial_j (\psi^{(2)} - \phi^{(2)}) + 2 \partial^i \varphi \partial_j \varphi + 4 \varphi \partial^i \partial_j \varphi -& \frac{1}{3} \delta^i_{\;j} \bigg[ \frac{1}{2}\nabla^2  (\psi^{(2)} - \phi^{(2)}) + 2  ( \nabla \varphi )^2  + 4\varphi \nabla^2 \varphi \bigg] \nonumber \\
=& \;8\pi a^2\bar{\rho} \;\big( v^{(1)i} v^{(1)}_{j} - \frac{1}{3}  \delta^i_{\;j}  v^{(1)2} \big) \label{cpt2tracefreeij} \;,
\end{align}

\noindent
Applying the operator $\partial_i\partial^j$ to this equation, it is then easy to show that (following \cite{Bartolo:2005kv})
\begin{align}
\psi^{(2)} - \phi^{(2)} = -4\varphi^2 + Q\;,
\end{align}
where 
\begin{align*}
\nabla^2 Q &\equiv - P + 3N \;, \\
\nabla^2 N &\equiv  \partial_i \partial^j P^i_{\;j}\;, \\
P^i_{\;j} &\equiv 2 \partial^i \varphi \partial_j \varphi + 8\pi a^2 \bar{\rho} v^{(1)i} v^{(1)}_{j} \; ,
\end{align*}
and $P = P^i_{\;i}$. These quantities are easily calculated using the first-order solutions given above, which in EdS results in \cite{Bartolo:2005kv}
%
\begin{align} \label{cpt2psiphiconstraintX}
\hspace{-5pt}
\psi^{(2)} - \phi^{(2)} = -4\varphi^2  -\frac{10}{3} \nabla^{-4} \bigg[ \nabla^2 (\nabla \varphi)^2  -  3  \partial_i \partial^j  \big( \partial^i \varphi \partial_j \varphi\big) \bigg].
\end{align}
This equation now enables us to write down and solve the evolution equation for the gravitational perturbation $\psi^{(2)}$, by eliminating all occurrences of $\phi^{(2)}$. 

\subsection{Initial conditions at second-order}

More care must be taken with regards to the specification of initial conditions in CPT, as compared to NPT  (where we were free to choose them arbitrarily), due to the existence of additional constraint equations. An excellent description of the procedure for fixing initial conditions is given in \cite{Bartolo:2003gh}, which we will summarise here.
 
Initial conditions are usually fixed at a time when the cosmological perturbations relevant for large-scale structure in the Universe today were well outside the horizon \cite{Bartolo:2001cw}. The statistical characteristics of the seeds of these density fluctuations can be predicted by theories of the very early Universe, such as inflation. In order to connect specific models to the initial conditions for structure formation, it is convenient to use the \textit{curvature perturbation of uniform density hypersurfaces}, which we expand as $\displaystyle \zeta = \zeta^{(1)} + \frac{1}{2}\zeta^{(2)} + \dots$. The first-order contribution to this quantity is given by 
\begin{align}
\zeta^{(1)} = -\psi^{(1)} - \mathcal{H} \frac{\updelta\rho^{(1)}}{\bar{\rho}'}\;,
\end{align}
while the second-order contribution is given by

\begin{align} \label{zeta2def}
\zeta^{(2)} = -\psi^{(2)} - \mathcal{H} \frac{\updelta\rho^{(2)}}{\bar{\rho}'} + 2\mathcal{H} \frac{\updelta\rho^{(1)'}}{\bar{\rho}'}\frac{\updelta\rho^{(1)}}{\bar{\rho}'} + 2 \frac{\updelta\rho^{(1)}}{\bar{\rho}'}(\psi^{(1)'} + 2\mathcal{H} \psi^{(1)}) - \bigg(\frac{\updelta\rho^{(1)}}{\bar{\rho}'} \bigg)^2 \bigg(\mathcal{H}\frac{\bar{\rho}''}{\bar{\rho}} - \mathcal{H}' - 2 \mathcal{H}^2\bigg)\;.
\end{align}
Critically, one finds that $\zeta$ is found to be constant on super-horizon scales (assuming isocurvature perturbations are not present), which makes it ideal for specifying initial conditions, such as primordial non-Gaussianity.

For example, in standard single field inflation on finds $\zeta^{(2)} \approx 2 \zeta^{(1)2}$. A standard way of parameterising the amount of primordial non-Gaussianity is therefore to introduce $a_{\Rm{nl}}$ and $f_{\Rm{nl}}$ via $\zeta^{(2)} = 2a_{\Rm{nl}} \zeta^{(1)2}$ and $\displaystyle f_{\Rm{nl}} = \frac{5}{3}( a_{\Rm{nl}} - 1)$ \cite{Bartolo:2005kv, Villa:2015ppa}. We will choose $a_{\Rm{nl}} = 1$ for the duration of this paper, to facilitate a direct comparison of 2PPT and CPT without having to worry about additional complications in correlation functions due to the presence of primordial non-Gaussianity. 

In an EdS universe, assuming the standard single field model of inflation, the first and second-order curvature perturbations are given by $\displaystyle \zeta^{(1)} = - \frac{5}{3} \varphi$ and $\displaystyle \zeta^{(2)} = \frac{50}{9} \varphi^2$. We can use these results with Eq. (\ref{zeta2def}) to eliminate $\updelta^{(2)}$ in favour of $\zeta^{(2)}$ in the $00$-field equation, which evaluated at an initial time deep in the matter-dominated era gives
\begin{align} 
\frac{1}{3}\nabla^2 \psi^{(2)} - \mathcal{H} \psi^{(2)'} - \mathcal{H}^2 \phi^{(2)} = \frac{4\pi a^2 \bar{\rho}}{3} \updelta^{(2)} + \frac{8\pi a^2 \bar{\rho}}{3}v^{(1)2} - (\nabla \varphi)^2 - 4\mathcal{H}^2 \varphi^2 - \frac{8}{3}\varphi \nabla^2 \varphi \,.
\end{align}
Using our equation for the second-order slip (\ref{cpt2psiphiconstraintX}), we can then write initial conditions for the gravitational potentials in EdS as 
\begin{align}
\phi^{(2)}_{0} &= 2 \bigg[ \varphi^2 + \nabla^{-4} \Big( \nabla^2 (\nabla \varphi)^2  -  3  \partial_i \partial^j  \big( \partial^i \varphi \partial_j \varphi\big) \Big) \bigg] \;, \\
\psi^{(2)}_{0} &= 2 \bigg[ -\varphi^2 - \frac{2}{3} \nabla^{-4} \Big( \nabla^2 (\nabla \varphi)^2  -  3  \partial_i \partial^j  \big( \partial^i \varphi \partial_j \varphi\big) \Big) \bigg]\;,
\end{align}
which satisfy both the $00$-field equation and Equation (\ref{cpt2psiphiconstraintX}). Let us now consider evolution of perturbations in EdS.

\subsection{Second-order scalar evolution} \label{cpt2evolution}

Substituing for $\psi^{(2)} - \phi^{(2)}$ from Eq. (\ref{cpt2psiphiconstraintX}) one find that the scalar evolution equation for gravitational perturbations can be written as an inhomogeneous ODE:
\begin{align} \label{inhomODE}
\psi^{(2)''} + 3\mathcal{H}\psi^{(2)'} = S(x) \;,
\end{align}
where $$\displaystyle S\equiv \frac{10}{3} \nabla^{-2} \partial_i \partial^j  \big( \partial^i \varphi \partial_j \varphi\big) - (\nabla \varphi)^2$$ has no $\tau$-dependence due to our restriction to EdS. The full solution to this equation should take the form $\psi^{(2)} = \psi^{(2)}_{0}(x) + \psi^{(2)}_P(\tau,x)$, where the second term is a particular solution.

We can proceed by supposing $\psi^{(2)}_P = \mathcal{B}(\tau) S(x)$, which, upon substituting back into Eq. (\ref{inhomODE}), gives 
$\mathcal{B}(\tau) = {a(\tau)}/{14}$ as a solution. We can therefore express the full second-order gravitational potentials in EdS as
\begin{align}
\phi^{(2)}(\tau,x) &= 2 \varphi^2 + 12  \, \mathbf{\Theta}_0 +a(\tau)\bigg[ \frac{1}{6} (\nabla \varphi)^2  - \frac{10}{21}\mathbf{\Psi}_0 \bigg] \;, \\
\psi^{(2)}(\tau,x) &= -2 \varphi^2 - 8 \, \mathbf{\Theta}_0 + a(\tau)\bigg[ \frac{1}{6} (\nabla \varphi)^2  - \frac{10}{21}\mathbf{\Psi}_0 \bigg] \;.
\end{align}
where we have written
\begin{align} \label{GRKernel}
\mathbf{\Theta}_0 &= \frac{1}{2} \nabla^{-4}\bigg[ \frac{1}{3} \nabla^2 (\partial^i \varphi \partial_i \varphi) - \partial_i\partial^j(\partial^i \varphi \partial_j \varphi)  \bigg] \;,
\\ \label{NewtonianKernel}
\mathbf{\Psi}_0 &= - \frac{1}{2} \nabla^{-2} \bigg[(\nabla^2 \varphi)^2 - \partial_i\partial_j \varphi \,\partial^i\partial^j \varphi \bigg] \;,
\end{align}
which are referred to as the \textit{GR} and \textit{Newtonian} kernels, respectively. In deriving these expressions one makes use of the identities $\nabla^2 \mathbf{\Theta}_0 = \mathbf{\Psi}_0 - \frac{1}{3} (\nabla \varphi)^2$ and $\nabla^{-2} \partial_i\partial^j \big(\partial^i \varphi \partial_j \varphi \big) = -2 \mathbf{\Psi}_0 + (\nabla \varphi)^2$ \cite{Villa:2015ppa}.

To obtain an expression for $\updelta^{(2)}$ we can use the two second-order constraint field equations to find
 \begin{align} \label{secondorderdelta}
 \updelta^{(2)} =&\; 4 \, \varphi^2 - 24 \,\mathbf{\Theta}_0 \\&+\bigg[ -\frac{22}{9\mathcal{H}^2}  (\nabla \varphi)^2 + \frac{8}{3\mathcal{H}^2}\varphi \nabla^2 \varphi +  \frac{16}{7 \mathcal{H}^2} \mathbf{\Psi}_0 \bigg] \nonumber \\
&+ \frac{4}{9 \mathcal{H}^4}  \bigg[ \frac{10}{7} (\nabla^2\varphi)^2 + 2 \nabla^2\partial_i \varphi \partial^i \varphi+ \frac{4}{7} \partial_i \partial_j \varphi \partial^i \partial^j \varphi \bigg] \;, \nonumber
\end{align}
where we have arranged the expression in powers of spatial derivatives, so that the scaling of each term is clear. 
%
A similar calculation can be performed for $v^{(2)i}$, and a full treatment in $\Lambda$CDM can be found in Ref. \cite{Villa:2015ppa}. Let us now consider how to use these solutions to derive statistics to describe structure in the Universe.

\subsection{Relativistic statistics}
 

We will again focus on the Feynman diagrams given in Fig. \ref{BispectrumFdiagram}, 
this time modifying the rules such that the vertices correspond to the relativistic kernels from CPT instead of Newtonian ones. We will also take each internal line to come with a factor of the linear power spectrum in Poisson gauge. Under these adjustments, we can proceed by writing down the following expression for the tree-level bispectrum:
\begin{align*}
B(k_1,k_2,k_3) = \mathcal{K}^{\updelta}_2(\Bf{k}_1,\Bf{k}_2)P_{P}(k_1)P_{P}(k_2) + \;\Rm{2}\;\Rm{cycl.}\; \Rm{perms} \;,
\end{align*}
where $P_{P}(k)$ is the matter power spectrum evaluated in Poisson gauge, and where  $\mathcal{K}^{\updelta}_2$ is the second-order relativistic CPT matter density kernel, defined by 
\begin{align*} \hspace{-5pt}
\updelta^{(2)}(\Bf{k},\tau) =\int \; \frac{\Rm{d}^3 q_1 \;\Rm{d}^3 q_2}{(2\pi)^3} \;\mathcal{K}^{\updelta}_2(\Bf{q}_1,\Bf{q}_2,\tau)\; \updelta^{(1)}(\Bf{q}_1,\tau)\;\updelta^{(1)}(\Bf{q}_2,\tau),
\end{align*}
where $\Bf{k}=\Bf{q}_1+\Bf{q}_2$. We have included time dependence in both the kernel and the density contrasts here, as the spatial and temporal dependencies are no longer separable. An expression for this kernel was calculated by Tram et al. in Ref. \cite{Tram:2016cpy} by Fourier transforming Equation (\ref{secondorderdelta});

\begin{align} \label{Tramkernel} \displaystyle
\mathcal{K}^{\updelta}_2(\Bf{k}_1,\Bf{k}_2,k) = \frac{ \big(\beta(k) - \alpha(k)\big) + \frac{\beta(k)}{2} \hat{\Bf{k}}_1 \cdot \hat{\Bf{k}}_2 \Big( \frac{k_1}{k_2} + \frac{k_2}{k_1} \Big) + \alpha(k) \Big(\hat{\Bf{k}}_1 \cdot \hat{\Bf{k}}_2\Big)^2 + \gamma(k) \Big( \frac{k_1}{k_2} - \frac{k_2}{k_1} \Big)^2}{\Big(1 + 3 \frac{\mathcal{H}^2}{k_1^2}\Big)\Big(1 + 3 \frac{\mathcal{H}^2}{k_2^2}\Big)} \;,
\end{align}

\noindent
where the coefficient functions $\alpha(k,\tau)$, $\beta(k,\tau)$ and $\gamma(k,\tau)$ are given by 
\begin{align}
\alpha(k,\tau) &= \frac{2}{7} + \frac{59 \mathcal{H}^2}{14k^2} + \frac{45 \mathcal{H}^4}{2k^4} \;, \\
\beta(k,\tau) &= 1  - \frac{ \mathcal{H}^2}{2k^2} + \frac{54 \mathcal{H}^4}{k^4} \;, \\
\gamma(k,\tau) &= - \frac{3 \mathcal{H}^2}{2k^2} + \frac{9 \mathcal{H}^4}{2k^4} \;,
\end{align}
for an EdS universe. In this form, it is easy to see that relativistic corrections scale as $\displaystyle \frac{\mathcal{H}^2}{k^2}$ and $\displaystyle \frac{\mathcal{H}^4}{k^4}$. For large $k$, these terms will be irrelevant and the expression reduces to the Newtonian kernel. However, as $k$ decreases the extra terms will become more and more significant.
 
%

\section{Solutions to two-parameter perturbation theory}
  
Back in Section \ref{Necessity}, we outlined the utility of making further approximations to the 2PPT equations in order to find analytic solutions. In effect, this will involve solving the Newtonian equations perturbatively, and then considering the knock-on effect on the cosmological quantities (a type of cosmological back-reaction, from nonlinear structures on to the large-scale perturbations). In this section we will use and develop the techniques from Sections \ref{npt} and \ref{cpt}  to find explicit solutions to the 2PPT Eqs. (\ref{evol})-(\ref{momentum}). 
 
This is achieved by inserting the perturbative expansions, as given in Eq. (\ref{Npertseries}), into Eqs. (\ref{evol})-(\ref{momentum}).  A first approximation to the 2PPT dynamics is made by neglecting any quadratic terms in $\varphi$, leading a homogeneous set of PDEs that can be solved to find the first approximation to the 2PPT solutions. This is then followed in standard fashion, by calculating second approximations to 2PPT solutions using quadratic products of the first approximations as inhomogeneous source terms, continuing {\it ad infinitum} to higher orders.

In this section we will start by finding the first approximation to the 2PPT equations, before moving on to consider the scalar constraints at second approximation. We will then solve the relevant evolution equations, and discuss suitable initial conditions. These results will then all be used to calculate the matter bispectrum, which will be compared to the corresponding quantity in NPT  and CPT.

\subsection{First approximation to 2PPT}

Linearising in $\varphi$, we find that the resulting system takes the following form:

\begin{align} 
&\frac{1}{3} \nabla^2 \phi^{(1)} + \mathcal{H}(\phi^{(1)\prime} + \psi^{(1)\prime} + 2 U^{(1)\prime}) + (\psi^{(1)\prime\prime} + U^{(1)\prime\prime}) + 2\mathcal{H}^{\prime}(\phi^{(1)}   + U^{(1)})
= \frac{\mathcal{H}^2}{2}\updelta^{(1)}  \; \label{00tree}  , \\
&\frac{1}{3} \nabla^2 \psi^{(1)} - \mathcal{H}(\psi^{(1)\prime}+U^{(1)\prime}) - \mathcal{H}^2(\phi^{(1)} + U^{(1)}) = \frac{\mathcal{H}^2}{2}\updelta^{(1)} \; \label{iitree},
\end{align}

\noindent
as well as
\begin{align}
&\nabla^2\big( \psi^{(1)\prime} + \mathcal{H}\phi^{(1)} \big) =- \frac{3\mathcal{H}^2}{2} \theta^{(1)}  \; ,\label{0itree} \\
&\nabla^4(\phi^{(1)} - \psi^{(1)}) = 0\;  \label{ijtree} .
\end{align}
These equations take a form familiar from linear CPT in conformal Newtonian gauge. Using Eq. (\ref{ijtree}) to eliminate $\phi^{(1)}$ for $\psi^{(1)}$, and then subtracting Eq. (\ref{iitree}) from Eq. (\ref{00tree}), we are left with the analogous evolution equation for the linearised scalar degree of freedom:
\begin{align} \label{potentialevol}
     \Big(\psi^{(1)\prime\prime} + U^{(1)\prime\prime} \Big) + 3\mathcal{H}\Big(\psi^{(1)\prime} + U^{(1)\prime}\Big) = 0 \;.
\end{align}
The form of this equation motivates a number of questions. In particular, is it consistent to consider $U^{(1)}$ and $\psi^{(1)}$ to be entirely separate degrees of freedom in this setup?  How should we interpret $\psi^{(1)}$? How should we connect the initial conditions, $\varphi(x)$, defined as a continuous function on all length scales, to $U^{(1)}$ and $\phi^{(1)}$, when each of them are defined only on the spatial scales where the expansion used to derive them applies?

To answer these questions, we take note of the fact that since Eqs. (\ref{00tree}) and (\ref{iitree}) are linear, one can always write the solutions in the form
\begin{align}
\psi^{(1)} = \psi^{(1)R} + \psi^{(1)N} \;,
\end{align}
where $\psi^{(1)N}$ satisfies the Newton-Poisson equation \textit{on large scales}, 
\begin{align} \label{longpoisson}
\nabla^2  \psi^{(1)N} = \frac{3\mathcal{H}^2}{2} \updelta^{(1)N} \;,
\end{align}
and where $\psi^{(1)R}$ is whatever is left (the relativistic contribution). The large-scale density contrast and velocity can also be split in a corresponding fashion:
\begin{align}
\updelta^{(1)} = \updelta^{(1)R} + \updelta^{(1)N} \;, \\
\theta^{(1)} = \theta^{(1)R} + \theta^{(1)N} \;.
\end{align}
We can then regard $\psi^{(1)N}$, $\updelta^{(1)N}$ and $\theta^{(1)N}$ to be the long-wavelength extension of the  quantities $U^{(1)}$, $\updelta^{(1)}_{\Rm{N}}$ and $\theta^{(1)}_{\Rm{N}}$. 

The evolution equation can then be written as 
\begin{align} \label{potentialevol}
&\Big(\psi^{(1)R\prime\prime} + \psi^{(1)N\prime\prime} + U^{(1)\prime\prime} \Big) \\+& 3\mathcal{H}\Big(\psi^{(1)R\prime} + \psi^{(1)N\prime} + U^{(1)\prime}\Big) = 0 \;. \nonumber
\end{align}
Since the Newtonian and cosmological perturbation theories have identical gravitational potentials on all scales at first-order, this motivates us to choose our initial conditions as
\begin{align}
\psi^{(1)N} + U^{(1)} = \varphi \;,
\end{align}
where $\varphi$ now has support on all spatial scales, making the extension of the  solution to large scales explicit. We can then consistently also choose $\psi^{(1)R} = 0$ at all times, which is equivalent to the statement that there is no leading-order large-scale correction to the Newtonian gravitational potential.

Whilst this discussion makes explicit the extension of the Newtonian solution to all scales, it is a notational inconvenience to persevere with so many different terms, especially when many of these terms always appear together alongside each other in our equations. We will therefore implement the following re-labellings:
\begin{align}
U^{(1)} + \psi^{(1)N} &\rightarrow U^{(1)} 
\;, \\
\updelta^{(1)N} + \updelta^{(1)}_{\Rm{N}} &\rightarrow \updelta^{(1)}_{\Rm{N}} 
\;,\\
\updelta^{(1)R} &\rightarrow \updelta^{(1)} \;, \\
\theta^{(1)N} + \theta^{(1)}_{\Rm{N}} &\rightarrow \theta^{(1)}_{\Rm{N}} \;,\\
\theta^{(1)R} &\rightarrow \theta^{(1)} \;.
\end{align}
This simply allows us to recharacterize $\psi^{(1)}$, $\updelta^{(1)}$ and $\theta^{(1)}$ as being the purely relativistic corrections to the all-scales Newtonian quantities $U^{(1)}$, $\updelta^{(1)}_{\Rm{N}}$ and $\theta_{\Rm{N}}^{(1)}$, without having to introduce any superfluous degrees of freedom.

Having done this, we can then use the field equations to work out the rest of the cosmological quantities. In particular, Eq. (\ref{iitree}) yields
\begin{align}
\updelta^{(1)} &= - 2  U^{(1)} = -2 \varphi\;,
\end{align}
which we recognise as the standard result for the relativistic correction to the density contrast in Poisson gauge, while Eq. (\ref{0itree}) guarantees
\begin{align}
\theta^{(1)} = 0\;.
\end{align}
The significance of this result is immediately apparent; If you linearise the short scale nonlinear structures in the two-parameter perturbation theory equations, you simply obtain the results for the relativistic corrections in standard first-order CPT in Poisson gauge. This is not at all surprising, given that the linear terms in the field equations satisfy equations of the same form \cite{Milillo:2015cva}. We will now proceed to the second approximation.

\subsection{Second approximation to the scalar constraints}

To obtain the second approximation to the combination $\psi^{(2)} - \phi^{(2)}$, it is necessary to consider the application of the operator $\partial_i\partial^j$ to Eq. (\ref{tracefreeij}). As described in Section \ref{Necessity}, this will mean that we have to consider terms that appear up to order $\sim{\eta^6}/{L_{\Rm{N}}^4}$ in order to capture the full dynamics, due to the possible change in size of terms when using the ``inverse Laplacian" operator on products. As is also discussed in Ref. \cite{Gallagher:2018bdl}, it is necessary to include any terms that may not be included in Eq. (\ref{tracefreeij}), but that may nonetheless end up contributing to the $\mathcal{O}({\eta^4}/{L_{\Rm{N}}^4})$, $\mathcal{O}({\eta^5}/{L_{\Rm{N}}^4})$ or $\mathcal{O}({\eta^6}/{L_{\Rm{N}}^4})$ expressions that result from the application of the $\partial_i \partial^j$ operator to (\ref{tracefreeij}). 


For our present purposes, we need to calculate $\displaystyle \frac{1}{3}\nabla^4(\psi^{(2)} - \phi^{(2)})$ up to $\mathcal{O}({\eta^6}/{L_{\Rm{N}}^4})$.  This is required to perform our study consistently, but it will also enable us to discuss the pros and cons of this approach, and highlight the areas where differences and benefits can occur between this formalism and standard CPT. Schematically, we can write
\begin{align}
\frac{1}{3}\nabla^4(\psi^{(2)} - \phi^{(2)}) = \mathcal{S}^{(2)}_4 + \mathcal{S}^{(2)}_5 + \mathcal{S}^{(2)}_6 \;, 
\end{align}
where

\begin{align}
\mathcal{S}^{(2)}_4 = &\; 16\pi a^2 \;\bar{\rho}\; v_{\Rm{N}i}^{(1)}\; \partial^i \theta^{(1)}_{\Rm{N}} + 8\pi a^2 \bar{\rho} (\theta^{(1)}_{\Rm{N}})^2 + 8\pi a^2 \bar{\rho}\; \partial_i  v_{\Rm{N}j}^{(1)}\;\partial^j  v_{\Rm{N}}^{(1)i} \nonumber \\
 &- \frac{2}{3} (\nabla^2 U^{(1)})^2 - \frac{14}{3} \partial_i \partial_j U^{(1)} \partial^i \partial^j U^{(1)}   - \frac{16\pi a^2 \bar{\rho}}{3}\; \partial_j  v_{\Rm{N}i}^{(1)}\;\partial^j  v_{\Rm{N}}^{(1)i} - \frac{16\pi a^2 \bar{\rho}}{3}\; v_{\Rm{N}i}^{(1)}\;\nabla^2  v_{\Rm{N}}^{(1)i} \nonumber \\
 & - 8 \partial_i \nabla^2 U^{(1)} \partial^i U^{(1)} - \frac{8}{3} U^{(1)} \nabla^4 U^{(1)}  - \frac{8}{3} \psi^{(1)} \nabla^4 U^{(1)}  
 \;,  \\
 \mathcal{S}^{(2)}_5 = &\; 16\pi a^2 \;\bar{\rho}\; v_{\Rm{N}i}^{(1)}\; \partial^i \theta^{(1)} - 8 \partial_i \nabla^2 U^{(1)} \partial^i \psi^{(1)} - \frac{16\pi a^2 \bar{\rho}}{3}\; v_{i}^{(1)}\;\nabla^2  v_{\Rm{N}}^{(1)i}   
 \;, \\
  \mathcal{S}^{(2)}_6 =  &\;  8\pi a^2 \bar{\rho} \;\theta^{(1)}_{\Rm{N}}\;\theta^{(1)} + 8\pi a^2 \bar{\rho}\; \partial_i  v_{\Rm{N}j}^{(1)}\;\partial^j  v^{(1)i}  - \frac{2}{3} \nabla^2 U^{(1)}\nabla^2 \psi^{(1)} \nonumber \\
  &- \frac{14}{3} \partial_i \partial_j U^{(1)} \partial^i \partial^j \psi^{(1)} - \frac{16\pi a^2 \bar{\rho}}{3}\; \partial_j  v_{\Rm{N}i}^{(1)}\;\partial^j  v^{(1)i}  
 \; ,
\end{align}
and where subscript $n$ indicates that the order of the quantity is $\sim \eta^n / L_{\Rm{N}}^4$. 

If we want to obtain a second approximation to the dynamics, we should take note of the following facts that considerably simplify the results of this calculation.
\begin{itemize}
\item[(i)]{The first-order large-scale relativistic corrections are all zero, apart from the density contrast $\updelta^{(1)}= -2 \varphi$, which receives a linear correction.}
\item[(ii)]{The density contrast only appears in a cubic product in Eq. (\ref{tracefreeij}).}
\item[(iii)]{Therefore, the only source terms that will contribute to the second approximation will be quadratic products of the Newtonian leading-order quantities.}
\item[(iv)]{Quadratic products of Newtonian quantities can be at maximum $\sim \eta^4$.}
\end{itemize}
It is therefore only necessary to consider the terms at $\mathcal{O}({\eta^4}/{L_{\Rm{N}}^4})$ to find the second approximation to this equation. 

The reader should note that the second approximation to cosmological quantities will obey \textit{different} equations to the second approximation to Newtonian quantities; We therefore expect $\psi^{(2)} \neq 0$, along with the rest of the second approximations to cosmological large-scale perturbations, even though the source terms are all Newtonian. Thus, when we come to consider the third approximation to Eq. (\ref{tracefreeij}), there will be contributions from terms like $\psi^{(2)}\nabla^4 U^{(1)}$, a coupling between an explicitly relativistic source term and a Newtonian one. Although the calculation is long, a third approximation to the dynamics of the two-parameter field equations is vastly preferable to full third-order CPT. 


Applying the logic presented above, we immediately see that $\mathcal{S}^{(2)}_5 =0 $ and $\mathcal{S}^{(2)}_6=0$  (although we note that $\mathcal{S}^{(3)}_5 \neq 0 $ and $\mathcal{S}^{(3)}_6 \neq 0$, due to the fact that $\psi^{(2)} \neq 0$). We are therefore left with
\begin{align}
\frac{1}{3}\nabla^4(\psi^{(2)} - \phi^{(2)}) =& \;16\pi a^2 \;\bar{\rho}\; v_{\Rm{N}i}^{(1)}\; \partial^i \theta^{(1)}_{\Rm{N}} + 8\pi a^2 \bar{\rho} (\theta^{(1)}_{\Rm{N}})^2 + 8\pi a^2 \bar{\rho}\; \partial_i  v_{\Rm{N}j}^{(1)}\;\partial^j  v_{\Rm{N}}^{(1)i} \nonumber \\ 
& - \frac{2}{3} (\nabla^2 U^{(1)})^2 - \frac{14}{3} \Big(\partial_i \partial_j U^{(1)} \Big)\Big(\partial^i \partial^j U^{(1)} \Big)
 - \frac{16\pi a^2 \bar{\rho}}{3}\; \partial_j  v_{\Rm{N}i}^{(1)}\;\partial^j  v_{\Rm{N}}^{(1)i} \nonumber \\
 &- \frac{16\pi a^2 \bar{\rho}}{3}\; v_{\Rm{N}i}^{(1)}\;\nabla^2  v_{\Rm{N}}^{(1)i} - 8 \Big(\partial_i \nabla^2 U^{(1)} \Big) \partial^i U^{(1)} - \frac{8}{3} U^{(1)} \nabla^4 U^{(1)}  
 \;.
\end{align}
We can now directly insert our  solutions, given here in terms of the initial gravitational potential fluctuation, $\varphi$;
\begin{align}
U^{(1)} = \varphi \; , \qquad \updelta_{\Rm{N}}^{(1)} = \frac{2 \nabla^2 \varphi}{3\mathcal{H}^2}  \;, 
\qquad {\rm and} \qquad
v_{\Rm{N}}^{(1)} = \frac{-2 \varphi}{3\mathcal{H}}\;.
\end{align}
Evaluating this, and using the identity $\nabla^4 (\varphi^2) = 2 \nabla^2 \varphi \nabla^2 \varphi + \varphi \nabla^4 \varphi + 4 \partial_i \partial_j \varphi \partial^i \partial^j \varphi + 8 \partial^j \nabla^2 \partial_j \varphi$, we establish
 \begin{align} \label{2ppt2psiphiconstraintX}
 \psi^{(2)} - \phi^{(2)} = -4\varphi^2  -\frac{10}{3} \nabla^{-4} \bigg[  \nabla^2(\nabla \varphi)^2  -  3  \partial_i \partial^j  \big( \partial^i \varphi \partial_j \varphi\big) \bigg]\;.
 \end{align}
This is the same constraint as we obtained in Eq. (\ref{cpt2psiphiconstraintX}) for CPT. 
This is a direct result of the fact that there are no relativistic corrections to any first-order quantities, apart from the density contrast.
 
The third approximation of this problem (although still involved) is \textit{considerably} easier than calculating results in full third-order cosmological perturbation theory, and it is easy to see that there will be interactions between quantities like $\psi^{(2)}$ and $U^{(1)}$ that explicitly demonstrate couplings between long-wavelength relativistic corrections and the linear Newtonian potential. Although such terms also arise naturally in third-order CPT (alongside many other such terms that are neglected in this scheme), in a realistic universe we expect that the terms that arise in the third approximation to 2PPT to be the largest and most relevant ones, as the full two-parameter perturbation theory equations are valid even in universes with highly nonlinear structures on short scales.

\subsection{Evolution of gravitational potentials in 2PPT}
 
The second approximation to the 2PPT evolution equation can be written as
\begin{align} 
\Big(\frac{1}{2}\psi^{(2)}+ \frac{1}{2}U^{(2)}\Big)'' + 3\mathcal{H}\Big(\frac{1}{2}\psi^{(2)} + \frac{1}{2}U^{(2)}\Big)'   =&\;  \frac{4\pi a^2\bar{\rho} }{3}(v_{\Rm{N}}^{(1)})^2 + \mathcal{H}\Big(\frac{1}{2}\psi^{(2)\prime}-\frac{1}{2}\phi^{(2)\prime}\Big) \nonumber \\&    + \frac{7}{6}(\nabla U^{(1)})^2 + \frac{2}{3}(\phi^{(1)} + \psi^{(1)} + 2U^{(1)})\nabla^2 U^{(1)} \nonumber \\
& + \frac{1}{6}\nabla^2 (\psi^{(2)} - \phi^{(2)})\;.  \label{2ndorderevol} 
 \end{align}
Inserting our first approximations, and taking note of the fact that $\psi^{(2)\prime} - \phi^{(2)\prime} = 0$, we recover
 \begin{align} 
(U^{(2)} + \psi^{(2)})'' + 3\mathcal{H}(U^{(2)} + \psi^{(2)})' &=\;  \frac{8\pi a^2\bar{\rho} }{3}(v_{\Rm{N}}^{(1)})^2 + \frac{1}{3}\nabla^2 (\psi^{(2)} - \phi^{(2)})  + \frac{7}{3}(\nabla \varphi)^2 + \frac{8}{3}\varphi \nabla^2 \varphi \;,  \label{2pt2evol} \\ \nonumber
&=  \frac{10}{3} \nabla^{-2} \partial_i \partial^j  \big( \partial^i \varphi \partial_j \varphi\big) - (\nabla \varphi)^2 \;,
 \end{align}
which is an inhomogeneous evolution equation for $(U^{(2)} + \psi^{(2)}) $, and which (given that the relationship between $\psi^{(2)}$ and $\phi^{(2)}$ is the same as in CPT) is of precisely the same form as the second-order CPT evolution equation (\ref{inhomODE}). 

Without assuming anything about $U^{(2)}$, we can solve directly for the combination $(U^{(2)} + \psi^{(2)})$, yielding the solution
\begin{align}
(U^{(2)} + \psi^{(2)})  = (U^{(2)} + \psi^{(2)})_{0} + (U^{(2)} + \psi^{(2)})_P\;,
\end{align}
where $(U^{(2)} + \psi^{(2)})_P$ is the particular solution, found via the same method detailed in Section \ref{cpt2evolution}) to be
\begin{align} \label{2pptSol}
(U^{(2)} + \psi^{(2)})_P = \frac{a(\tau)}{14} \bigg[ \frac{10}{3} \nabla^{-2} \partial_i \partial^j  \big( \partial^i \varphi \partial_j \varphi\big) - (\nabla \varphi)^2 \bigg]\;,
\end{align}
and where $\psi^{(2)}_{0}$ is an initial condition. We can now use the fact that the leading-order 2PPT field equations for $U^{(2)}$ are identical to NPT to find a separate expression for $U^{(2)}$.

Performing a Fourier transform, and using the identity $\nabla^2 \big( \partial^i \varphi \partial_i \varphi \big) = 2 \partial^i\partial^j \varphi \;\partial_i\partial_j \varphi + 2\nabla^2\partial_i \varphi \partial^i \varphi$, it is possible to show that the RHS of Eq. (\ref{2pptSol}) is, in fact, precisely equal to $U^{(2)}$ (as calculated in NPT, using the second-order Newton-Poisson equation to relate $\updelta_{\Rm{N}}^{(2)}(k)$ to $U^{(2)}(k)$). We are therefore left with
\begin{align} 
U^{(2)} = a(\tau)\bigg[ \frac{1}{6} (\nabla \varphi)^2  - \frac{10}{21}\mathbf{\Psi}_0 \bigg] \, ,\qquad \psi^{(2)} =  \psi^{(2)}_{0}\;, \qquad {\rm and} \qquad \phi^{(2)} = \phi^{(2)}_{0} \;,
\end{align}
where $\phi^{(2)}_{0}$ can be obtained from $\psi^{(2)}_{0} $ using the constraint from Equation (\ref{2ppt2psiphiconstraintX}), and $\mathbf{\Psi}_0 $ is the quantity defined in Eq. (\ref{NewtonianKernel}) . This result demonstrates that purely relativistic effects only arise as a result of second-order initial conditions in the second approximation to 2PPT. 

The second approximation to the remaining 2PPT constraint equation,
\begin{align}
\frac{1}{3} \nabla^2 \psi^{(2)}_{0} - \mathcal{H}(\psi_{0}^{(2)\prime} + U^{(2)\prime}) - \mathcal{H}^2(\phi_{0}^{(2)} +U^{(2)} ) &= \frac{4\pi a^2 \bar{\rho}}{3}\updelta^{(2)}  + \frac{8\pi a^2 \bar{\rho}}{3}(v_{\Rm{N}}^{(1)})^2 - (\nabla \varphi)^2 - \frac{8}{3} \varphi \nabla^2 \varphi \;,
\end{align}
allows us to write the second approximation to the density contrast, $\updelta^{(2)}$, in terms of the initial conditions to the potentials $\psi^{(2)}_{0}$ and $\phi^{(2)}_{0}$ as follows:
\begin{align}
\updelta^{(2)} = 2\bigg(\frac{1}{3\mathcal{H}^2}\nabla^2 \psi^{(2)}_{0} - \phi^{(2)}_{0} \bigg) + \frac{10}{9\mathcal{H}^2}(\nabla \varphi)^2 + \frac{16}{3\mathcal{H}^2} \varphi \nabla^2 \varphi - 4 a(\tau) \bigg[ \frac{1}{6} (\nabla \varphi)^2 - \frac{10}{21}\mathbf \Psi_0 \bigg] \;.
\end{align}
Having identified the importance of second-order initial conditions, let us now turn to how these should be calculated.

\subsection{Initial conditions in 2PPT}

In standard cosmological perturbation theory, initial conditions for the growth of structure are usually specified using the curvature perturbation on uniform density hypersurfaces, $\zeta$, which can be connected to the output of various theories of the early Universe (e.g. inflationary models). Different models lead to different parameterisations of the second-order curvature perturbation in terms of the first, which can be written as $\zeta^{(2)} = 2 a_{\Rm{NL}} \zeta^{(1)2}$. One can also calculate $\zeta^{(2)}$ directly, which using the Einstein equations and energy conservation equation gives
\begin{align} \label{cptzetarelation}
\phi^{(2)}_{0} =  - \frac{3}{5} \zeta^{(2)} + \frac{16}{3} \varphi^2 + 2 \nabla^{-4} \bigg[ \nabla^2 (\nabla \varphi)^2  -  3  \partial_i \partial^j  \big( \partial^i \varphi \partial_j \varphi\big) \bigg] \;.
\end{align}
However, when working with 2PPT we must take note of the fact that the second approximation to the 2PPT equations do not have precisely the same structure as the second-order CPT equations. In particular, the $00$-field equation in second-order CPT contains the quadratic source term $-4\mathcal{H}^2 \varphi^2$, which is absent in the second approximation to the corresponding 2PPT equation (due to it being of order $\sim{\eta^6}/{L_{\Rm{N}}^2}$). We must therefore take some care in interpreting quantities like $\psi^{(2)}_{0}$ and $\phi^{(2)}_{0}$, as having a different $00$-field equation implies that Eq. (\ref{cptzetarelation}) is no longer true, and correspondingly the initial conditions may have to be modified.

Our physical interpretation of solving the 2PPT equations in the way outlined in this paper is that it systematically highlights which higher-order terms from regular perturbation theory should amplified by the presence of nonlinear structures at late times in the universe. This means that the second approximation to 2PPT will not contain all terms that appear in full second-order CPT, as certain terms were never present in the full 2PPT system to begin with. We suggest that these terms are the ones that are not prone to being amplified by the presence of nonlinear structures (at least, not ones that can be described using post-Newtonian expansions). As the linearly evolving parts of both the second-order CPT metric scalars and the 2PPT metric scalars are identical, and equal to the second-order Newtonian gravitational potential, the question of appropriate initial conditions for the 2PPT metric scalars would then appear to be most appropriately specified by simply using the initial conditions from second-order CPT. As second-order CPT should be accurate up until the formation of nonlinear structures, any terms that exist within the second-order initial conditions that are ``too small" (in terms of the 2PPT counting scheme) will be sub-dominant to those that 2PPT identifies will be amplified by the presence of nonlinear structure.

Let us formalise this choice. We choose a moment in conformal time, $\tau_{\Rm{cross}}$, which should be deep in the matter-dominated era, but before significant growth of nonlinear structure, which we will refer to as the ``crossover time". At that moment, we switch from using the second-order CPT equations to using the second approximation to the 2PPT equations, which formally allows for some traditional second-order terms to become larger. Our choice of initial conditions is automatically consistent with the second approximation to the 2PPT scalar constraint on $\psi^{(2)}$ and $\phi^{(2)}$, as that constraint is identical to the one in regular cosmological perturbation theory, and has the benefit of ensuring that the metric is continuous at the crossover time. Our choice can therefore be written as
\begin{align}
\phi^{(2)}_{0}(\tau_{\Rm{cross}}) 
&= 2 \bigg[ \varphi^2 + \nabla^{-4} \Big( \nabla^2 (\nabla \varphi)^2  -  3  \partial_i \partial^j  \big( \partial^i \varphi \partial_j \varphi\big) \Big) \bigg] = 2\varphi^2 + 12 \,\mathbf{\Theta}_0 \;, \\
\psi^{(2)}_{0}(\tau_{\Rm{cross}}) 
&= 2 \bigg[ -\varphi^2 - \frac{2}{3} \nabla^{-4} \Big( \nabla^2 (\nabla \varphi)^2  -  3  \partial_i \partial^j  \big( \partial^i \varphi \partial_j \varphi\big) \Big) \bigg] = -2\varphi^2 - 8 \,\mathbf{\Theta}_0  \;,
\end{align}
where $\mathbf{\Theta}_0$ is the quantity defined in Eq. (\ref{GRKernel}). This choice ensures that $g_{\mu\nu}(\tau_{\Rm{cross}})^{\Rm{2PPT}} =  g_{\mu\nu}(\tau_{\Rm{cross}})^{\Rm{CPT}}$, at the expense of the appearance of a negligible discontinuity in the second approximation to the 2PPT dark matter density contrast and peculiar velocity. 

Directly calculating the second approximation to the 2PPT dark matter density contrast, using the second approximation to the 2PPT field equations, we obtain
\begin{align} \label{delta2ppt}
\updelta^{(2)} = -4\varphi^2 - 24\; \mathbf{\Theta}_0  +\bigg[ -\frac{22}{9\mathcal{H}^2}  (\nabla \varphi)^2 + \frac{8}{3\mathcal{H}^2}\varphi \nabla^2 \varphi +  \frac{16}{7 \mathcal{H}^2} \mathbf{\Psi}_0 \bigg] \;,
\end{align}
which is very similar (but not identical) to the relativistic correction to the second-order density contrast in regular CPT. The difference arises due to the $00$-field equation in the second approximation to 2PPT and second-order CPT taking different forms; Specifically, the term $-4\mathcal{H}^2 \varphi^2$ in the regular second-order CPT $00$-field equation is no longer present in the second approximation to 2PPT, which results in a net change in sign for the term $-4\varphi$ in Eq. (\ref{delta2ppt}), as compared to Eq. (\ref{secondorderdelta}). We note that this may not be the only choice available for initial conditions in 2PPT, and that it may be possible that other choices could arise from repeating the calculation of second-order initial conditions performed in \cite{Bartolo:2003gh}, but using the second approximation to the 2PPT equations instead of the full second-order Einstein equations. We leave this calculation to a follow-up paper \cite{Gallagher:2019lcd}.

\subsection{Statistics in 2PPT}
 
We now arrive at the question of calculating statistics using 2PPT. We will use the intrinsic dark matter bispectrum as our example statistic, as it is one of the easiest to calculate. One can write the expression for this as
\begin{align}
(2\pi)^3 \delta^{(3)}(\Bf{k}_1+ \Bf{k}_2+ \Bf{k}_3) B_{\Rm{2PPT}}(k_1,k_2,k_3) 
&\simeq \bigg\langle \big(\updelta_{\Rm{N}} + \updelta \big) \big(\updelta_{\Rm{N}} + \updelta \big)\big(\updelta_{\Rm{N}} + \updelta \big)\bigg\rangle \nonumber  \\
&\simeq \bigg\langle \big(\updelta_{\Rm{N}}^{(1)} + \updelta_{\Rm{N}}^{(2)} + \dots  + \updelta^{(1)} + \updelta^{(2)}   + \dots \big)^3 \bigg\rangle \;.
\end{align}
It is easy to see that at leading order this reduces to
\begin{align}
(2\pi)^3 \delta^{(3)}(\Bf{k}_1+ \Bf{k}_2+ \Bf{k}_3) B_{\Rm{2PPT}}(k_1,k_2,k_3) =  (2\pi)^3 \delta^{(3)}(\Bf{k}_1+ & \Bf{k}_2+ \Bf{k}_3) \times \nonumber \\
&\Big( B_{\rm N}(k_1,k_2,k_3) + B_{\rm R}(k_1,k_2,k_3) \Big)\;,
\end{align}
where $B_{\rm N}$ is the Newtonian bispectrum, $B_{\rm R}$ is a relativistic correction. $B_{\Rm{2PPT}}$ is given by
\begin{align}
(2\pi)^3 \delta^{(3)}(\Bf{k}_1+ \Bf{k}_2+ \Bf{k}_3) B_{\Rm{2PPT}}(k_1,k_2,k_3) =  \big \langle \updelta_{\rm 2PPT}^{(1)}(\Bf{k}_1)\updelta_{\rm 2PPT}^{(1)}(\Bf{k}_2)\updelta_{\rm 2PPT}^{(2)}(\Bf{k}_3) \big \rangle + \;\;\Rm{2}\;\Rm{cycl.}\;\Rm{perms} \;,
\end{align}
where $\updelta_{\rm 2PPT}^{(n)}\equiv \updelta_{\Rm{N}}^{(n)} + \updelta^{(n)}$. Doing this allows us to calculate the modified 2PPT kernel, defined implicitly by 
\begin{align}
\updelta_{\Rm{2PPT}}^{(2)}(k) =  \int \;\frac{\Rm{d}^3 q_1\Rm{d}^3 q_2}{(2\pi)^3} \;\mathcal{K}^{(2)}_{\rm 2PPT}(\Bf{q}_1,\Bf{q}_2,\tau)\; \updelta^{(1)}_{\Rm{2PPT}}(\Bf{q}_1,\tau)\;\updelta^{(1)}_{\Rm{2PPT}}(\Bf{q}_2,\tau)\;,
\end{align}
which can be used in a modified set of Feynman rules, where instead of using the relativistic kernel at each vertex we use $\mathcal{K}^{(2)}_{\rm 2PPT}$. We can then just read off the following expression from Figure \ref{BispectrumFdiagram},
%
%
\begin{align}
B_{\rm 2PPT}(k_1,k_2,k_3) = \mathcal{K}^{(2)}_{\rm 2PPT}(\Bf{k}_1,\Bf{k}_2)P_{\Rm{2PPT}}(k_1)P_{\Rm{2PPT}}(k_2) + \;\Rm{2}\;\Rm{cycl.}\; \Rm{perms} \dots \;,
\end{align}
where $P_{\Rm{2PPT}}(k)$ is the tree level two-parameter matter power spectrum, defined in the usual way by
\begin{align}
 (2\pi)^3 \delta^{(3)}(\Bf{k}_1+ \Bf{k}_2) P_{\Rm{2PPT}}(k_1) =  \Big \langle \big(\updelta_{\Rm{N}}^{(1)}(\Bf{k}_1) + \updelta^{(1)}(\Bf{k}_1)\big)  \big(\updelta_{\Rm{N}}^{(1)}(\Bf{k}_2) + \updelta^{(1)}(\Bf{k}_2)\big) \Big \rangle  \;.
\end{align}

All that remains is then to directly calculate the 2PPT kernel for the dark matter density contrast. Starting with Eq. (\ref{delta2ppt}), we take a Fourier transform to obtain
\begin{align}
\updelta^{(2)}_{\rm 2PPT}(k) = \int \Rm{d}^3q_1 &\,\Rm{d}^3q_2 \,\delta^{(3)}(\Bf{k}-\Bf{q}_1 - \Bf{q}_2) \, \varphi(\Bf{q}_1) \, \varphi(\Bf{q}_2)  \,\nonumber \\ \times\Bigg[& - 4 - \frac{4 \, \Bf{q}_1\cdot \Bf{q}_2}{k^2} + \frac{22\,\Bf{q}_1\cdot \Bf{q}_2}{9\mathcal{H}^2}  - \frac{4}{3\mathcal{H}^2}(q_1^2 + q_2^2) 
 \nonumber \\ &  + \frac{12}{k^4} \Big( q_1^2q_2^2  + (q_1^2 + q_2^2)(\Bf{q}_1\cdot \Bf{q}_2) + (\Bf{q}_1\cdot \Bf{q}_2)^2\Big)+ \frac{8}{7\mathcal{H}^2k^2}\big(q_1^2q_2^2 - (\Bf{q}_1\cdot \Bf{q}_2)^2\big) 
 \nonumber \\ &+ \frac{4}{9\mathcal{H}^4}\Big(\frac{10}{7} q_1^2 q_2^2 +  (q_1^2 + q_2^2)(\Bf{q}_1\cdot \Bf{q}_2)  + \frac{4}{7}(\Bf{q}_1\cdot \Bf{q}_2)^2 \Big)    \Bigg]\;.
\end{align}
In order to get this expression into the required form, we can use
\begin{align} \displaystyle
\updelta^{(1)}_{\rm 2PPT} = \updelta_{\Rm{N}}^{(1)} + \updelta^{(1)} = \frac{2 \nabla^2 \varphi}{3\mathcal{H}^2} - 2\varphi \;, 
\end{align}
which implies
\begin{align}
\varphi(k)  =& -\frac{3\mathcal{H}^2}{2k^2}\bigg( 1 + \frac{3\mathcal{H}^2}{k^2}\bigg)^{-1} \updelta^{(1)}_{\rm 2PPT}(k) \;, 
\end{align}
to relate $\varphi(q_1)$ and $\varphi(q_2)$  to $\updelta^{(1)}_{\rm 2PPT}(q_1) $ and  $\updelta^{(1)}_{\rm 2PPT}(q_2) $, yielding the final expression for the second-order 2PPT matter density kernel
\begin{align}
\mathcal{K}^{(2)}_{\rm 2PPT}(\Bf{q}_1,\Bf{q}_2,k,\tau) = \frac{9\mathcal{H}^4}{4\,q_1^2q_2^2}&\bigg( 1 + \frac{3\mathcal{H}^2}{q_1^2}\bigg)^{-1} \bigg( 1 + \frac{3\mathcal{H}^2}{q_2^2}\bigg)^{-1}  \\
 \times\Bigg[& - 4 - \frac{4 \, \Bf{q}_1\cdot \Bf{q}_2}{k^2} + \frac{22\,\Bf{q}_1\cdot \Bf{q}_2}{9\mathcal{H}^2}  - \frac{4}{3\mathcal{H}^2}(q_1^2 + q_2^2)  
 \nonumber \\ &+ \frac{12}{k^4} \Big( q_1^2q_2^2  + (q_1^2 + q_2^2)(\Bf{q}_1\cdot \Bf{q}_2) + (\Bf{q}_1\cdot \Bf{q}_2)^2\Big)  \nonumber \\& + \frac{8}{7\mathcal{H}^2k^2}\big(q_1^2q_2^2 - (\Bf{q}_1\cdot \Bf{q}_2)^2\big)
 \nonumber \\ &+ \frac{4}{9\mathcal{H}^4}\Big(\frac{10}{7} q_1^2 q_2^2 +  (q_1^2 + q_2^2)(\Bf{q}_1\cdot \Bf{q}_2)  + \frac{4}{7}(\Bf{q}_1\cdot \Bf{q}_2)^2 \Big)    \Bigg] \;.
\end{align}
This equation can be written more compactly as
\begin{align} \label{Tram2pkernel}
\mathcal{K}^{(2)}_{\rm 2PPT}(\Bf{q}_1,\Bf{q}_2,k,\tau) = &\; \frac{1}{\Big(1 + 3 \frac{\mathcal{H}^2}{q_1^2}\Big)\Big(1 + 3 \frac{\mathcal{H}^2}{q_2^2}\Big)} \Bigg[\Big( \beta_{\rm 2PPT}(k,\tau) - \alpha_{\rm 2PPT}(k,\tau)\Big)  \nonumber \\&+ \frac{\beta_{\rm 2PPT}(k,\tau)}{2} \hat{\Bf{q}}_1 \cdot \hat{\Bf{q}}_2 \Big( \frac{q_1}{q_2} + \frac{q_2}{q_1} \Big)+ \alpha_{\rm 2PPT}(k,\tau) \Big(\hat{\Bf{q}}_1 \cdot \hat{\Bf{q}}_2\Big)^2  \nonumber \\&+ \gamma_{\rm 2PPT}(k,\tau) \Big( \frac{q_1}{q_2} - \frac{q_2}{q_1} \Big)^2\Bigg] \;,
\end{align}
where
\begin{align}
\alpha_{\rm 2PPT} &= \frac{2}{7} + \frac{59 \, \mathcal{H}^2}{14\, k^2} - \frac{27 \, \mathcal{H}^4}{14\, k^4}  \;, \\
\beta_{\rm 2PPT} &= 1 - \frac{ \mathcal{H}^2}{2\, k^2} - \frac{18 \, \mathcal{H}^4}{ k^4} \;,  \\ 
\gamma_{\rm 2PPT} & = - \frac{3 \, \mathcal{H}^2}{2\, k^2} - \frac{9 \, \mathcal{H}^4}{2\, k^4}  \; .
\end{align}
In deriving these expressions, which are now in a form similar to the ones used by Tram et. al. in Ref. \cite{Tram:2016cpy}, we have made use of the following identities:
\begin{align}
\frac{1}{q_1q_2} =&\; \frac{1}{k^2}\bigg(\frac{q_1}{q_2} + \frac{q_2}{q_1} \bigg) +  \frac{2}{k^2} \hat{\Bf{q}}_1 \cdot \hat{\Bf{q}}_2 \;, \\
\frac{1}{q^2_1q^2_2} = & \; \frac{1}{k^4}\Bigg(4+ \bigg(\frac{q_1}{q_2} - \frac{q_2}{q_1}\bigg)^2 +4 \, \hat{\Bf{q}}_1 \cdot \hat{\Bf{q}}_2\bigg(\frac{q_1}{q_2} + \frac{q_2}{q_1}\bigg) + 4 (\hat{\Bf{q}}_1 \cdot \hat{\Bf{q}}_2)^2  \Bigg)   \nonumber \;.
\end{align}
The functions, $\alpha_{\rm 2PPT}$,  $\beta_{\rm 2PPT}$ and  $\gamma_{\rm 2PPT}$ encode relativistic corrections in powers of $ \displaystyle \frac{\mathcal{H}^2}{k^2}$. Comparison of the 2PPT coefficient functions with those from standard second-order CPT and reveals that differences arise at $\displaystyle \mathcal{O}\Big(\frac{\mathcal{H}^4}{k^4}\Big)$, i.e at extremely large scales.

\begin{figure}
\centering
\includegraphics[width=0.8\linewidth]{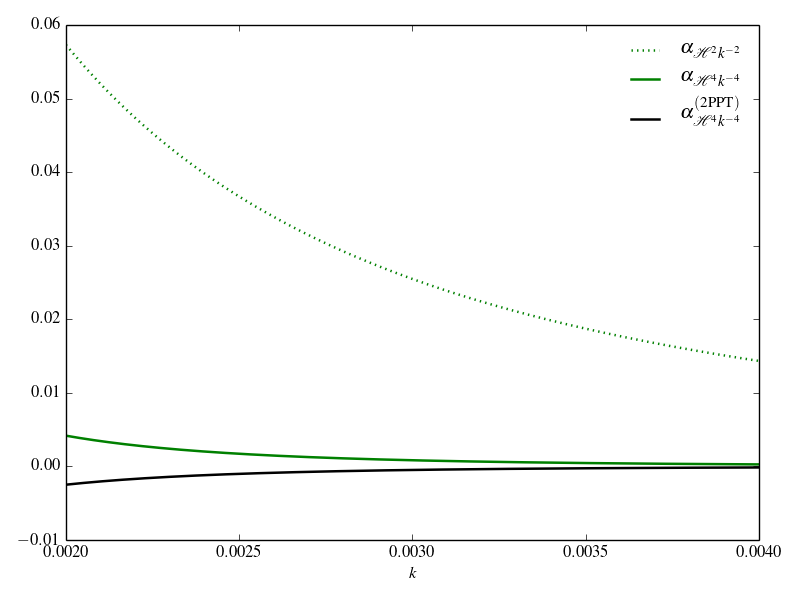}
\caption{
Comparison of terms in $\alpha$ and $\alpha_{\rm 2PPT}$ which scale as $\mathcal{H}^2k^{-2}$ and $\mathcal{H}^4k^{-4}$ respectively at the scale of interest. }
\label{AlphaComparison}
\end{figure}

\begin{figure}
\centering
\includegraphics[width=0.8\linewidth]{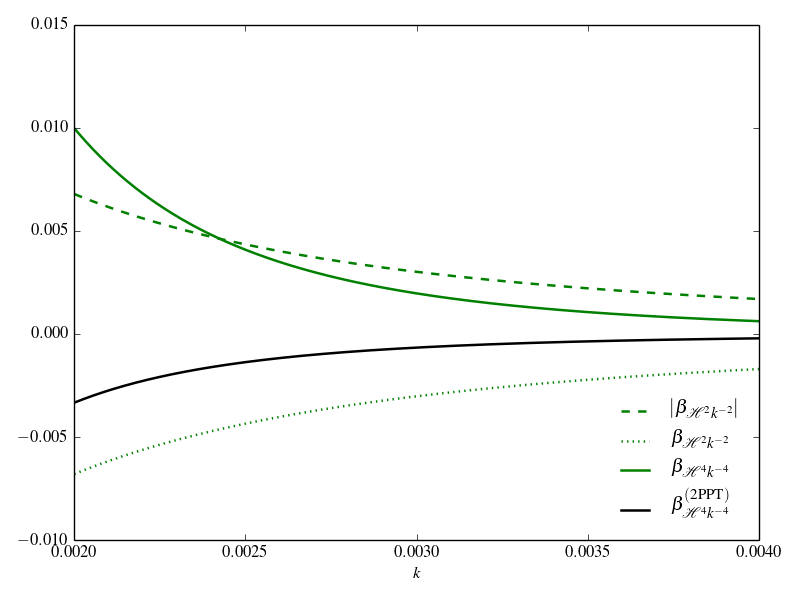}
\caption{
Comparison of terms in $\beta$ and $\beta_{\rm 2PPT}$ which scale as $\mathcal{H}^2k^{-2}$ and $\mathcal{H}^4k^{-4}$ respectively at the second scale of interest. }
\label{BetaComparison}
\end{figure}

\begin{figure}
\centering
\includegraphics[width=0.8\linewidth]{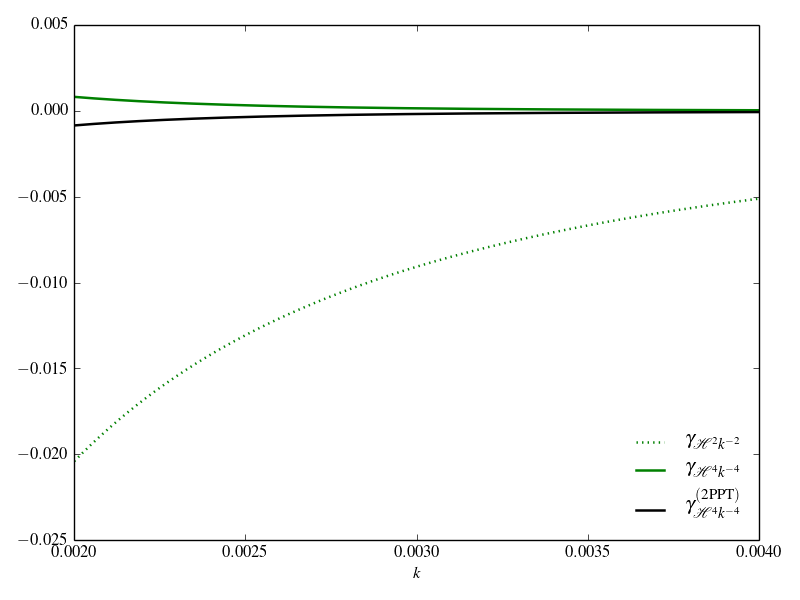}
\caption{
Comparison of terms in $\gamma$ and $\gamma_{\rm 2PPT}$ which scale as $\mathcal{H}^2k^{-2}$ and $\mathcal{H}^4k^{-4}$ respectively at the scale of interest. }
\label{GammaComparison}
\end{figure}

The scale dependent terms in these functions are plotted in Figure \ref{AlphaComparison}, Figure \ref{BetaComparison} and Figure \ref{GammaComparison} and are compared to the equivalent terms in $\alpha$,  $\beta$ and  $\gamma$ in second order relativistic perturbation theory.
It is clear from Figures \ref{AlphaComparison} and \ref{GammaComparison} that terms scaling as $\frac{\mathcal{H}^2}{k^2}$ (plotted as green dots) remain an order of magnitude larger than those scaling as $\frac{\mathcal{H}^4}{k^4}$ (plotted as the solid green line for $\alpha$ and as the solid black line for $\alpha_{\rm 2PPT}$ in both $\alpha$ and $\alpha_{\rm 2PPT}$, at least down to scales of $k \sim 0.003 \; \Rm{Mpc}^{-1}$. We can therefore confidently expect the difference in $\alpha$ and $\alpha_{\rm 2PPT}$ to be extremely small, at least down to these scales, and consequently that the approximation of this function is very good.

In the case of the Figure \ref{BetaComparison}, we see that the magnitude of the terms in $\beta$ that scale as  $\frac{\mathcal{H}^4}{k^4}$ (the green solid line) becomes larger than the magnitude of the terms that scale as $\frac{\mathcal{H}^2}{k^2}$ (the green dashed line) below scales of $k \sim 0.0026$. Accordingly, below these scales, $\beta_{\rm 2PPT}$ will begin to significantly mis-estimate $\beta$. 
It is worth noting though, that the terms scaling as $\frac{\mathcal{H}^2}{k^2}$ in the functions $\beta$ and $\beta_{\rm 2PPT}$ at the scale $k \sim 0.003 \; \Rm{Mpc}^{-1}$ are still an order of magnitude smaller than the terms scaling as $\frac{\mathcal{H}^2}{k^2}$ in the functions $\alpha$, $\alpha_{\rm 2PPT}$ and $\gamma$, $\gamma_{\rm 2PPT}$, hence the resulting error in the density kernel $\mathcal{K}^{(2)}_{\rm 2PPT}(\Bf{k}_1,\Bf{k}_2,\Bf{k}_3)$ should still be extremely small, provided that none of the magnitudes of the arguments $k_1$, $k_2$, $k_3$ are smaller than  $k \sim 0.003 \; \Rm{Mpc}^{-1}$.

\begin{figure}
\centering
\includegraphics[width=0.8\linewidth]{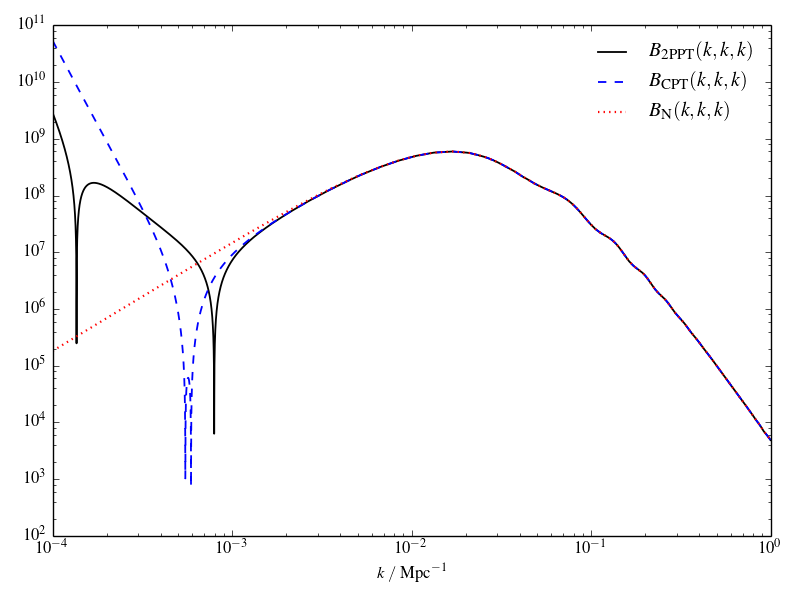}
\caption{
The absolute value of the tree-level bispectrum induced by gravity for the equilateral configuration $B(k,k,k)$, in 2PPT, CPT and NPT.
}
\label{2pBispectrum1}
\centering
\includegraphics[width=0.8\linewidth]{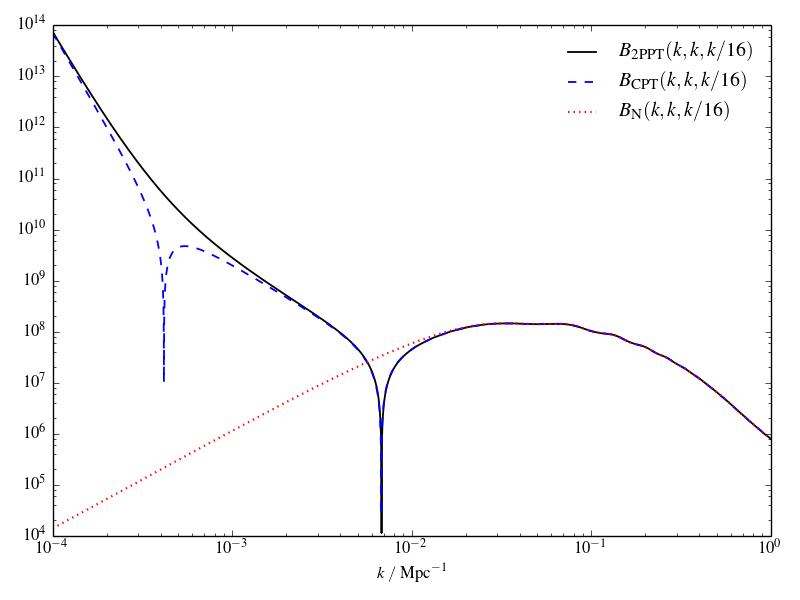}
\caption{
The absolute value of the tree-level bispectrum induced by gravity for the squeezed configuration $B(k,k,k/16)$, in 2PPT, CPT and NPT.
}
\label{2pBispectrum2}
\end{figure}

\begin{figure}
\centering
\includegraphics[width=0.8\linewidth]{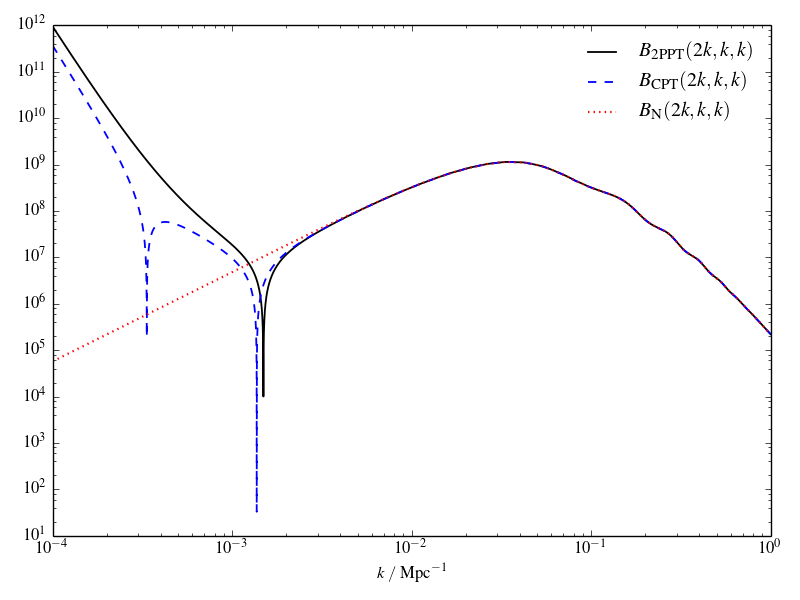}
\caption{
The absolute value of the tree-level bispectrum induced by gravity for the flattened configuration $B(k,k/2,k/2)$, in 2PPT, CPT and NPT.
}
\label{2pBispectrum3}
\end{figure}

\begin{figure}
\centering
\includegraphics[width=0.8\linewidth]{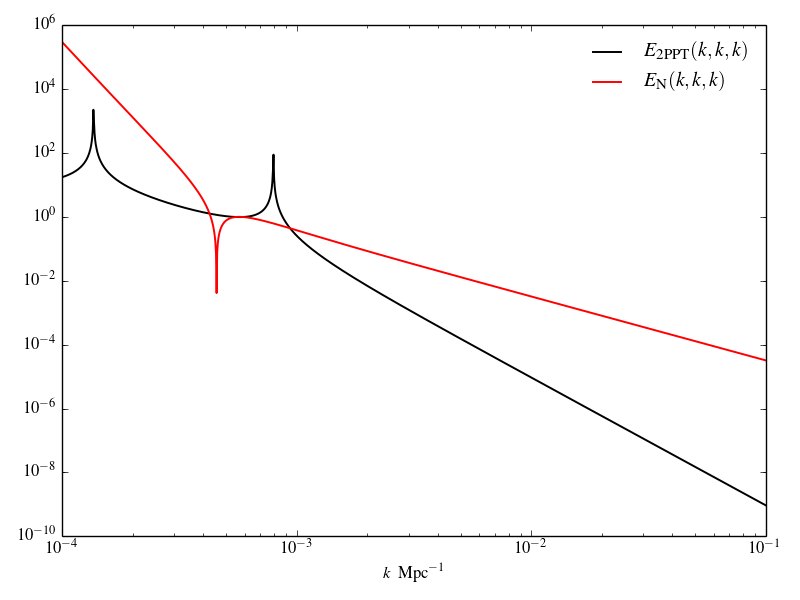}
\caption{The difference statistics $E_{\rm 2PPT}(k,k,k)$ and $E_{\rm N}(k,k,k)$, for equilateral configuration.}
\label{2pError1}
\end{figure}

\begin{figure}
\centering
\includegraphics[width=0.8\linewidth]{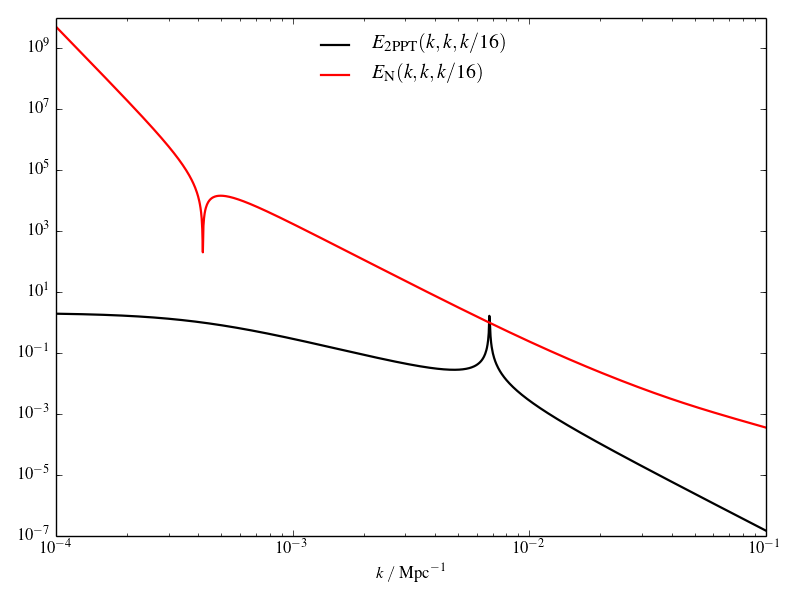}
\caption{
The difference statistics $E_{\rm 2PPT}(k,k,k/16)$ and $E_{\rm N}(k,k,k/16)$, for the squeezed configuration.}
\label{2pError2}
\end{figure}

\begin{figure}
\centering
\includegraphics[width=0.8\linewidth]{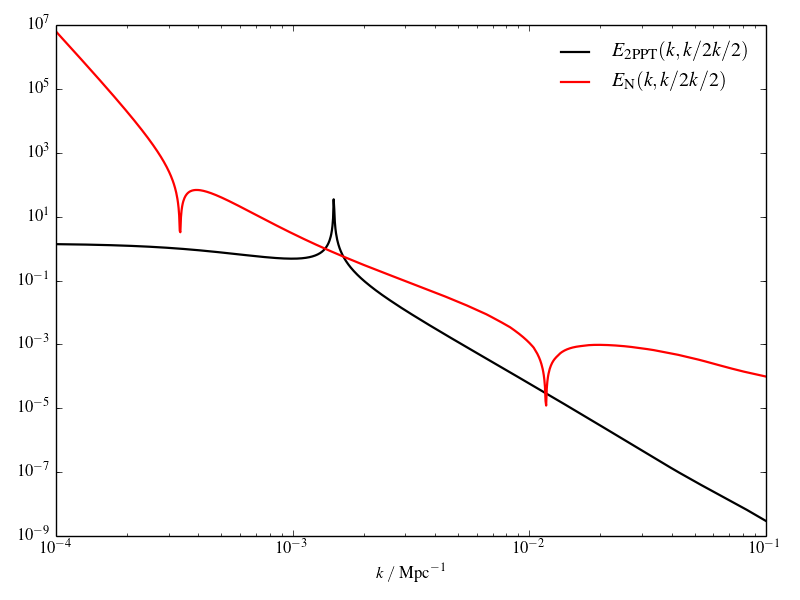}
\caption{
The difference statistics $E_{\rm 2PPT}(k,k/2,k/2)$ and $E_{\rm N}(k,k/2,k/2)$, for the flattened configuration.}
\label{2pError3}
\end{figure}

 The bispectra for equilateral, squeezed and flattened configurations are shown in Figs. \ref{2pBispectrum1}-\ref{2pBispectrum3}, respectively, along with the results from second-order CPT and NPT.
We can define the scale-dependent relative difference between the CPT bispectrum and the bispectra found using 2PPT as
\begin{align}
E_{\rm 2PPT}(\Bf{k}_1,\Bf{k}_2,\Bf{k}_3) &= \Bigg|\frac{B_{\rm CPT}(\Bf{k}_1,\Bf{k}_2,\Bf{k}_3) - B_{\rm 2PPT}(\Bf{k}_1,\Bf{k}_2,\Bf{k}_3)}{B_{\rm 2PPT}(\Bf{k}_1,\Bf{k}_2,\Bf{k}_3) } \Bigg| \;,
\end{align}
and also the relative difference between the CPT bispectrum and NPT bispectrum,
\begin{align}
E_{\Rm{N}}(\Bf{k}_1,\Bf{k}_2,\Bf{k}_3) &= \Bigg|\frac{B_{\rm CPT}(\Bf{k}_1,\Bf{k}_2,\Bf{k}_3) - B_{\rm N}(\Bf{k}_1,\Bf{k}_2,\Bf{k}_3)}{B_{\rm N}(\Bf{k}_1,\Bf{k}_2,\Bf{k}_3) } \Bigg| \, .
\end{align}
These functions tell us how close the 2PPT and NPT predictions are to the standard results from CPT - the smaller the value of $E_{\Rm{2PPT}}(\Bf{k}_1,\Bf{k}_2,\Bf{k}_3)$ and $E_{\Rm{N}}(\Bf{k}_1,\Bf{k}_2,\Bf{k}_3)$, the closer the result is to second-order CPT. These differences are shown in Figs. \ref{2pError1}, \ref{2pError2} and \ref{2pError3} for the equilateral, squeezed and flattened configurations.

\begin{figure}
\centering
\includegraphics[width=0.8\linewidth]{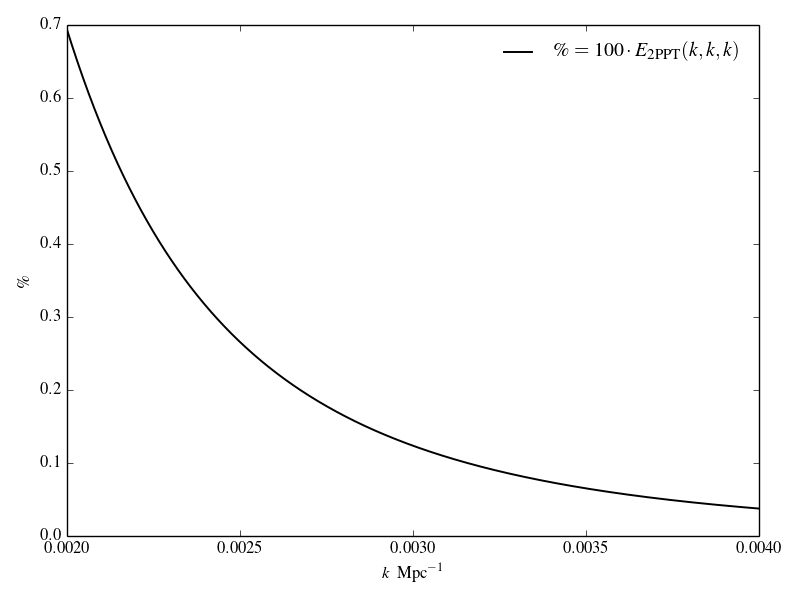}
\caption{
The percentage error compared to cosmological perturbation theory for the equilateral configuration in the regime of interest.}
\label{EqPc}
\end{figure}

Figures \ref{2pError1},  \ref{2pError2} and  \ref{2pError3} span many orders of magnitude. It is useful focus on some particular scales of interest, in particular to highlight where we can expect sub-percent accuracy can be achieved. For this we compute the percentage difference between the 2PPT bispectrum and the CPT bispectrum in various $k$ regimes and configurations, defined as
\begin{align}
\% = E_{\rm 2PPT}(\Bf{k}_1,\Bf{k}_2,\Bf{k}_3) \cdot 100 \;.
\end{align}
Figure \ref{EqPc} demonstrates that the theoretical error remains at sub-percent levels above scales of $k \sim 0.002 \;\Rm{Mpc}^{-1}$ in the equilateral configuration. This is to be contrasted with the Newtonian approximation, for which the theoretical error is orders of magnitude higher at these scales.

\begin{figure}
\centering
\includegraphics[width=0.8\linewidth]{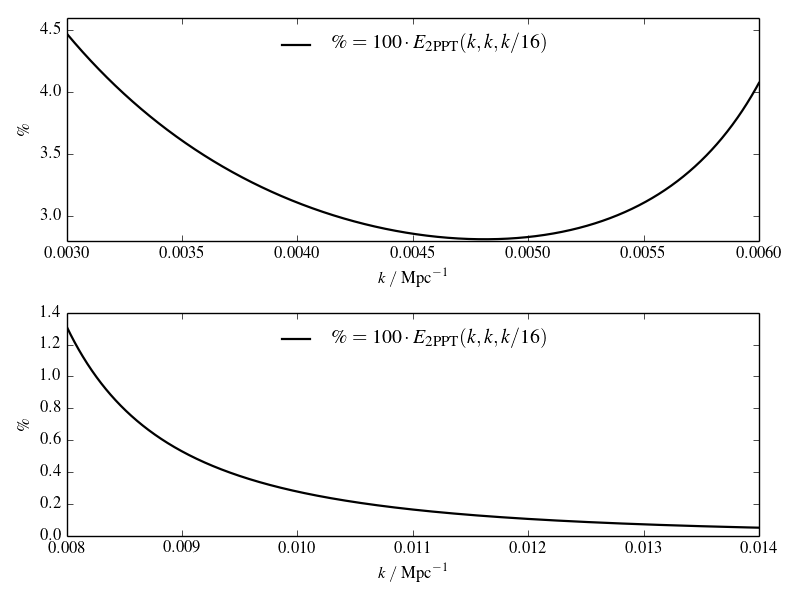}
\caption{
The percentage error compared to cosmological perturbation theory for a moderately squeezed configuration in the two regime of interest, corresponding to scales of  $ 0.003 \; \Rm{Mpc}^{-1} < k <  0.006 \; \Rm{Mpc}^{-1}$ and $k \sim 0.01 \; \Rm{Mpc}^{-1}$, for the top and bottom panels, respectively. }
\label{SqPc}
\end{figure}

The bottom panel of Figure \ref{SqPc} demonstrates that the theoretical error remains at sub-percent levels above scales of $k \sim 0.01 \; \Rm{Mpc}^{-1}$ in the squeezed configuration. This is to be contrasted with the Newtonian approximation, for which the theoretical error is orders of magnitude higher at these scales. The presence of a zero-crossing means that analysing the theoretical error is difficult on larger scales due to the divergence in the denominator.

 

We also focus on the percentage error compared to cosmological perturbation theory for the squeezed configuration in the regime $ 0.003 \; \Rm{Mpc}^{-1} < k <  0.006 \; \Rm{Mpc}^{-1}$ in the top panel of Figure \ref{SqPc}. Whilst the rise in the theoretical error on smaller scales is due to the divergence, the rise in the error on large scales is fundamentally due to the fact that one of the arguments of the bispectrum, $k/16 \sim 0.003/16 \; \Rm{Mpc}^{-1}\sim 0.0002\; \Rm{Mpc}^{-1} $, is being evaluated far outside the regime of applicability of the 2PPT approximation we considered by comparing the sizes of differently scaling terms in the coefficient functions $\alpha_{\rm 2PPT}$,  $\beta_{\rm 2PPT}$,  $\gamma_{\rm 2PPT}$. We therefore conclude that 2PPT will only be useful constructing an order of magnitude estimate of the cosmological perturbation theory bispectrum in the range $0.001 \; \Rm{Mpc}^{-1} < k <  0.01 \; \Rm{Mpc}^{-1}$. The Newtonian approximation however completely fails to predict the correct order of magnitude for the bispectrum in this range of scales, and so 2PPT significantly improves on the results of pure Newtonian perturbation theory.

\begin{figure}
\centering
\includegraphics[width=0.8\linewidth]{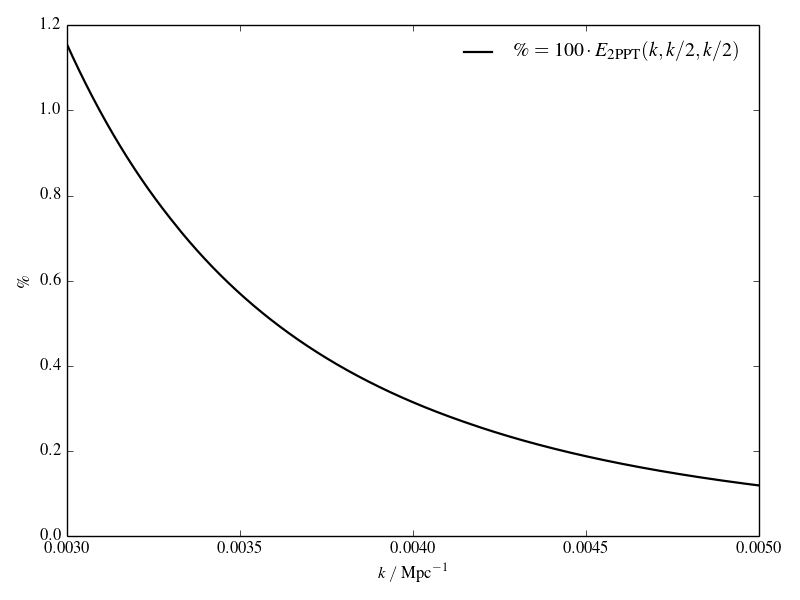}
\caption{
The percentage error compared to cosmological perturbation theory for a flattened configuration in the regime of interest.}
\label{FLPc}
\end{figure}

Figure \ref{FLPc} demonstrates that the theoretical error remains at sub-percent levels above scales of $k \sim 0.0032 \, h \, \Rm{Mpc}^{-3}$ in the flattened configuration. This is to be contrasted with the Newtonian approximation, again for which the theoretical error is orders of magnitude higher at these scales. The 2PPT approximation performs slightly less well than in the case of the equilateral configuration, again due to the presence of smaller values in the argument of the bispectrum, resulting in theoretical error in the values of coefficient function $\beta_{\rm 2PPT}$, compared to $\beta$.

\clearpage

\section{Discussion}

Figures \ref{2pBispectrum1}-\ref{2pBispectrum3} demonstrate that there are significant differences between full second-order CPT and the second approximation to 2PPT. We emphasise that {this is to be expected}, as the field equations in each approach are different. The philosophy of the 2PPT approach is fundamentally different to cosmological perturbation theory, and this difference is illustrated in Figure \ref{flowchart}. Rather than directly trying to approximate the full Einstein equation via linearisation, as is done in regular perturbation theory, the two-parameter expansion is used to derive a different set equations for describing physics on multiple different length scales. The full 2PPT equations, however, contain nonlinear and inhomogeneous elements that make analytic progress difficult, so we have investigated using linearisation of the 2PPT equations to try and learn something about the physics they contain.

We believe it is important to stress that the relative success of the 2PPT scheme should \textit{not} be judged on its proximity to second-order cosmological perturbation theory, since they are approximations to fundamentally different equations. Rather, this paper is an attempt to introduce the reader to a methodology for approximating solutions to the 2PPT equations that is similar to the methodology used in standard cosmological perturbation theory. The reader can be assured that we will always find that full second-order CPT results for large scales can be recovered simply by considering the $\mathcal{O}(\epsilon^2)$ 2PPT quantities, solving for them in an analogous fashion as we have done for the $\mathcal{O}(\epsilon)$ quantities here, and adding the results together. Rather, the second approximation to the 2PPT quantities that we have calculated here should themselves be understood as approximations to the \textit{first-order} equations of cosmological perturbation theory, with corrections due to the existence of nonlinear structure on small scales.

It can be seen in Figures \ref{2pError1}-\ref{2pError3} that differences between the 2PPT bispectrum and the second-order CPT bispectrum are quite small for any scales that are not ultra-large (i.e $k > 10^{-3}\,\Rm{Mpc}^{-1}$), with the possible exception of the flattened case. This is to be expected - the second approximation to 2PPT captures most, but not all of, the terms present in the full second-order field equations, because spatial derivatives of Newtonian terms appear at lower orders than they normally would, but time derivatives do not. We therefore expect that 2PPT will capture most of the interesting relativistic dynamics occurring on intermediate length scales ($0.003\,\Rm{Mpc}^{-1} < k < 10^{-2}\,\Rm{Mpc}^{-1}$). In the flattened case, one can see that differences between the 2PPT result and the second-order CPT result can become quite large around the $k \sim 10^{-3}\, \Rm{Mpc}^{-1}$ scale. This is due to the zero-crossing occurring at slightly different values of $k$ in each case. Both the squeezed and equilateral configurations yield remarkably similar results in 2PPT to second-order CPT down to $k \gtrsim 0.003\, \Rm{Mpc}^{-1}$.  In particular, Figures \ref{2pError1}-\ref{2pError3} shows that 2PPT is at least an order of magnitude closer to the full relativistic second-order CPT result than NPT alone in the region $ 0.003\, \Rm{Mpc}^{-1}< k < 10^{-2}\, \Rm{Mpc}^{-1}$. 

In order to recover the terms most relevant at ultra-large scales, one would need to consider the 2PPT field equations for quantities of order $\sim \epsilon^2$ (i.e. at the level of second-order cosmological perturbations), and solve these equations in a fashion similar to the manner described in the previous sections. If we have to go through this convoluted setup procedure to calculate results that are relatively easily calculated in regular second-order CPT, without all of the information on the ultra-large scales, one might be tempted to ask why 2PPT is necessary at all. The answer is that 2PPT may yield significant advantages when trying to approximate quantities that require \textit{third-order} calculation, at least on intermediate scales. In particular, calculating $P(k)_{\Rm{1-loop}} = P(k)_{22} + P(k)_{13}$ in full relativistic CPT is an extremely challenging proposal, requiring a third-order calculation of $\updelta^{(3)}$ in Poisson gauge. Whilst such calculations have been carried out in comoving gauges, for example in Refs. \cite{Yoo:2016tcz, Jeong:2010ag}, we are not aware of any explicit third order solutions in Poisson gauge in the literature. Of course, formal solutions could be found by performing a third-order gauge transformation on the results of Ref. \cite{Yoo:2016tcz}, but we would like also to emphasise the strength of working directly in a longitudinal gauge when considering nonlinear structures on short scales, as illustrated in Ref. \cite{Clifton:2020oqx}.  2PPT naturally provides a framework in which only the most relevant terms from third-order Poisson gauge theory are included. This may enable an estimation of relativistic effects in the Poisson gauge $P(k)_{\Rm{1-loop}}$ down to $k \sim 0.003 \,\Rm{Mpc}^{-1} $, the scales likely to be accessed by next-generation surveys.

 In order to calculate a third approximation to the full solutions to Equations (\ref{evol}), (\ref{tracefreeij}), (\ref{genPoisson}), (\ref{momentum}), the first quantity we would have to consider would be 
\begin{align}
\frac{1}{3}\nabla^4 \big( \psi^{(3)} - \phi^{(3)}\big) = \mathcal{S}_{4}^{(3)} + \mathcal{S}_{5}^{(3)} + \mathcal{S}_{6}^{(3)} \;.
\end{align}
Equipped with the knowledge that all first approximations to relativistic corrections (apart from $\updelta^{(1)} = - 2\varphi$) are zero, we can simplify the forms of the quadratic source terms $\mathcal{S}^{(3)}_4$, $\mathcal{S}^{(3)}_5$ and $\mathcal{S}^{(3)}_6$ to containing only products of $\{ U^{(2)}, \updelta^{(2)}_N, \theta_N^{(2)}, \psi^{(2)}, \phi^{(2)}, \theta^{(2)}, \updelta^{(2)} \}$ with \\* $\{ U^{(1)}, \updelta^{(1)}_N, \theta_N^{(1)}, \updelta^{(1)}\}$, and the cubic term $\updelta_N^{(1)}v_N^{(1)^2}$.
This situation is simpler than the one arising in full third order Poisson gauge relativistic perturbation theory, where a third order calculation of all field equations and source terms is required. There are a significantly larger number of these terms, rendering the calculation more cumbersome and difficult. 

There are other difficulties associated with the calculation of relativistic effects in the 1-loop power spectrum aside from simply the number of terms involved. In Newtonian perturbation theory, Galilean invariance is known to guarantee the cancellation of separate IR divergences that appear in the convolution integrals contributing to $P_{22}$ and $P_{13}$.  This cancellation is not guaranteed to happen in relativistic perturbation theory, indeed as is found by the authors in Ref. \cite{Martinez-Carrillo:2019nqt} when considering this calculation in the context of the gradient expansion approach. Without performing the calculation directly we cannot comment on the specifics of IR divergences, however we do anticipate being able implement (at least in a worst-case scenario) a hard cut-off at very small values of $k$.

One must remember that we have calculated only an approximation to 2PPT, in the regime where where linearisation of the Newtonian equations is feasible. In the real late universe, significant nonlinearity should be present on the shortest scales, and one would expect precisely the terms highlighted by 2PPT to be dominant when looking at their effect on intermediate scales. In such a universe, the (all orders) 2PPT equations should be expected to provide a {better} prediction than pure CPT equations (if they could be solved at all), as the 2PPT approach was designed with exactly this problem in mind. One approach to probe the physics described by 2PPT in such a situation would be to use Newtonian N-body simulations to provide accurate nonlinear solutions to all orders for the short-scale physics, and then to solve the resulting 2PPT system numerically. Alternatively, one could look to analytic approximations like HALOFIT to provide expressions for nonlinear leading-order quantities, and then try to solve the resulting 2PPT equations. Going beyond general relativity, we believe the approach based on the 2PPT equations is also much more susceptible to being formulated in terms of parameterized alternative theories of gravity \cite{Sanghai:2016tbi,Clifton:2018cef}. However, we will save these approaches for future studies.

\section{Conclusions}
In this work we have presented a blueprint for finding analytic approximations to the 2PPT equations in an Einstein-de Sitter universe. By taking the well-known formal solutions from NPT in EdS, inserting them into the 2PPT equations, and assuming a corresponding perturbative expansion in the quantities in the 2PPT equations themselves, we are able to perturbatively construct order-by-order approximations to the full 2PPT dynamics, in analogy to the way that regular perturbation theory constructs order-by-order approximations to the dynamics described by the fully nonlinear Einstein equations. It is shown that the evolving part of the second-order solution is described completely by the second-order NPT result, and that any second-order relativistic corrections arise only in the form of initial conditions. 
  
Providing the Newtonian equations can be linearised, we find that the second approximations to solutions of the 2PPT equations are very similar to those of standard second-order cosmological perturbation theory, with difference only arising on ultra-large scales (due to the different structure of the field equations). To highlight the differences between each approach, we have focused on the tree-level bispectrum, one of the simplest statistics to calculate in perturbation theory. We find reasonable agreement for scales with $k \gtrsim 10^{-3} \,\Rm{Mpc}^{-1}$ between the second approximation to 2PPT and second-order CPT, indicating that 2PPT does well at approximating a universe with exclusively linear fluctuations at these scales. Differences arise at ultra-large scales due the fact that we have only considered 2PPT large-scale fluctuations up to order $\sim\epsilon$.
 
Significant further work remains to be done on the subject of two-parameter perturbation theory. In particular, the extension of this linearisation scheme to a full $\Lambda$CDM universe is an immediate priority. We expect greater differences at large scales between second-order cosmological perturbation theory and the second approximation to 2PPT dynamics, due to the larger number of missing terms in the 2PPT equations involving time derivatives. Of direct interest is the calculation of the second-order 2PPT peculiar velocity kernel, which (unlike the density contrast) comes directly into the calculation of the bispectrum of observed galaxy number counts, an important relativistic observable. Furthermore, calculations of induced vector and tensors can be performed using these techniques, analogously to the way in which similar such calculations were performed in Refs. \cite{Villa:2015ppa, Lu:2009cbv, Hwang:2017oxa}. A full 2PPT calculation of the initial conditions, following Ref. \cite{Bartolo:2003gh}, would shed light on the question of appropriate initial conditions to use in 2PPT dynamics, and a two-parameter version of the consistency relation between $n$-point and $(n+1)$-point statistics needs to be derived, and checked \cite{Koyama:2018ttg, Abolhasani:2018gyz, Creminelli:2004yq}.

The most pressing calculation, however, is that of $\updelta_{\rm 2PPT}^{(3)}$. Using the results presented here, such a calculation is eminently feasible, and should be expected to provide a reasonable prediction for the relativistic $P(k)_{\Rm{1-loop}} = P(k)_{22} + P(k)_{13}$, at least down to $k \sim 10^{-3} \Rm{Mpc}^{-1}$. This quantity is of great interest to upcoming ultra-large surveys, since it is one of the easiest quantities to actually measure, and the 2PPT approach shows great promise in providing a framework in which a calculation of this quantity could realistically be performed. Although some large scale information would be missing from such a calculation, the fact that the real late universe has significant nonlinear structures means that the terms highlighted by 2PPT could be larger than anticipated in normal approaches to perturbation theory.

A further quantity of significant interest is $v_{\rm 2PPT}^{(2)}$, as it appears directly in the calculation of the observed galaxy number count overdensity at second-order \cite{Fuentes:2019nel, Bertacca:2014dra, Bertacca:2014wga,  Bertacca:2014hwa, DiDio:2015bua, DiDio:2014lka,Yoo:2014sfa}, and consequently affects cosmological observables such as the bispectrum of observed galaxy number counts \cite{Umeh:2016nuh, Jolicoeur:2017nyt, Jolicoeur:2017eyi, Jolicoeur:2018blf, DiDio:2015bua, DiDio:2014lka}. It would be extremely interesting to compute the percentage effect that using 2PPT rather than second-order perturbation theory gave on strictly relativistic observables such as the dipole in observed galaxy number counts \cite{Clarkson:2018dwn}. Of course, to replicate such a calculation, one would have to carry out a full 2PPT analysis of lightcone projection effects, which will be necessary in the future as comparison with observables is, of course, a highly desirable objective.

More generally, the extension of the techniques described in this paper to include mixed-order quantities like $\updelta^{(1,1)}$ and post-Newtonian quantities like $\updelta^{(0,4)}$ is conceivable, as is modifying the NPT sector to incorporate the renormalised perturbation theory of Crocce and Scoccimaro \cite{Crocce:2005xy} or the EFTofLSS \cite{Carrasco:2012cv}. Comparison to analytic weak-field results like those in Ref. \cite{Castiblanco:2018qsd} and to simulations like gevolution \cite{Adamek:2016zes} may yield insight into both the nature of two-parameter perturbation theory, but also to weak-field general relativity, a scheme close in spirit to that of two-parameter perturbation theory. In particular, it may be necessary to renormalize the background quantities as is done in Ref. \cite{Castiblanco:2018qsd}. Directly solving the full 2PPT equations using numerical techniques may also be a realistic possibility, considering the work of Ref. \cite{Macpherson:2016ict, MacPherson:2018btl}, and the relative simplicity of the 2PPT equations compared to the all-orders Einstein equations, and again, direct comparison to the weak-field results from gevolution should prove most illuminating. Finally, we would like to emphasise that extending the two-parameter setup to modified theories of gravity is also entirely feasible. The existence of consistent parameterizations of alternative theories on both small and large scales \cite{Sanghai:2016tbi,Clifton:2018cef} means that it may be possible to construct a two-parameter setup {\it without} specifying a full set of field equations at all. The approach discussed in this paper would then allow relativistic corrections on cosmological statistics to be inferred in a much wider class of theories.

In conclusion, we hope that two-parameter perturbation theory will provide a new arena in which questions about the nonlinearity of general relativity and its effect on large-scale structure can be effectively investigated. The approach presented in this paper will enable us to start using two-parameter perturbation theory to perform practical calculations, analogous to those that have already been performed using more traditional methods. We hope that the methodology may also influence those working on similar problems using different approaches, such as the weak-field approximation to general relativity, and that the formal derivations of two-parameter perturbation theory may highlight more clearly the assumptions that go into such schemes. The era of precision cosmology is only in its infancy, and new techniques and approaches will become increasingly necessary as the data from next-generation surveys starts to be collected. We hope that two-parameter perturbation theory can find its place amongst the variety of new techniques that will be used.

\section*{Acknowedgements}

We are grateful to Karim Malik, Alkistis Pourtsidou, Julien Larena and Roy Maartens for useful discussions. CG, TC and CC are supported by the STFC under grant ST/P000592/1.

\clearpage

\appendix

\section{Field equations in gauge-invariant variables} \label{FieldEquationsGaugeInvariantVariables}

This appendix contains the full 2PPT field equations in terms of the gauge-invariant variables, including vectors and tensors, as derived in Ref. \cite{Goldberg:2016lcq}. The choice of relations between $\epsilon$, $\eta$, $L_{\Rm{C}}$ and $L_{\Rm{N}}$ is as given in Eq. (\ref{choice}). One can convert between the variables used here and those used in the main body of the paper by using the substitutions:
\bea
\label{new1}
U &\equiv& - {\textstyle \frac{1}{2}} \left( \Phi^{(0,2)} + \Phi^{(1,1)} \right) \;, \\
\phi &\equiv& - {\textstyle \frac{1}{2}} \left( \Phi^{(1,0)} + \Phi^{(1,2)} +{\textstyle \frac{1}{2}}\Phi^{(0,4)} \right) \;, \\
\psi &\equiv& {\textstyle \frac{1}{2}} \left(\Psi^{(1,0)} + \Psi^{(1,2)} + {\textstyle \frac{1}{2}}\Psi^{(0,4)} \right) \;, \\
 S_j &\equiv& -\left( \textbf{B}_j^{(1,0)} +\textbf{B}_j^{(0,3)}+ \textbf{B}_j^{(1,2)} \right) \;, \\
 h_{ij} &\equiv& {\textstyle \frac{1}{4}} \left( \textbf{h}_{ij}^{(1,0)} + \textbf{h}_{ij}^{(1,2)} +{\textstyle \frac{1}{2}}\textbf{h}_{ij}^{(0,4)} \right) \, , 
\eea
and
\bea
\updelta \rho_{\rm N} &\equiv& \delta \mathbf{\rho}^{(0,2)} + \mathbf{\rho}^{(1,1)} \;, \\
\updelta \rho &\equiv& \mathbf{\rho}^{(1,0)} +\mathbf{\rho}^{(1,2)} +{\textstyle \frac{1}{2}}\mathbf{\rho}^{(0,4)} \;, \\
{\rm v}_{{\rm N}i} &\equiv& \mathbf{v}^{(0,1)}_i  \;,\\
{\rm v}_{i} &\equiv& \mathbf{v}^{(1,0)}_i \, .
\label{new10}
\eea
For the situation considered in this paper, we will neglect vector and tensor degrees of freedom, and we will also neglect mixed and post-Newtonian quantities (i.e. those of the form $Q^{(1,1)}$ and $Q^{(0,4)}$).

\subsection{Background-order potentials}

\noindent
The trace-free part of the $ij$-equations at $\mathcal{O}(\eta^2L_{\Rm{N}}^{-2})$ gives
\bea
D_{ij}\left(\Phi^{(0,2)} + \Psi^{(0,2)} \right) - \frac{1}{2}\nabla^2 \mathbf{h}_{ij}^{(0,2)} =0 \, , \label{FINALijnottrace02}
\eea
which implies 
\be
\Phi^{(0,2)} = - \Psi^{(0,2)} \qquad {\rm and} \qquad \mathbf{h}^{(0,2)}_{ij}= 0 \, . \label{psiphihij02}
\ee
The $00$-field equation at $\mathcal{O}(\eta^2L_{\Rm{N}}^{-2})$ can be written as
\bea
 \frac{\ddot{a}}{a} + \frac{1}{6a^2}\nabla^2 \Phi^{(0,2)} = - \frac{4 \pi}{3}  {\mathbf \rho}^{(0,2)} \, , \label{FINAL0002} 
\eea
and the trace of the $ij$-equation at $\mathcal{O}(\eta^2L_{\Rm{N}}^{-2})$ gives
\bea
&\ & \left( \frac{\dot{a}}{a} \right)^2 - \frac{1}{3 a^2} \nabla^2 \Phi^{(0,2)} = \frac{8 \pi}{3} {\mathbf \rho}^{(0,2)} \, ,
\label{FINALij02}
\eea
where we have substituted in the results from Eq. (\ref{FINALijnottrace02}). These equations govern the leading-order part of the gravitational field, at $\mathcal{O}(\eta^2L_{\Rm{N}}^{-2})$. 

\vspace{0.5cm}
\subsection{Vector potentials}

\noindent
The $0i$-field equations at $\mathcal{O}(\eta^3L_{\Rm{N}}^{-2})$ give

\bea
&\ & \nabla^2 {\mathbf B}^{(0,3)}_{i} + {2 }\left(a \dot{\Phi}^{(0,2)} + {\dot{a}}\Phi^{(0,2)}\right)_{,i} = 16 \pi a^2  {\mathbf \rho}^{(0,2)} {\mathbf v}^{(0,1)}_i \, .  \label{FINAL0i03} 
\eea
Although ${\mathbf B}^{(0,3)}_{i}$ is a divergenceless vector, Eq. (\ref{FINAL0i03}) has a divergenceless vector and scalar part, which can be separated out with a derivative. At $\mathcal{O}(\eta^4L_{\Rm{N}}^{-2})$ the $0i$-field equations give
\begin{align} \nonumber
&\nabla^2 \left({\mathbf B}^{(1,0)}_{i} +{\mathbf B}^{(1,2)}_{i}\right) +2 \big( a \left(\Phi^{(1,1)}-\Psi^{(1,0)} \dot{\Big)\;}+ \dot{a}\left(\Phi^{(1,1)}+ \Phi^{(1,0)} \right)   \right)_{,i} \nonumber \\
& -2 \left( 2 {\dot{a}^2} + a {\ddot{a}} \right) {\mathbf B}^{(1,0)}_{i} - {\mathbf B}^{(1,0)}_{j}\Phi^{(0,2)}_{,ij} \label{FINAL0i04} \nonumber \\[5pt]
= & \; 8 \pi a^2  \left(2 \mathbf{\rho}^{(0,2)} \mathbf{v}^{(1,0)}_i +  2\mathbf{\rho}^{(1,1)}\mathbf{v}^{(0,1)}_i + \mathbf{\rho}^{(0,2)}  \mathbf{B}^{(1,0)}_{i} \right)  \, , \; \; \; \; \; \; 
\end{align}

\noindent
which can also be split into scalar and divergenceless vector part using a derivative. The reader may note that the quadratic term, which includes the lower-order potential $\Phi^{(0,2)}$, does not source the vector part of Eq. (\ref{FINAL0i04}).

\subsection{Higher-order scalar potentials}

\noindent
The $00$-field equation and the trace of the $ij$-field equation at $\mathcal{O}(\epsilon \eta L_{\Rm{N}}^{-2})$ gives 
\bea
&\ &   \nabla^2 \Phi^{(1,1)} = - 8 \pi a^2  {\mathbf \rho}^{(1,1)}\, , \label{FINAL0011} 
\eea
which implies
\be
\Phi^{(1,1)} = - \Psi^{(1,1)} \, . \label{condition11} 
\ee
Using the $00$-field equation at $\mathcal{O}(\eta^4L_{\Rm{N}}^{-2})$ gives
\begin{align}
& \nabla^2 \left( \Phi^{(1,0)} + \frac{1}{2}\Phi^{(0,4)} + \Phi^{(1,2)}\right) + \left(\nabla \Phi^{(0,2)}\right)^2
+{3 a \dot{a}}\big(3\Phi^{(0,2)} + \Phi^{(1,0)} -2 \Psi^{(1,0)}\dot{\big)\;} \nonumber \\[5pt]
& + {3 a^2}\big(\Phi^{(0,2)}- \Psi^{(1,0)} \ddot{\big)\;} 
 -\nabla^2\Phi^{(0,2)}\left( \Phi^{(0,2)} -\Psi^{(1,0)}\right) - \frac{1}{2} \Phi^{(0,2)}_{,ij}\mathbf{h}^{(1,0)}_{ij} \nonumber \\[5pt]
=& -8 \pi a^2  \left[ \mathbf{\rho}^{(1,0)}  + \mathbf{\rho}^{(1,2)} + \frac{1}{2}\mathbf{\rho}^{(0,4)} - \mathbf{\rho}^{(0,2)}  \left(\Phi^{(1,0)} + \Phi^{(0,2)} \right) \right]  -16\pi a^2 \left( \mathbf{v}^{(0,1)}_i \right)^2  \mathbf{\rho}^{(0,2)}   \, ,
\label{FINAL0004} 
\end{align}
while the trace of the $ij$-field equation at $\mathcal{O}(\eta^4L_{\Rm{N}}^{-2})$ gives
\begin{align}
&- 2 \nabla^2 \left(\Psi^{(1,0)} + \Psi^{(1,2)} + \frac{1}{2}\Psi^{(0,4)}\right) 
- 3\left(2 \dot{a}^2 + a \ddot{a}\right)\left(\Phi^{(1,0)} - \Psi^{(1,0)} + 2\Phi^{(0,2)}\right) 
+ 6\dot{a}a\big(\Psi^{(1,0)} -\Phi^{(0,2)} \dot{\big)\;}
 \nonumber \\[5pt]
=&\; -16 \pi a^2 \left[   \mathbf{\rho}^{(1,0)} +  \frac{1}{2}\mathbf{\rho}^{(0,4)} + \mathbf{\rho}^{(1,2)} + \mathbf{\rho}^{(0,2)}\left( \mathbf{v}^{(0,1)}_i \right)^2  \right]  -4 \pi a^2 \left[ 2\Phi^{(0,2)} \mathbf{\rho}^{(0,2)} - \mathbf{\rho}^{(0,2)}\left( \Phi^{(1,0)} +3\Psi^{(1,0)} \right) \right] \nonumber \\ &+ \mathcal{A}\;,\qquad \label{FINALijtrace04}
\end{align}
where
\bea
\mathcal{A} &\equiv& \nabla^2 \Phi^{(0,2)}\left(3\Phi^{(0,2)} + \frac{1}{2}\Phi^{(1,0)} - \frac{5}{2}\Psi^{(1,0)}\right) + \frac{3}{2}\left(\nabla \Phi^{(0,2)}\right)^2 + \frac{1}{2}\Phi^{(0,2)}_{,ij}\mathbf{h}^{(1,0)}_{ij} \, . 
\label{FINALA}
\eea
These are all of the scalar equations that exist up to $\mathcal{O}(\eta^4L_{\Rm{N}}^{-2})$.

\subsection{Tensor potentials}

\noindent
The trace-free part of the $ij$-field equation at $\mathcal{O}(\epsilon \eta L_{\Rm{N}}^{-2})$ is
\bea
D_{ij}\left(\Phi^{(1,1)} + \Psi^{(1,1)} \right) - \frac{1}{2}\nabla^2 \mathbf{h}_{ij}^{(1,1)} =0 \, , \label{FINALij11}
\eea
which implies 
\be
\Phi^{(1,1)} = - \Psi^{(1,1)} \qquad {\rm and} \qquad \mathbf{h}^{(1,1)}_{ij}= 0 \, . \label{psiphihij11}
\ee
Unlike in the case of $\Psi^{(0,2)}$ and $\Phi^{(0,2)}$, the first of these conditions has already been given by the $00$-field equation and the trace of the $ij-$field equations (\ref{condition11}). Finally, the $\mathcal{O}(\eta^4L_{\Rm{N}}^{-2})$ part of the $ij$-field equation can be used to write
\bea
&& - D_{ij}\left(\Phi^{(1,0)} + \Phi^{(1,2)} + \frac{1}{2}\Phi^{(0,4)} + \Psi^{(1,0)} + \Psi^{(1,2)} + \frac{1}{2}\Psi^{(0,4)}\right) 
 + \frac{1}{2} \nabla^2 \left(\mathbf{h}_{ij}^{(1,0)} + \mathbf{h}_{ij}^{(1,2)} + \frac{1}{2}\mathbf{h}_{ij}^{(0,4)}\right) \nonumber \\[5pt]
&& +\frac{2}{a}\left[ a^2\left(\mathbf{B}_{(i,j)}^{(0,3)} + \mathbf{B}_{(i,j)}^{(1,0)}  \right) \right]\dot{} - \left(2\dot{a}^2 + a\ddot{a}\right)\mathbf{h}^{(1,0)}_{ij} 
 -\frac{3}{2} a\dot{a}\dot{\mathbf{h}}_{ij}^{(1,0)} - \frac{1}{2} a^2  \ddot{\mathbf{h}}_{ij}^{(1,0)}  \nonumber \\[5pt]
& =&  -4\pi a^2 \left[  {\mathbf \rho}^{(0,2)}  \mathbf{h}^{(1,0)}_{ij}+ 4 \mathbf{\rho}^{(0,2)} {\mathbf v}^{(0,1)}_{\langle i} {\mathbf v}^{(0,1)}_{j\rangle} \right] + \mathcal{B}_{ij}  \, , \; \; \; 
\label{FINALijtracefree04}
\eea
where 
\bea
\mathcal{B}_{ij} &\equiv& D_{ij}\Phi^{(0,2)}\left(2\Phi^{(0,2)} + \Phi^{(1,0)} -\Psi^{(1,0)}\right)  
+  \Phi^{(0,2)}_{,\langle i}\Phi^{(0,2)}_{, j \rangle} -  \Phi^{(0,2)}_{,k \langle i }\mathbf{h}^{(1,0)}_{j\rangle k} \, ,
\label{FINALBij}
\eea
and where we have used Eq. (\ref{psiphihij11}). Note that, unlike standard cosmological perturbation theory,  $\Phi^{(1,0)} \neq -\Psi^{(1,0)}$ and $\mathbf{h}_{ij}^{(1,0)} \neq 0$. Furthermore, scalar, vector and tensor modes do not decouple at linear order in cosmological perturbations.

\section{Side-by-side comparison of the second approximation to 2PPT and second-order CPT} \label{appendixlong}

In this section we will present a direct comparison between second-order CPT equations and the second approximation to the 2PPT field equations, obtained by neglecting vectors, tensors, mixed and post-Newtonian quantities in the field equations presented in Appendix \ref{FieldEquationsGaugeInvariantVariables}, and applying the substitutions from Eqs. (\ref{new1})-(\ref{new10}), followed by expanding all quantities in a series in the initial condition, $\varphi$.

\subsection{Second-order Poisson gauge cosmological perturbation theory equations}

The second-order Poisson gauge CPT equations can be calculated directly using the Mathematica package \textit{xPand} \cite{Pitrou:2013hga}. They are:
 \begin{align} 
&\psi^{(2)\prime\prime} + 3\mathcal{H}\psi^{(2)'} =\;  \frac{8\pi a^2\bar{\rho} }{3}v_1^2 + \mathcal{H}(\psi^{(2)'}-\phi^{(2)'})  + \frac{1}{3}\nabla^2 (\psi^{(2)} - \phi^{(2)})  + \frac{7}{3}(\nabla \varphi)^2 + \frac{8}{3}\varphi \nabla^2 \varphi \;,  \label{cpt2evol} 
\\
&\frac{1}{2}\partial^i \partial_j (\psi^{(2)} - \phi^{(2)}) + 2 \partial^i \varphi \partial_j \varphi + 4 \varphi \partial^i \partial_j \varphi - \frac{1}{3} \delta^i_{\;j} \bigg[ \frac{1}{2}\nabla^2  (\psi^{(2)} - \phi^{(2)}) + 2  ( \nabla \varphi )^2  + 4\varphi \nabla^2 \varphi \bigg] \nonumber \\
&\quad\qquad \qquad=\; 8\pi a^2\bar{\rho} \;\big( v^{(1)i} v_{j}^{(1)} - \frac{1}{3}  \delta^i_{\;j}  v^{(1)2} \big) \label{cpt2tracefreeij} \;,
\\
 \label{cpt2psiphiconstraint}
& \psi^{(2)} - \phi^{(2)} = -4\varphi^2  -\frac{10}{3} \nabla^{-4} \bigg[ \nabla^2 (\nabla \varphi)^2  -  3  \partial_i \partial^j  \big( \partial^i \varphi \partial_j \varphi\big) \bigg]\;, 
\\
 \label{cpt200}
 &\frac{1}{3}\nabla^2 \psi^{(2)} - \mathcal{H} \psi^{(2)\prime} - \mathcal{H}^2 \phi^{(2)} = \frac{4\pi a^2 \bar{\rho}}{3} \updelta^{(2)} + \frac{8\pi a^2 \bar{\rho}}{3}v^{(1)2} - (\nabla \varphi)^2 - 4\mathcal{H}^2 \varphi^2 - \frac{8}{3}\varphi \nabla^2 \varphi \;,
 \end{align}

\subsection{Second approximation to the 2PPT field equations}

The second approximation to the 2PPT field equations can be obtained by inserting the $\varphi$ series decomposition into the full 2PPT field equations, cancelling away the first approximation, and neglecting cubic and higher-order products. This gives
\begin{align} 
\Big(\frac{1}{2}\psi^{(2)}+ \frac{1}{2}U^{(2)}\Big)'' + 3\mathcal{H}\Big(\frac{1}{2}\psi^{(2)} + \frac{1}{2}U^{(2)}\Big)' =&\;  \frac{4\pi a^2\bar{\rho} }{3}(v_{\Rm{N}}^{(1)})^2 + \mathcal{H}\Big(\frac{1}{2}\psi^{(2)\prime}-\frac{1}{2}\phi^{(2)\prime}\Big)  + \frac{1}{6}\nabla^2 (\psi^{(2)} - \phi^{(2)}) \nonumber \\ 
 &    + \frac{7}{6}(\nabla U^{(1)})^2 + \frac{2}{3}(\phi^{(1)} + \psi^{(1)} + 2U^{(1)})\nabla^2 U^{(1)} \;.  \label{2ndorderevol} 
 \end{align}
To find the second approximation to the constraint on $\psi - \phi$, it is necessary to consider not just the second approximation to the trace free $ij$-field equation, but to first take two divergences, making sure to include any higher-order terms that appear, and only then insert the series in $\varphi$. Performing this procedure, the result takes the form
\begin{align}
\frac{1}{3}\nabla^4(\psi^{(2)} - \phi^{(2)}) = \mathcal{S}^{(2)}_4 + \mathcal{S}^{(2)}_5 + \mathcal{S}^{(2)}_6 \;, 
\end{align}
 where each of the source terms can be written
\begin{align}
\mathcal{S}^{(2)}_4 = &\; 16\pi a^2 \;\bar{\rho}\; v_{\Rm{N}i}^{(1)}\; \partial^i \theta^{(1)}_{\Rm{N}} + 8\pi a^2 \bar{\rho} (\theta^{(1)}_{\Rm{N}})^2 + 8\pi a^2 \bar{\rho}\; \partial_i  v_{\Rm{N}j}^{(1)}\;\partial^j  v_{\Rm{N}}^{(1)i} \nonumber \\
 &- \frac{2}{3} (\nabla^2 U^{(1)})^2 - \frac{14}{3} \partial_i \partial_j U^{(1)} \partial^i \partial^j U^{(1)}   - \frac{16\pi a^2 \bar{\rho}}{3}\; \partial_j  v_{\Rm{N}i}^{(1)}\;\partial^j  v_{\Rm{N}}^{(1)i} - \frac{16\pi a^2 \bar{\rho}}{3}\; v_{\Rm{N}i}^{(1)}\;\nabla^2  v_{\Rm{N}}^{(1)i} \nonumber \\
 & - 8 \partial_i \nabla^2 U^{(1)} \partial^i U^{(1)} - \frac{8}{3} U^{(1)} \nabla^4 U^{(1)}  - \frac{8}{3} \psi^{(1)} \nabla^4 U^{(1)}  
 \;,  \\
 \mathcal{S}^{(2)}_5 = &\; 16\pi a^2 \;\bar{\rho}\; v_{\Rm{N}i}^{(1)}\; \partial^i \theta^{(1)} - 8 \partial_i \nabla^2 U^{(1)} \partial^i \psi^{(1)} - \frac{16\pi a^2 \bar{\rho}}{3}\; v_{i}^{(1)}\;\nabla^2  v_{\Rm{N}}^{(1)i}   
 \;, \\
  \mathcal{S}^{(2)}_6 =  &\;  8\pi a^2 \bar{\rho} \;\theta^{(1)}_{\Rm{N}}\;\theta^{(1)} + 8\pi a^2 \bar{\rho}\; \partial_i  v_{\Rm{N}j}^{(1)}\;\partial^j  v^{(1)i}  - \frac{2}{3} \nabla^2 U^{(1)}\nabla^2 \psi^{(1)} \nonumber \\
  &- \frac{14}{3} \partial_i \partial_j U^{(1)} \partial^i \partial^j \psi^{(1)} - \frac{16\pi a^2 \bar{\rho}}{3}\; \partial_j  v_{\Rm{N}i}^{(1)}\;\partial^j  v^{(1)i}  
 \; ,
\end{align}

 It is demonstrated in the main body of the paper that the constraint reduces to:
 \begin{align} \label{2ppt2psiphiconstraint}
 \psi^{(2)} - \phi^{(2)} = -4\varphi^2  -\frac{10}{3} \nabla^{-4} \bigg[  \nabla^2(\nabla \varphi)^2  -  3  \partial_i \partial^j  \big( \partial^i \varphi \partial_j \varphi\big) \bigg]\;.
 \end{align}
The second approximation to the $00$-field equation is given by
 \begin{align}
\frac{1}{3} \nabla^2 \psi^{(2)} - \mathcal{H}(\psi^{(2)\prime} + U^{(2)\prime}) - \mathcal{H}^2(\phi^{(2)}+U^{(2)}) &= \frac{4\pi a^2 \bar{\rho}}{3}\updelta^{(2)} + \frac{4\pi a^2 \bar{\rho}}{3}(v^{(1)}_{\Rm{N}})^2 \nonumber \\& - \frac{1}{2} (\nabla U^{(1)})^2 - \frac{4}{3}(\psi^{(1)} + U^{(1)}) \nabla^2 U^{(1)} \;.
\end{align}

\subsection{Poisson gauge CPT solutions}

\subsubsection{First-order solutions}
The first-order solutions are
\begin{align}
\phi^{(1)} = \psi^{(1)} = \varphi \;, \\
\updelta^{(1)} = \frac{2 \nabla^2\varphi}{3\mathcal{H}^2} - 2\varphi \;, \\
v^{(1)} =   -\frac{2 \nabla^2\varphi}{3\mathcal{H}}\;.
\end{align}

\subsubsection{Second-order solutions}

The second-order solutions for the potential and density contrast are
\begin{align}
\phi^{(2)} =& 2 \varphi^2 + 12  \, \mathbf{\Theta}_0 +\frac{4}{\mathcal{H}^2}\bigg[ \frac{1}{6} (\nabla \varphi)^2  - \frac{10}{21}\mathbf{\Psi}_0 \bigg] \;, \\
\psi^{(2)} =& -2 \varphi^2 - 8 \, \mathbf{\Theta}_0 + \frac{4}{\mathcal{H}^2}\bigg[ \frac{1}{6} (\nabla \varphi)^2  - \frac{10}{21}\mathbf{\Psi}_0 \bigg] \;,
\\
 \updelta^{(2)} =& 4 \, \varphi^2 - 24 \,\mathbf{\Theta}_0 \\&+\bigg[ -\frac{22}{9\mathcal{H}^2}  (\nabla \varphi)^2 + \frac{8}{3\mathcal{H}^2}\varphi \nabla^2 \varphi +  \frac{16}{7 \mathcal{H}^2} \mathbf{\Psi}_0 \bigg] \nonumber \\
 &+ \frac{4}{9 \mathcal{H}^4}  \bigg[ \frac{10}{7} (\nabla^2\varphi)^2 + 2 \nabla^2\partial_i \varphi \partial^i \varphi+ \frac{4}{7} \partial_i \partial_j \varphi \partial^i \partial^j \varphi \bigg] \;, \nonumber
 \end{align}
where we have used the following useful shorthand quantities:
\begin{align}
\mathbf{\Theta}_0 = \frac{1}{2} \nabla^{-4}\bigg[ \frac{1}{3} \nabla^2 (\partial^i \varphi \partial_i \varphi) - \partial_i\partial^j(\partial^i \varphi \partial_j \varphi)  \bigg] \;,
\end{align}
and
\begin{align}
\mathbf{\Psi}_0 = - \frac{1}{2} \nabla^{-2} \bigg[(\nabla^2 \varphi)^2 - \partial_i\partial_j \varphi \,\partial^i\partial^j \varphi \bigg] \;.
\end{align}

\subsection{Approximations to 2PPT solutions}

\subsubsection{First approximations}

The first approximations to the solutions are
\begin{align}
U^{(1)} &= \varphi \;, \\
\phi^{(1)} &= \psi^{(1)} = 0 \;, \\
\updelta^{(1)}_{\Rm{N}} &= \frac{2 \nabla^2\varphi}{3\mathcal{H}^2} \;, \\
\updelta^{(1)} &= -2\varphi \;, \\
v_{\Rm{N}}^{(1)} &=   -\frac{2 \nabla^2\varphi}{3\mathcal{H}}\; \\
v^{(1)} &= 0 
\end{align}

\subsubsection{Second approximations}

The second approximation to the solutions are

\begin{align}
U^{(2)} &= \frac{4}{\mathcal{H}^2}\bigg[ \frac{1}{6} (\nabla \varphi)^2  - \frac{10}{21}\mathbf{\Psi}_0 \bigg] \;, \\
\phi^{(2)} &= 2 \varphi^2 + 12  \, \mathbf{\Theta}_0  \;, \\
\psi^{(2)} &= -2 \varphi^2 - 8 \, \mathbf{\Theta}_0  \;, \\
 \updelta_{\Rm{N}}^{(2)} =& \frac{4}{9 \mathcal{H}^4}  \bigg[ \frac{10}{7} (\nabla^2\varphi)^2 + 2 \nabla^2\partial_i \varphi \partial^i \varphi+ \frac{4}{7} \partial_i \partial_j \varphi \partial^i \partial^j \varphi \bigg] \;, \\
 \updelta^{(2)} =& -4 \, \varphi^2 - 24 \,\mathbf{\Theta}_0 \\&+\bigg[ -\frac{22}{9\mathcal{H}^2}  (\nabla \varphi)^2 + \frac{8}{3\mathcal{H}^2}\varphi \nabla^2 \varphi +  \frac{16}{7 \mathcal{H}^2} \mathbf{\Psi}_0 \bigg] \;. \nonumber
 \end{align}

\section{Determining the NPT Kernels} \label{PTkernels}

In order to determine the perturbation theory kernels, we Fourier transform the nonlinear equations to get
\begin{align} 
\updelta_{\Rm{N}}^{\prime}(\Bf{k},\tau) + \theta_{\Rm{N}}(\Bf{k}, \tau) = &- \int \frac{\Rm{d}^3 p_1\;\Rm{d}^3 p_2}{(2\pi)^3}\;\delta^{(3)}( \Bf{k} - \Bf{p}_1 - \Bf{p}_2)  \; \times \nonumber \\ &\bigg[ \alpha(\Bf{p}_1,\Bf{p}_2) \; \updelta_{\Rm{N}}(\Bf{p}_1,\tau) \; \theta_{\Rm{N}}(\Bf{p}_2,\tau) \bigg]\label{fspacecont}\;,\\
\theta_{\Rm{N}}^{\prime}(\Bf{k},\tau) + \mathcal{H}\theta_{\Rm{N}} (\Bf{k}, \tau) + \frac{3}{2}\mathcal{H}^2 \updelta_{\Rm{N}}(\Bf{k}, \tau) = & - \int \frac{\Rm{d}^3 p_1\;\Rm{d}^3 p_2 }{(2\pi)^3} \;\delta^{(3)}(\Bf{k} - \Bf{p}_1 - \Bf{p}_2) \; \times \nonumber \\ &\bigg[\beta(\Bf{p}_1,\Bf{p}_2) \;\theta_{\Rm{N}}(\Bf{p}_1,\tau) \; \theta^{(1)}_{\Rm{N}}(\Bf{p}_2,\tau) \bigg]\label{fspaceeuler} \;,
\end{align}
where $\alpha$ and $\beta$ are defined in Section \ref{npt}. We can now insert our perturbation series in the form of Eqs. (\ref{deltaNsplit}) and (\ref{thetaNsplit}) into Eqs. (\ref{fspacecont}) and (\ref{fspaceeuler}). This results in the following (somewhat cumbersome) expressions for the continuity equation:

\begin{align}
&\sum_{n= 1}^{\infty}  \frac{ a^n \mathcal{H}}{n!} \Bigg\{\int \bigg( \prod_{i=1}^n \frac{\Rm{d}^3 k_i}{(2\pi)^{3i}} \updelta_{0}^{(1)}(\Bf{k}_i)   \bigg)(2\pi)^3 \delta^{(3)}\bigg(\Bf{k} - \sum_{i=1}^n \Bf{k}_i\bigg)\;\times \nonumber \\& \qquad \qquad \bigg[n F_{n}(\Bf{k}_1,\Bf{k}_2,...,\Bf{k}_n) - G_n(\Bf{k}_1,\Bf{k}_2,...,\Bf{k}_n)\bigg] \Bigg\} \nonumber \\
=&  \sum_{m= 1}^{\infty} \sum_{l= 1}^{\infty} \frac{ a^m a^l \mathcal{H}}{m! l!}  \Bigg\{ \int \frac{\Rm{d}^3 p_1\;\Rm{d}^3 p_2}{(2\pi)^3}\;\delta^{(3)}( \Bf{k} - \Bf{p}_1 - \Bf{p}_2)  \bigg( \prod_{j=1}^m \frac{\Rm{d}^3 q_{1j}}{(2\pi)^{3j}} \updelta_{0}^{(1)}(\Bf{q}_{1j}) \bigg) \bigg(  \prod_{k=1}^l \frac{\Rm{d}^3 q_{2k}}{(2\pi)^{3k}} \updelta_{0}^{(1)}(\Bf{q}_{2k}) \bigg) \times \nonumber \\
 &(2\pi)^6 \delta^{(3)}\bigg(\Bf{p}_1 - \sum_{j=1}^m \Bf{q}_{1j}\bigg)\delta^{(3)}\bigg(\Bf{p}_2 - \sum_{k=1}^l \Bf{q}_{2k}\bigg) F_m(\Bf{q}_{11},\Bf{q}_{12},...,\Bf{q}_{1m}) G_l(\Bf{q}_{21},\Bf{q}_{22},...,\Bf{q}_{2l})\; \alpha(\Bf{p}_1, \Bf{p}_2) \Bigg\}\;, 
 \end{align}
 and the Euler equation:
 \begin{align}
 &\sum_{n= 1}^{\infty}  \frac{ a^n \mathcal{H}}{n!} \Bigg\{\int \bigg( \prod_{i=1}^n \frac{\Rm{d}^3 k_i}{(2\pi)^{3i}} \updelta_{0}^{(1)}(\Bf{k}_i)   \bigg)(2\pi)^3 \delta^{(3)}\bigg(\Bf{k} - \sum_{i=1}^n \Bf{k}_i\bigg)\; \times \nonumber \\& \qquad \qquad\bigg[(2n+1) G_{n}(\Bf{k}_1,\Bf{k}_2,...,\Bf{k}_n) - 3F_n(\Bf{k}_1,\Bf{k}_2,...,\Bf{k}_n)\bigg] \Bigg\} \nonumber \\
=&  \sum_{m= 1}^{\infty} \sum_{l= 1}^{\infty} \frac{ a^m a^l \mathcal{H}}{m! l!}  \Bigg\{ \int \frac{\Rm{d}^3 p_1\;\Rm{d}^3 p_2}{(2\pi)^3}\;\delta^{(3)}( \Bf{k} - \Bf{p}_1 - \Bf{p}_2)  \bigg( \prod_{j=1}^m \frac{\Rm{d}^3 q_{1j}}{(2\pi)^{3j}} \updelta_{0}^{(1)}(\Bf{q}_{1j}) \bigg) \bigg(  \prod_{k=1}^l \frac{\Rm{d}^3 q_{2k}}{(2\pi)^{3k}} \updelta_{0}^{(1)}(\Bf{q}_{2k}) \bigg) \times \nonumber \\
 &(2\pi)^6 \delta^{(3)}\bigg(\Bf{p}_1 - \sum_{j=1}^m \Bf{q}_{1j}\bigg)\delta^{(3)}\bigg(\Bf{p}_2 - \sum_{k=1}^l \Bf{q}_{2k}\bigg) G_m(\Bf{q}_{11},\Bf{q}_{12},...,\Bf{q}_{1m}) G_l(\Bf{q}_{21},\Bf{q}_{22},...,\Bf{q}_{2l})\; \beta(\Bf{p}_1, \Bf{p}_2) \Bigg\}\;.
\end{align}
Evaluating the integrals over $\Bf{p}_1$ and $\Bf{p}_2$, and selecting the $n^{\Rm{th}}$ term from each expression, it is easy to see that by relabelling the integration variables $\Bf{q}_{1j}$ and $\Bf{q}_{2k}$ as $\Bf{k}_i$, that one can equate the two integrands, and therefore be left with the following purely algebraic expressions for the $n^{\Rm{th}}$-order kernels in terms of products of lower-order kernels:
\begin{align}
n F_n(\Bf{k}_{1...n} ) - G_n(\Bf{k}_{1...n})  &= \sum_{m=1}^{m=n-1} \frac{n!}{m!(n-m)!} \alpha(\Bf{k}_{1:m},\Bf{k}_{m:n}) F_m(\Bf{k}_{1...m})G_{n-m}(\Bf{k}_{m...n}) \; , \\
(2n+1)G_nn(\Bf{k}_{1...n} )  - 3F_n(\Bf{k}_{1...n} ) &= \sum_{m=1}^{m=n-1} \frac{n!}{m!(n-m)!} 2 \beta(\Bf{k}_{1:m},\Bf{k}_{m:n}) G_m(\Bf{k}_{1...m})G_{n-m}(\Bf{k}_{m...n}) \;,
\end{align}
It is important to note that we are free to relabel the integration variables in any manner we choose. This implies that we should symmetrise on the wavevectors $\Bf{k}_i$ since each permutation corresponds to a different relabelling of the integration variables, all of which are equivalent. Generally, it is easiest to perform this procedure at the end of the calculation, so we will leave it until then. It is easy to solve these algebraic equations for $F_n$ and $G_n$ (the unsymmetrised kernels). The resulting expressions are
\begin{align}
F_n(\Bf{k}_{1...n} ) =& \sum_{m=1}^{m=n-1} {n \choose m} \frac{G_{n-m}(\Bf{k}_{m...n}) }{(2n+3)(n-1)}\bigg\{ (2n+1) \alpha(\Bf{k}_{1:m},\Bf{k}_{m:n}) F_m(\Bf{k}_{1...m}) \nonumber \\
&\;\;\;\;\;\;\;\;\;\;\;\;\;\;+2 \beta(\Bf{k}_{1:m},\Bf{k}_{m:n})G _m(\Bf{k}_{1...m})\bigg\}\; , \\
G_n(\Bf{k}_{1...n} )  =& \sum_{m=1}^{m=n-1} {n \choose m} \frac{G_{n-m}(\Bf{k}_{m...n}) }{(2n+3)(n-1)}\bigg\{ 3 \alpha(\Bf{k}_{1:m},\Bf{k}_{m:n}) F_m(\Bf{k}_{1...m}) \nonumber \\
&\;\;\;\;\;\;\;\;\;\;\;\;\;\;+2n \beta(\Bf{k}_{1:m},\Bf{k}_{m:n})G _m(\Bf{k}_{1...m})\bigg\}\; .
\end{align}
The reader will notice an additional factor of ${n \choose m}$ compared to the standard expressions in the literature. These factors come from our choice to include factors on $\displaystyle \frac{1}{n!}$ in the perturbation expansion, so as to match up with the expansions in traditional relativistic perturbation theory. This normalisation choice is purely arbitrary and has no effect on the physics.

\bibliography{1-LoopEdS2}

\begin{thebibliography}{10}

\bibitem{Bardeen:1980gic}
J.~M. Bardeen.
\newblock Gauge-invariant cosmological perturbations.
\newblock {\em Phys. Rev. D}, 22:1882--1905, 1980.

\bibitem{Malik:2008im}
K.~A. Malik and D.~Wands.
\newblock {Cosmological perturbations}.
\newblock {\em Phys. Rept.}, 475:1--51, 2009.

\bibitem{Carlson:2009clc}
J.~Carlson, M.~White, and N.~Padmanabhan.
\newblock {Critical look at cosmological perturbation theory techniques}.
\newblock {\em Phys. Rev.}, D80:043531, 2009.

\bibitem{SKA}
{SKA} telescope.
\newblock \url{www.skatelescope.org}.

\bibitem{Euclid}
{Euclid} satellite.
\newblock \url{www.sci.esa.int/euclid}.

\bibitem{LSST}
{LSST} collaboration.
\newblock \url{www.lsst.org}.

\bibitem{Clifton:2010fr}
T.~Clifton.
\newblock {Cosmology Without Averaging}.
\newblock {\em Class. Quant. Grav.}, 28:164011, 2011.

\bibitem{Clarkson:2011zq}
C.~Clarkson, G.~Ellis, J.~Larena, and O.~Umeh.
\newblock {Does the growth of structure affect our dynamical models of the
  universe? The averaging, backreaction and fitting problems in cosmology}.
\newblock {\em Rept. Prog. Phys.}, 74:112901, 2011.

\bibitem{Baumann:2010tm}
D.~Baumann, A.~Nicolis, and M.~Zaldarriaga. L.~Senatore, Leonardo.
\newblock {Cosmological Non-Linearities as an Effective Fluid}.
\newblock {\em JCAP}, 1207:051, 2012.

\bibitem{Carrasco:2012cv}
J.~J.~M. Carrasco, M.~P. Hertzberg, and L.~Senatore.
\newblock {The Effective Field Theory of Cosmological Large Scale Structures}.
\newblock {\em JHEP}, 09:082, 2012.

\bibitem{Will:1981}
C.~M. Will.
\newblock {\em Theory and experiment in gravitational physics}.
\newblock Cambridge University Press, 1981.

\bibitem{Poisson:2014}
E.~Poisson and C.~M. Will.
\newblock {\em Gravity: Newtonian, Post-Newtonian, Relativistic}.
\newblock Cambridge University Press, 2014.

\bibitem{Goldberg:2016lcq}
S.~R. Goldberg, T.~Clifton, and K.~A. Malik.
\newblock {Cosmology on all scales: a two-parameter perturbation expansion}.
\newblock {\em Phys. Rev.}, D95(4):043503, 2017.

\bibitem{Goldberg:2017gsm}
S.~R. Goldberg, C.~S. Gallagher, and T.~Clifton.
\newblock {Perturbation theory for cosmologies with nonlinear structure}.
\newblock {\em Phys. Rev.}, D96(10):103508, 2017.

\bibitem{Gallagher:2018bdl}
C.~S. Gallagher and T.~Clifton.
\newblock {Relativistic Euler equations in cosmologies with nonlinear
  structures}.
\newblock {\em Phys. Rev.}, D98(10):103516, 2018.

\bibitem{GoldbergThesis}
S.~R. Goldberg.
\newblock {\em {Two-parameter Perturbation Theory for Cosmologies with
  Non-linear Structure}}.
\newblock PhD thesis, Queen Mary, U. of London (main), 2017.

\bibitem{Clifton:2020oqx}
T.~Clifton, C.~S. Gallagher, S.~R. Goldberg, and K.~A. Malik.
\newblock {Viable Gauge Choices in Cosmologies with Non-Linear Structures}.
\newblock 2020.

\bibitem{Gallagher:2019lcd}
C.~S. Gallagher, T.~Clifton, and C.~Clarkson.
\newblock {Multi-Scale Perturbation Theory II: $\Lambda$CDM universes (in
  preparation)}.
\newblock 2019.

\bibitem{Villa:2015ppa}
E.~Villa and C.~Rampf.
\newblock {Relativistic perturbations in $\Lambda$CDM: Eulerian \& Lagrangian
  approaches}.
\newblock {\em JCAP}, 1601(01):030, 2016.
\newblock [Erratum: JCAP1805,no.05,E01(2018)].

\bibitem{Nakamura:2007}
K.~Nakamura.
\newblock {Second-order gauge invariant cosmological perturbation theory:
  Einstein equations in terms of gauge invariant variables}.
\newblock {\em Prog. Theor. Phys.}, 117:17--74, 2007.

\bibitem{Umeh:2016nuh}
O.~Umeh, S.~Jolicoeur, R.~Maartens, and C.~Clarkson.
\newblock {A general relativistic signature in the galaxy bispectrum: the local
  effects of observing on the lightcone}.
\newblock {\em JCAP}, 1703(03):034, 2017.

\bibitem{Jolicoeur:2017nyt}
S.~Jolicoeur, O.~Umeh, R.~Maartens, and C.~Clarkson.
\newblock {Imprints of local lightcone \ projection effects on the galaxy
  bispectrum. Part II}.
\newblock {\em JCAP}, 1709(09):040, 2017.

\bibitem{Jolicoeur:2017eyi}
O.~Umeh S.~Jolicoeur, R.~Maartens, and C.~Clarkson.
\newblock {Imprints of local lightcone projection effects on the galaxy
  bispectrum. Part III. Relativistic corrections from nonlinear dynamical
  evolution on large-scales}.
\newblock {\em JCAP}, 1803(03):036, 2018.

\bibitem{Jolicoeur:2018blf}
S.~Jolicoeur, A.~Allahyari, C.~Clarkson, J.~Larena, O.~Umeh, and R.~Maartens.
\newblock {Imprints of local lightcone projection effects on the galaxy
  bispectrum IV: Second-order vector and tensor contributions}.
\newblock {\em JCAP}, 1903:004, 2019.

\bibitem{Castiblanco:2018qsd}
L.~Castiblanco, R.~Gannouji, J.~Norena, and C.~Stahl.
\newblock {Relativistic cosmological large scale structures at one-loop}.
\newblock {\em JCAP}, 1907(07):030, 2019.

\bibitem{Goroff:1986ep}
M.~H. Goroff, B.~Grinstein, S.~J. Rey, and M.~B. Wise.
\newblock {Coupling of Modes of Cosmological Mass Density Fluctuations}.
\newblock {\em Astrophys. J.}, 311:6--14, 1986.

\bibitem{Bernardeau:2001qr}
F.~Bernardeau, S.~Colombi, E.~Gaztanaga, and R.~Scoccimarro.
\newblock {Large scale structure of the universe and cosmological perturbation
  theory}.
\newblock {\em Phys. Rept.}, 367:1--248, 2002.

\bibitem{Scoccimarro:1996jy}
R.~Scoccimarro.
\newblock {Cosmological perturbations: Entering the nonlinear regime}.
\newblock {\em Astrophys. J.}, 487:1, 1997.

\bibitem{McEwen:2016fjn}
J.~E. McEwen, X.~Fang, C.~M. Hirata, and J.~A. Blazek.
\newblock {FAST-PT: a novel algorithm to calculate convolution integrals in
  cosmological perturbation theory}.
\newblock {\em JCAP}, 1609(09):015, 2016.

\bibitem{Bartolo:2005kv}
N.~Bartolo, S.~Matarrese, and A.~Riotto.
\newblock {The full second-order radiation transfer function for large-scale
  cmb anisotropies}.
\newblock {\em JCAP}, 0605:010, 2006.

\bibitem{Bartolo:2003gh}
N.~Bartolo, S.~Matarrese, and A.~Riotto.
\newblock {Enhancement of non-Gaussianity after inflation}.
\newblock {\em JHEP}, 04:006, 2004.

\bibitem{Bartolo:2001cw}
N.~Bartolo, S.~Matarrese, and A.~Riotto.
\newblock {Nongaussianity from inflation}.
\newblock {\em Phys. Rev.}, D65:103505, 2002.

\bibitem{Tram:2016cpy}
T.~Tram, C.~Fidler, R.~Crittenden, K.~Koyama, G.~W. Pettinari, and D.~Wands.
\newblock {The Intrinsic Matter Bispectrum in $\Lambda$CDM}.
\newblock {\em JCAP}, 1605(05):058, 2016.

\bibitem{Milillo:2015cva}
I.~Milillo, D.~Bertacca, M.~Bruni, and A.~Maselli.
\newblock {Missing link: A nonlinear post-Friedmann framework for small and
  large scales}.
\newblock {\em Phys. Rev.}, D92(2):023519, 2015.

\bibitem{Yoo:2016tcz}
J.~Yoo and J.~Gong.
\newblock {Exact analytic solution for non-linear density fluctuation in a
  $\Lambda$CDM universe}.
\newblock {\em JCAP}, 1607(07):017, 2016.

\bibitem{Jeong:2010ag}
D.~Jeong, J.~Gong, H.~Noh, and J.~Hwang.
\newblock {General relativistic effects on non-linear power spectra}.
\newblock {\em Astrophys. J.}, 727:22, 2011.

\bibitem{Martinez-Carrillo:2019nqt}
Rebeca Martinez-Carrillo, Josue De-Santiago, Juan~Carlos Hidalgo, and Karim~A.
  Malik.
\newblock {Relativistic and non-Gaussianity contributions to the one-loop power
  spectrum}.
\newblock 2019.

\bibitem{Sanghai:2016tbi}
V.~A.~A. Sanghai and T.~Clifton.
\newblock {Parameterized Post-Newtonian Cosmology}.
\newblock {\em Class. Quant. Grav.}, 34(6):065003, 2017.

\bibitem{Clifton:2018cef}
T.~Clifton and V.~A.~A. Sanghai.
\newblock {Parametrizing Theories of Gravity on Large and Small Scales in
  Cosmology}.
\newblock {\em Phys. Rev. Lett.}, 122(1):011301, 2019.

\bibitem{Lu:2009cbv}
T.~Lu, K.~Ananda, C.~Clarkson, and R.~Maartens.
\newblock The cosmological background of vector modes.
\newblock {\em Journal of Cosmology and Astroparticle Physics}, 2009, 01 2009.

\bibitem{Hwang:2017oxa}
J.~Hwang, D.~Jeong, and H.~Noh.
\newblock {Gauge dependence of gravitational waves generated from scalar
  perturbations}.
\newblock {\em Astrophys. J.}, 842(1):46, 2017.

\bibitem{Koyama:2018ttg}
K.~Koyama, O.~Umeh, R.~Maartens, and D.~Bertacca.
\newblock {The observed galaxy bispectrum from single-field inflation in the
  squeezed limit}.
\newblock {\em JCAP}, 1807(07):050, 2018.

\bibitem{Abolhasani:2018gyz}
A.~A. Abolhasani and M.~Sasaki.
\newblock {Single-field consistency relation and $\delta N$-formalism}.
\newblock {\em JCAP}, 1808(08):025, 2018.

\bibitem{Creminelli:2004yq}
P.~Creminelli and M.~Zaldarriaga.
\newblock {Single field consistency relation for the 3-point function}.
\newblock {\em JCAP}, 0410:006, 2004.

\bibitem{Fuentes:2019nel}
J.~L. Fuentes, J.~C. Hidalgo, and K.~A. Malik.
\newblock {Galaxy number counts at second order: an independent approach.
  arXiv:1908.08400}, 2019.

\bibitem{Bertacca:2014dra}
D.~Bertacca, R.~Maartens, and C.~Clarkson.
\newblock {Observed galaxy number counts on the lightcone up to second order:
  I. Main result}.
\newblock {\em JCAP}, 1409(09):037, 2014.

\bibitem{Bertacca:2014wga}
D.~Bertacca, R.~Maartens, and C.~Clarkson.
\newblock {Observed galaxy number counts on the lightcone up to second order:
  II. Derivation}.
\newblock {\em JCAP}, 1411(11):013, 2014.

\bibitem{Bertacca:2014hwa}
D.~Bertacca.
\newblock {Observed galaxy number counts on the light cone up to second order:
  III. Magnification bias}.
\newblock {\em Class. Quant. Grav.}, 32(19):195011, 2015.

\bibitem{DiDio:2015bua}
E.~Di Dio, R.~Durrer, G.~Marozzi, and F.~Montanari.
\newblock {The bispectrum of relativistic galaxy number counts}.
\newblock {\em JCAP}, 1601:016, 2016.

\bibitem{DiDio:2014lka}
E.~Di Dio, R.~Durrer, G.~Marozzi, and F.~Montanari.
\newblock {Galaxy number counts to second order and their bispectrum}.
\newblock {\em JCAP}, 1412:017, 2014.
\newblock [Erratum: JCAP1506,no.06,E01(2015)].

\bibitem{Yoo:2014sfa}
Jaiyul J.~Yoo and M.~Zaldarriaga.
\newblock {Beyond the Linear-Order Relativistic Effect in Galaxy Clustering:
  Second-Order Gauge-Invariant Formalism}.
\newblock {\em Phys. Rev.}, D90(2):023513, 2014.

\bibitem{Clarkson:2018dwn}
C.~Clarkson, E.~de~Weerd, S.~Jolicoeur, R.~Maartens, and O.~Umeh.
\newblock {The dipole of the galaxy bispectrum}.
\newblock {\em Mon. Not. Roy. Astron. Soc.}, 486(1):L101--L104, 2019.

\bibitem{Crocce:2005xy}
M.~Crocce and R.~Scoccimarro.
\newblock {Renormalized cosmological perturbation theory}.
\newblock {\em Phys. Rev.}, D73:063519, 2006.

\bibitem{Adamek:2016zes}
J.~Adamek, D.~Daverio, R.~Durrer, and M.~Kunz.
\newblock {gevolution: a cosmological N-body code based on General Relativity}.
\newblock {\em JCAP}, 1607(07):053, 2016.

\bibitem{Macpherson:2016ict}
H.~Macpherson, P.~Lasky, and D.~J. Price.
\newblock {Inhomogeneous Cosmology with Numerical Relativity}.
\newblock {\em Phys. Rev.}, D95(6):064028, 2017.

\bibitem{MacPherson:2018btl}
H.~Macpherson, D.~J. Price, and P.~Lasky.
\newblock {Einstein's Universe: Cosmological structure formation in numerical
  relativity}.
\newblock {\em Phys. Rev.}, D99(6):063522, 2019.

\bibitem{Pitrou:2013hga}
C.Pitrou, X.~Roy, and O.~Umeh.
\newblock {xPand: An algorithm for perturbing homogeneous cosmologies}.
\newblock {\em Class. Quant. Grav.}, 30:165002, 2013.

\end{thebibliography}

\end{document}